\documentclass[11pt,a4paper]{article}
\pdfoutput=1
\usepackage{jheppub}

\usepackage[normalem]{ulem}
\usepackage{amsmath}
\usepackage{amssymb}
\usepackage{amscd}
\usepackage{enumerate}
\usepackage{amsfonts}
\usepackage{epsfig}
\usepackage{mathtools}
\usepackage{yfonts}
\usepackage{bbold}
\usepackage{breqn}

\definecolor{davecolor}{rgb}{0.95,  0.5,  0.2}

\definecolor{darkgreen}{rgb}{0,0.5,0}
\definecolor{darkblue}{rgb}{0,0,0.6}
\definecolor{purple}{rgb}{0.4,0.15,0.21}
\definecolor{black}{rgb}{.2,.2,.2}

\usepackage{dsfont}

\usepackage{graphicx}
\usepackage{bm}
\usepackage{xcolor}

\definecolor{davecolor}{rgb}{0.95,  0.5,  0.2}

\def\({\left(}
\def\){\right)}
\def\[{\left[}
\def\]{\right]}
\def\<{\langle}
\def\>{\rangle}

\def \f{\frac}

\newcommand{\be}{\begin{equation}}
\newcommand{\ee}{\end{equation}}
\newcommand{\bea}{\begin{eqnarray}}
\newcommand{\eea}{\end{eqnarray}}
\newcommand{\bwt}{\begin{widetext}}
\newcommand{\ewt}{\end{widetext}}

\newcommand{\bi}{\begin{itemize}}
\newcommand{\ei}{\end{itemize}}
\newcommand{\ben}{\begin{enumerate}}
\newcommand{\een}{\end{enumerate}}
\newcommand{\bca}{\begin{cases}}
\newcommand{\eca}{\end{cases}}
\newcommand{\bln}{\begin{align}}
\newcommand{\eln}{\end{align}}
\newcommand{\bst}{\begin{split}}
\newcommand{\est}{\end{split}}

\renewcommand{\Im}{\textrm{Im}\,}
\renewcommand{\Re}{\textrm{Re}\,}

\title{
A General Proof of the Quantum Null Energy Condition}
\author[a]{Srivatsan Balakrishnan, Thomas Faulkner, Zuhair U. Khandker,} 
\author[a,b]{and Huajia Wang}
\affiliation[a]{Department of Physics, University of Illinois, 1110 W. Green St., Urbana IL 61801-3080, U.S.A.} 
\affiliation[b]{Kavli Institute for Theoretical Physics, University of California, Santa Barbara, CA 93106, U.S.A. \vspace{5mm}}

\emailAdd{sblkrsh2@illinois.edu}
\emailAdd{tomf@illinois.edu}
\emailAdd{zuhair@illinois.edu}
\emailAdd{huajia@kitp.ucsb.edu}

\abstract{
We prove a conjectured lower bound on $\left< T_{--}(x) \right>_\psi$ in any state $\psi$ of a relativistic QFT dubbed the Quantum Null Energy Condition (QNEC). The bound is given by the second order shape deformation, in the null direction, of the geometric entanglement entropy of an entangling cut passing through $x$. 
Our proof involves a combination of the two independent methods that were used recently to prove the weaker Averaged Null Energy Condition (ANEC). In particular the properties of modular Hamiltonians under  shape deformations for the state $\psi$ play an important role, as do causality considerations.
We study the two point function of a ``probe'' operator $\mathcal{O}$ in the state $\psi$ and use a lightcone limit to evaluate this correlator. 
Instead of causality in time we consider \emph{causality in modular time} for the modular evolved probe operators, which we constrain using Tomita-Takesaki theory as well as certain generalizations pertaining to the theory of modular inclusions. 
The QNEC follows from very similar considerations to the derivation of the chaos bound and the causality sum rule.
We use a kind of defect Operator Product Expansion to apply the replica trick to these modular flow computations, and the displacement operator plays an important role. Our approach was inspired by the AdS/CFT proof of the QNEC which follows from properties of the Ryu-Takayanagi (RT)  surface near the boundary of AdS, combined with the requirement of entanglement wedge nesting.
Our methods were, as such, designed as a precise probe of the RT surface close to the boundary of a putative gravitational/stringy dual of \emph{any}  QFT 
with an interacting UV fixed point. 
We also prove a higher spin version of the QNEC. }

\arxivnumber{1706.09432}

\begin{document}
\maketitle

\section{Introduction and summary}

\label{sec:intro}

Bounds on the stress tensor $T_{--}$ of a QFT have important consequences for the semi-classical limit of gravity - using these bounds we can rule out pathological spacetimes that might arise when coupling gravity to matter in the form of this QFT. Typically these pathologies have their root in some form of causality violation of the resulting spacetime. However naive bounds that apply to classical field theory, like the local Null Energy Condition (NEC) $T_{--}>0$, are violated quantum mechanically. The  NEC was central to the classical proofs of the black hole area law \cite{hawking1971gravitational}, singularity theorems \cite{penrose1965gravitational}, topological cencorship \cite{Friedman:1993ty}, etc \cite{hawking1973the}. In order to generalize these proofs to the quantum regime several  new energy conditions on $T_{--}$ have been conjectured with various degrees of non-locality \cite{Borde:1987qr,Klinkhammer:1991ki,wald1991general,verch:1999nt,Ford:1994bj,Fewster:2012yh,Wall:2009wi,C:2013uza,Wall:2011hj}. Despite their origin in gravitational physics these generalizations often have a limit which applies directly to the QFT in curved space, and it is furthermore interesting to study their validity and consequences even in flat Minkowski space \cite{Hofman:2008ar,Hofman:2009ug}. Perhaps most excitingly their validity almost always relates to bounds on the behavior and manipulation of quantum information in the QFT \cite{Casini:2008cr,Blanco:2013lea}, further strengthening the important connection between gravity and quantum information.

Recently two different proofs of the Averaged Null Energy Condition (ANEC) in any UV complete QFT have appeared in the literature \cite{Faulkner:2016mzt,Hartman:2016lgu}. In Minkowski space this is the positivity constraint:
\be
\label{intro:anec}
\int_{-\infty}^{\infty} dx^- \left< T_{--}(x^-,x^+=0,y) \right>_\psi \geq 0
\ee
The proof \cite{Faulkner:2016mzt} works in Minkowski space and for null integrals along a complete null geodesic generator of the horizon of a static black hole. This is sufficient to rule out using these black holes as traversable wormholes \cite{Gao:2016bin,Graham:2007va}. The proof of the ANEC in \cite{Hartman:2016lgu} was based on causality considerations applied to the two point function of a probe operator $\mathcal{O}$ evaluated in the state $\left| \psi \right>$. In \cite{Faulkner:2016mzt} the proof was based on monotonicity of relative entropy under shape deformations of an entangling region $A$ which is taken to be a null deformed cut of the Rindler horizon $x^+=0$. This in turn imposed a condition on the negativity of the shape deformations of vacuum modular Hamiltonians for these entangling cuts which was then proven to be related to the ANEC operator in \eqref{intro:anec}. 
Indeed it does not take much more work to use these modular Hamiltonians combined with monotonicity of relative entropy to prove the quantum half-ANEC:
\be
\label{halfANEC}
2\pi\int_{\partial A}^{\infty} dx^- \left< T_{--}(x^-,x^+=0,y) \right>_\psi +\frac{\delta S_{EE}(A)}{\delta x^-(y)} \geq 0
\ee
which is then an important ingredient in the semi-classical proof of the Generalized Second Law for black hole mechanics \cite{Wall:2011hj}. It turns out however that we need an even more local constraint in order to generalize other classical gravitational theorems, such as the Bousso/covariant entropy bound \cite{Bousso:1999xy}, to the semi-classical regime. 
One such constraint is the Quantum Null Energy Condition (QNEC) \cite{Bousso:2015mna,Bousso:2015wca,Wall:2017blw} which is logically more general than the ANEC and the half-ANEC, implying both of these if it is true. For certain special cases the QNEC is the functional derivative $\delta/\delta x^-(y)$, or shape deformation, of \eqref{halfANEC}. Since we don't expect a second order shape deformation of relative entropy to be constrained in sign for a general quantum system, if we are to prove the QNEC it will involve an essentially new ingredient beyond monotonicity. As we will see the QNEC follows from a more fine grained notion of causality compared to the results in \cite{Hartman:2016lgu}, where we make use of the probe operator $\mathcal{O}$, but at the same time add the action of modular Hamiltonians into the mix.

We will prove a (slight) generalization of the QNEC. This is essentially an integrated version:
\be
\label{introq}
Q_-(A,B;y) \equiv \int_{\partial A}^{\partial B} d x^-\left< T_{--}(x^-,x^+=0,y) \right>_\psi  - \frac{1}{2\pi}\frac{\delta S_{EE}(B)}{ \delta x^-(y) } +\frac{1}{2\pi} \frac{\delta S_{EE}(A)}{ \delta x^-(y) } \geq 0
\ee
where $A$ and $B$ are two spatial regions of a fixed Cauchy slice. They should satisfy the inclusion property $\mathcal{D}(B) \subset \mathcal{D}(A)$ where $\mathcal{D}(C)$ is the domain of
dependence of $C$.  The $x^-$ parameterizes a null line passing from the entangling surfaces $\partial A$ to $\partial B$ located at fixed $(x^+=0,y)$ and $y$ locally labels the coordinates along the entangling surface with $x^\pm$ labeling null coordinates transverse to the surface. We will require the surfaces $\partial A , \partial B$ to be locally stationary at the point $y$, which means the extrinsic curvature in one of the null directions $\mathcal{K}^+_{ij} = 0 $ vanishes at $y$ as well as a  sufficient number of its $y$ derivatives. Other than this we only require that the domains of dependences $\mathcal{D}(B)$
and  $\mathcal{D}(\bar{A})$ are non timelike separated, so for example $A,B$ could have multiple disconnected components with non trivial topology etc.

\begin{figure}[h!]
\centering
\includegraphics[scale=1.15]{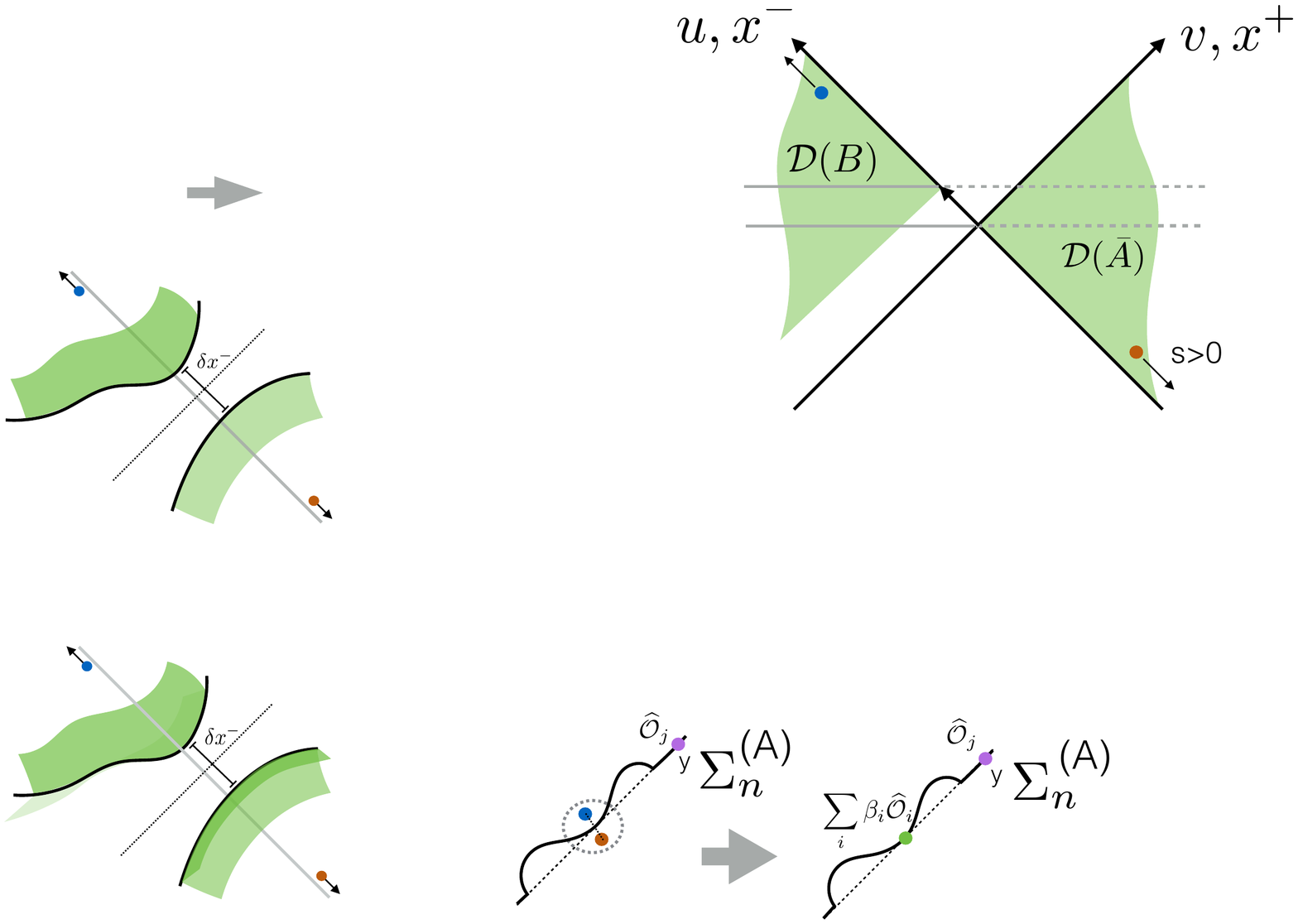}  \hspace{.1cm}
\includegraphics[scale=.78]{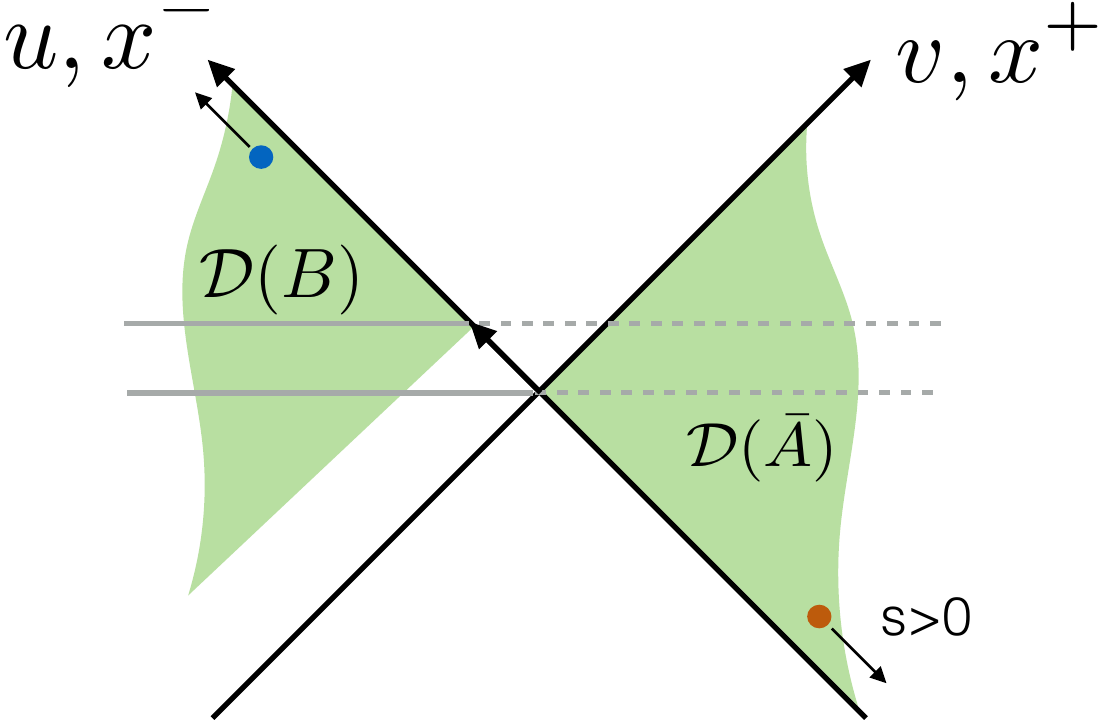}
\caption{Our setup involves two causally disconnected regions of Minkowski space, the domains
of dependence of $B$ and $\bar{A}$ (shaded green regions). These become close to null separated along a null line (gray curve in the left figure) along which we would like to prove the QNEC. The null separation along this line is the coordinate length $\delta x^-$. We insert two operators (blue and red dots) in these respective regions in a lightcone limit close to the continuation of this null line. 
\label{fig:setup}
}
\end{figure}

The essential idea for the proof is to study the following correlator:\footnote{If the state of interest is not pure, then $\left| \psi \right>$ represents the purifcation of the state in a doubled Hilbert space. } 
\be
\label{f}
f(s) = \frac{ \left< \psi \right| \mathcal{O}_B e^{  - i s K_B} e^{ i s K_A} \mathcal{O}_{\bar{A}} \left| \psi \right>}{ \left< \Omega \right| \mathcal{O}_B e^{  - i s K^0_B} e^{ i s K^0_A}  \mathcal{O}_{\bar A} \left| \Omega \right> }
\ee
where the two probe operators $\mathcal{O}_B, \mathcal{O}_{\bar{A}}$ are inserted in the region $\mathcal{D}(B)$ and $\mathcal{D}(\bar{A})$ respectively. We then act on these operators with modular flow $\mathcal{O}_B \rightarrow e^{ i s K_B} \mathcal{O}_B e^{-is K_B}$ using the (full) modular Hamiltonians $K_B,K_A$ defined for the sub regions $B,A$ respectively and for the state $\left| \psi \right>$. The modular Hamiltonian can be defined abstractly (with some technical assumptions on the state $\left| \psi \right>$) with respect to the algebra of bounded operators within the region, and Tomita-Takesaki theory \cite{tt,haag2012local} guarantees that the modular flowed operators for real $s$ are still contained within the algebra of operators of that region. More constructively the modular Hamiltonian is related to $2\pi H_A \equiv - \ln \rho_A$ the reduced density matrix  of $\left| \psi \right>$ restricted to $A$ and is sometimes referred to as the entanglement Hamiltonian, and modular flow simply involves time evolution using this Hamiltonian. In this paper we will mostly be interested in the full modular Hamiltonian which is $K_A = H_A \otimes \mathbb{1}_{\bar{A}} - \mathbb{1}_A \otimes H_{\bar{A}}$, and it is important to note that $K_{A,B} \left| \psi \right> =0$ for the defining state.

For some special cases modular evolution can be local, as for example the case where $A \rightarrow A_0$ is a half-space/Rindler cut in Minkowski space and for $\left| \psi \right> \rightarrow \left| 0 \right>$ the vacuum \cite{Bisognano:1976za}. The action of $K^0_A$ is then a boost holding fixed $\partial A_0$. This is our definition of the term in the denominator of $f(s)$ \eqref{f} where $A_0$ and $B_0$ are half space cuts such that $ \partial A_0$ and $\partial B_0$ are parallel to $\partial A$ and $\partial B$ at $y$ respectively.  Later in the paper we will slightly refine this definition of the denominator to  allow for $\partial A_0, \partial B_0$ to be null cuts of the Rindler horizon agreeing locally with $\partial A, \partial B$. In this case the denominator can be constrained using the so called theory of half-sided modular inclusions \cite{wiesbrock1993half,borchers1996half,araki2005extension,borchers1992cpt,borchers1995use} - the computation of which has some overlap with the recent paper \cite{Casini:2017roe}. 

Since the probe operators are initially spacelike separated they commute, and since $\mathcal{D}(B)$ and  $\mathcal{D}(\bar{A})$ are spacelike seperated the modular evolved operators will also commute:
\be
\label{comm}
\left[ e^{ i s K_B} \mathcal{O}_B e^{-is K_B}, e^{ i s K_A} \mathcal{O}_{\bar A} e^{-is K_A}\right] = 0
\ee
for real $s$. This fact, pertaining to causality,
translates into a statement about analyticity of $f(s)$ in the complex $s$ plane. Indeed a generalization of Tomita-Takesaki's modular theory \cite{buchholz1990nuclear,wiesbrock1993half,zbMATH00845667,araki2005extension} establishes the analytic extension of $f(s)$ in the strip $ -\pi < {\rm Im} s < \pi$. 

We will use this analyticity to prove the QNEC. Roughly speaking, if the QNEC were violated the modular evolved operators could exit their respective causal domains and cause a branch cut along ${\rm Im} s =0$ giving a non-zero commutator \eqref{comm}. Since we are using modular evolved operators this is a subtle violation of causality, but one which makes sense in the context of AdS/CFT. Indeed it was recently shown that modular evolved operators give a way to reconstruct bulk operators localized within the entanglement wedge associated to some boundary sub region such as $A$ \cite{Jafferis:2015del,Faulkner:2017vdd}. The entanglement wedge is believed to be the largest region containing information reconstructible using operators acting on the sub Hilbert space $\mathcal{H}_A$ in the QFT \cite{Czech:2012bh,Headrick:2014cta,Almheiri:2014lwa,Dong:2016eik}, and so these bulk regions are causally constrained by the boundary theory. Additionally the QNEC was proven for theories with an AdS/CFT dual using exactly this causality requirement \cite{Koeller:2015qmn,Akers:2016ugt,Fu:2017evt} - more specifically the entanglement wedge nesting (EWN) requirement. As we will explain there is a very precise sense in which this paper can be thought of as studying subtle QFT causality requirements via the causality properties of a gravitational dual with an emergent radial direction \cite{maldacena1999large}. Most QFTs do not have a classical gravity dual, but in some sense since we will only be studying properties of the gravitational system close to its boundary, the real stringy/strongly interacting nature of the dual gravitational system is suppressed.  Our results thus identify \eqref{comm} with the QFT equivalent of EWN.

Of course having setup the problem it may seem hard to compute $f(s)$ in any useful way, that is, retaining full generality over the state $\psi$ as well as the generality of the entanglement cuts $\partial A$ and $\partial B$. This is because $K_{A,B}$ are complicated non-local operators.  
We will manage to make progress here using a lightcone limit for the operators $\mathcal{O}_{B,\bar{A}}$, as pictured in the setup of Figure~\ref{fig:setup}, where the operators are separated in the $(x^+,x^-)$ direction by an amount $(\Delta v, \Delta u)$ and $\Delta v \rightarrow 0$ as $\Delta u$ is held fixed. This is a very similar limit to that considered in the causality ANEC proof \cite{Hartman:2016lgu} although now in the presence of entanglement cuts through points collinear with the operators. In this limit  we can use the replica trick to compute properties of the general modular Hamiltonians $K_{A,B}$, coupled with a defect lightcone OPE argument. The defect is the non-local co-dimension 2 twist operator of the $n$-replicated theory and a large part of our computations involve controlling the spectrum of local operators on the defect (referred to as defect operators) in the $n \rightarrow 1$ limit.

For large $s$ (but not too large as to move us out of the lightcone limit) we find the small but growing correction term:
\be
\label{ex}
f(s) = 1 -  e^s z^{d }   \frac{16\pi G_N\Delta_{\mathcal{O}} }{d(- \Delta v)}Q_- (A,B;y) + \ldots\,,
\qquad z^2 \equiv- \frac{\Delta v (\Delta u- \delta x^-)}{4}
\ee 
where $Q_-$ is the QNEC object  in \eqref{introq}. Note that we have introduced a quantity we call $G_N$  via its usual relation to $G_N \propto 1/c_T$ in holographic theories \eqref{gnct} in units where $R_{AdS}=1$. There is no need for $G_N$ to be small. We have also defined $z \ll 1$ in terms of the kinematics of the operator insertions which exactly plays the role of the radial $z$ coordinate in an emergent AdS. 
Here $\delta x^-$ is the coordinate distance between $\partial A$ and $\partial B$ and $\Delta u > \delta x^-$ must be true. 
All we have to do now, taking inspiration from the chaos bound \cite{Maldacena:2015waa} and causality bound \cite{Hartman:2015lfa} stories, is prove that ${\rm Re} f < 1$ along the lines ${\rm Im} s = \pm \pi/2$ in the complex $s$-strip. Analyticity then allows us to extract $Q_-$ from \eqref{ex} as an integral over $(1-{\rm Re} f)$ along these same lines which is then constrained to be positive thus proving the QNEC. 

Let us add one more comment on the meaning of $f(s)$. The modular flows generated by $K_{B_0,A_0}$ simply boost the operators $\mathcal{O}_{B,\bar{A}}$ and the denominator is explicitly given by the two-point function for the boosted operators. The numerator generally lacks such a simple picture, and it involves complicated interactions between the operators and the defect. However, in the light-cone limit we can interpret the numerator approximately as a two-point function for local operators. The leading correction to $f(s)$ in (\ref{ex}) can be viewed as giving a shift for the separation $\left(-\Delta v\right)>0$ in the numerator relative to that of the denominator:
\be
\label{vshift}
- \Delta v \rightarrow - \Delta v +  e^{s} z^{d} \frac{16\pi G_N}{d} Q_-
\ee
This is suggestive of a tendency towards shifting the branch cut singularity in the two point function onto the real $s$ axis if $Q_- <0$ for large enough $s$. This is not a precise argument since the shift is only important when the small correction in $f(s)$ competes with $1$, but that's okay since the precise argument was given above. However it allows us to identify the gravitational time delay/advance that we should look for in the bulk, and we will identify this by (very slightly) generalizing the arguments of \cite{Koeller:2015qmn} which proved the QNEC for holographic theories using entanglement wedge nesting (EWN) of the RT\cite{Ryu:2006bv}/HRT\cite{Hubeny:2007xt}/quantum extremal\cite{engelhardt2015quantum} surface near the boundary.

Many of the properties of the function $f(s)$ are the same as the function $f(t)$ defined in \cite{Maldacena:2015waa} for studying chaos using an  out of time order four point function for two different operators $W,V(t)$ in a thermal state. The analogy is strengthened by taking the thermal state to be that of the Rindler state, and time $t$ to be generated by boosts using the Rindler Hamiltonian.  This is then the same setup as \cite{Hartman:2016lgu} for proving the ANEC using causality - although the equivalent function is constrained in different kinematic regimes - determined by how large $t$ is and whether the operators are in a lightcone limit for the causality bound (large $t$) or the Regge limit (even larger $t$, but not as large as the scrambling time $\sim \ln c_T$) for the chaos bound. The analog of our setup would evolve $V$ not with the Rindler Hamiltonian but with the complicated modular Hamiltonian of the state $W \left| \Omega \right>$ and now reduced to two different entangling regions. From this point of view our paper represents a generalization of the chaos and causality bound setup, however we have not yet explored the extent to which we can apply this setup usefully to non-relativistic quantum systems and generic perturbed thermal states. We also do not have much to say about the equivalent ``even larger'' $s$ regime of $f(s)$ in a large-$N$ theory analogous to the Regge limit leaving these fascinating generalizations for future work. 

The paper is organized as follows. In Section~\ref{sec:background} we start with background on various known results that will be useful to us, including a discussion of the holographic proof of the QNEC as well as a discussion of geometric modular Hamiltonians and their action on local operators. We note an interesting relation 
to the well studied theory of half-sided modular inclusions.  In Section~\ref{sec:replica} we discuss the use of the replica trick to compute properties of general modular Hamiltonians. We then consider the defect OPE which is necessary to carry out the replica trick computation. This includes a discussion of possible local defect operators that arise when $n \approx 1$.  In Section~\ref{sec:modham} we compute the matrix elements of the modular Hamiltonian in the state excited by $\mathcal{O}_{B,\bar{A}}$ in the lightcone limit. In Section~\ref{sec:modflow} we use this result to find the action of modular flow in a perturbative expansion with respect to the lightcone limit which gives the result \eqref{ex}. In Section~\ref{sec:f} we detail the general properties of $f(s)$ which lead to the QNEC. In Section~\ref{sec:loose} we discuss some loose ends, including an understanding of local geometric contributions to entanglement entropy that can contaminate the QNEC quantity $Q_-$ and thus invalidate the bound for non-stationary entanglement cuts. We also discuss a higher spin
version of the QNEC, generalizing the higher spin version of the ANEC proven in \cite{Hartman:2016lgu}. 
We conclude in Section~\ref{sec:discussion} with several possible extensions. Some computations and details are relegated to the Appendices. 

\textbf{Note added: } In this version of the paper, we are introducing a minor but technically important modification to the operators $\mathcal{O}_{B,\bar{A}}$ appearing in $f(s)$ compared to the previous pre-print. In particular, we are inserting them at positions that are $s$-dependent in a way that we will specify later. The additional $s$-dependence is small at large $s$, and only affects the analytic properties of $f(s)$ in a controlled way that decouples from our main arguments. The reason behind this modification is to eliminate contributions to $f(s)$ that could potentially contaminate the relation between the bound for $f(s)$ and the QNEC statement $Q_-\geq 0$ in the light-cone limit. These modifications were understood in \cite{Ceyhan:2018zfg} where they naturally arose from relative modular flow. More details will be explained as we lay out the actual proof.  

\section{Background}

\label{sec:background}

In this section we collect some known results from the literature that we will make use of throughout the paper. We will bring our own perspective to these results relevant to our discussion. Let us set the stage by setting up the problem we wish to study more precisely than in the introduction.

\subsection{Setup and conventions}
We take the metric to be flat:
\be
ds^2 = - du dv + \delta_{ij} dy^i dy^j 
\ee 
where we use $u = x^- = t - x$ and $ v = x^+ = t + x$ for null coordinates adapted to 
an entangling surface $\partial A$ which passes through the point $u = v = y^i = 0$ (we have set $y=0$ relative to the introduction!). We use both $v,u$ and $x^\pm$ to maximize our variable options.
Wick rotation is given by $ \tau = i t$ such that  $-u = z = x + i \tau$ and $v = \bar{z} = x - i \tau $.

The first entangling surfaces will be defined close to $y=0$ via 
\be
\partial A: \, v = X_A^+(y), \quad u = X_A^-(y) \,; \qquad X_A^\pm(0)=0
\ee
such that $A$ is a space like region ending on $\partial A$ to the ``left'' - roughly speaking within the wedge $u >0, v<0$ close to $y=0$.  The other entangling surface is displaced in the $u$ direction at $y=0$: 
\be
\partial B: \, v = X_B^+(y), \quad u = X_B^-(y) \,; \qquad X_B^+(0) =0\,, \,\, X_B^-(0) = \delta x^-
\ee 
where again $B$ is a space-like region to the ``left'' of this cut.  This description could break down far from the null line passing through both $\partial A$ and $\partial B$ at $y=0$, but the details far away will not play a role in our computations. For now we will be agnostic to the exact shape of the entangling regions, except to require that $\mathcal{D}(B) \subset \mathcal{D}(A)$. We will later discover that some further local conditions are required in order to claim  the QNEC bound - these are similar conditions to those discussed in \cite{Koeller:2015qmn}, that locally the entangling cuts should be stationary with $(\partial_y)^p X_{A,B}^+(0) =0$ for sufficiently many derivatives. These further conditions contain the special case where $A$ and $B$ are general null cuts of the Rindler horizon $v=0$ such that $X_{A,B}^+=0$ and $X_{A,B}^-(y)$ are left arbitrary (except for the inclusion requirement
$X_{B}^-(y)>X_{A}^-(y)$ for all $y$ .)  However our results are much more general than this. 

To lighten the notation we will use the following:
\begin{align}
\mathcal{O}_B \equiv \mathcal{O}( u_B, v_B,y=0) \,,  &  \qquad \mathcal{O}_{\bar A}  \equiv \mathcal{O}( u_{\bar A}, v_{\bar{A}},y=0) \\ 
 K_B  = H_{B}^\psi - H_{\bar{B}}^\psi \,, & \qquad K_A = H_A^\psi - H_{\bar{A}}^\psi 
\\
  K_B^0  = H_{B_0}^\Omega - H_{\bar{B}_0}^\Omega\,, & \qquad K_A^0 = H_{A_0}^\Omega - H_{\bar{A}_0}^\Omega
\end{align}
where $u_B >0,v_B < 0$ and $u_{\bar A} < 0, v_{\bar{A}} >0$ and we have made explicit which state the modular Hamiltonian refers to. For example $H^\Omega_{B_0} = - (2\pi)^{-1} \ln {\rm Tr}_{\bar{B}_0} \left| \Omega \right> \left< \Omega \right|$ where $\left| \Omega \right>$ is the CFT vacuum. 
We will often suppress the $y=0$ label on the operator insertions. We also suppress
$K_{B_0}^0 \rightarrow K_B^0$ which should be understood from the superscript label. 
For most of the paper, except in Section~\ref{sec:loose} and below, we will take the undeformed regions $\partial A_0, \partial B_0$ to be flat Rindler cuts that agree with $\partial A$ and $\partial B$ at $y=0$. We turn now to a description of the modular Hamiltonians for these regions in vacuum.

\subsection{Vacuum modular Hamiltonians and modular inclusions} 

\label{modinclmain}

We start by describing a special class of modular Hamiltonians, these are the so called local modular Hamiltonians which apply for relativistic vacuum states and for simple flat Rindler cuts. 
Modular flow in this case is just a local boost around the entangling surface and the modular Hamiltonians for two flat cuts of the same Rindler horizon form an algebra which is the one that naturally arrises in the theory of half-sided modular inclusions \cite{wiesbrock1993half}.  This case applies to $K_{A,B}^0$ the modular Hamiltonians for the two uniform Rindler cuts $\mathcal{D}(B_0) \subset \mathcal{D}(A_0)$ in vacuum which are important for evaluating $f(s)$ in the lightcone limit. 
The action is simple:
\be
e^{ i K^0_A s }\mathcal{O}(u,v) e^{ - i K^0_A s } = \mathcal{O}(e^s u, e^{-s} v)\, ,
\quad i \left[ K^0_A , \mathcal{O}(u,v) \right] = \left( u \partial_{u} - v \partial_v \right) \mathcal{O}(u,v)
\ee
and for $B$:
\be
e^{ i K^0_B s }\mathcal{O}(u,v) e^{ - i K^0_B s } = \mathcal{O}(e^s (u- \delta x^-) + \delta x^-, e^{-s} v)\, ,
\quad i \left[ K^0_B - K^0_A, \mathcal{O}(u,v) \right] =  - \delta x^-\partial_{u} \mathcal{O}(u,v)
\ee
Furthermore these satisfy an algebra:
\be
\label{algincl}
\left[ K^0_B , K^0_A \right] =   i ( K_A^0 - K^0_B) \equiv  i \delta x^- P_-
\ee
where $P_- = 1/2 (H+ P_x)$ is the translation operator in the $x^-$ direction $i P_- = \partial_-$. 
This algebra is 2 dimensional and isomorphic to the algebra associated with the affine group $u \rightarrow a u + b$. For the pattern of modular flow in the correlator $f(s)$ we find:
\be
\label{incl}
e^{- i K_B^0 s } e^{ i K_A^0 s } = U( (1- e^{-s}) \delta x^-) \equiv e^{ i(1 - e^{-s}) \delta x^- P_-}
 \qquad 
\ee
where the $U(b)$ generates a translation in the null direction $u \rightarrow u + b$. Note that $P_-$ is  clearly a positive operator via vacuum stability, which it must have been due to the negativity constraint on modular Hamiltonians under shape deformations \cite{Blanco:2013lea}.  

In this paper we will work in a limit where the $\psi$ modular Hamiltonians are well approximated by these modular Hamiltonians
plus computable corrections. In Section~\ref{sec:loose} we will find that in order to account for some of these corrections it is useful to consider a more general class of vacuum modular Hamiltonians, the form of which was only recently elucidated \cite{Faulkner:2016mzt,Koeller:2017njr,Casini:2017roe}.\footnote{These modular Hamiltonians for general QFTs are consistent with those of free theories which were worked out by A. Wall \cite{Wall:2011hj} based on light front quantization.} This class derives from arbitrarily shaped null cuts of the Rindler horizon in vacuum. An important result now comes from the theory of half-sided modular inclusions which can be used to prove that the algebra \eqref{algincl}, suitably generalized, continues to apply in this more general case. 

One proceeds in two steps, the details of which are given in Appendix~\ref{app:modincl}. Firstly we recall that half sided modular inclusions apply to the case where $B_0$ is an arbitrary null cut of the Rindler horizon $v=0$ satisfying $\mathcal{D}(B_0) \subset \mathcal{D}(A_0)$ where $A_0$ is a uniform Rindler cut ending on $\partial A_0: u = 0, v=0$ with an associated local modular Hamiltonian. The region $B_0$ then has the nesting property that 
\be
\label{nest0}
 e^{ i s K^0_A } \mathcal{D}(B_0)  e^{ - i s K^0_B } \subset  \mathcal{D}(B_0)\,, \qquad s>0
 \ee 
These conditions are enough to prove the result that the algebra defined in \eqref{algincl} continues to hold with the replacements:
\be
\label{algincl2}
\left[ K^0\{ X^-_B \}  , K^0_A \right] =  i (K_A^0 -  K^0\{ X^-_B \}) 
\ee
where the notation $K^0\{ \cdot \}$ defines the modular Hamiltonian as a functional of the specific entangling cut of the Rindler horizon (recall that $X^+_{A,B}=0$ and $X^-_A=0$ for now.) Intriguingly one way to prove this is by studying
a very similar correlation function to that which appears in $f(s)$ namely:
\be
j (s) \equiv \left< 0 \right| \mathcal{O}_B e^{- i s K^0\{X_B^-\} }  e^{ - i e^s \left( K_A^0 -  K^0\{ X^-_B \} \right) }
e^{  i s K^0\{X_B^-\} }  \mathcal{O}_{\bar{A}} \left| 0 \right>
\ee
Applying the nesting property \eqref{nest0} and positivity properties of $K_A^0 -  K^0\{ X^-_B \}$ 
\cite{wiesbrock1993half} argued for an analytic extension that is periodic and holomorphic in the thermal $s$ strip: $-\pi< \Im s < \pi$ with $j( s + i \pi) = j(s-i \pi )$.
Similarly $j(s)$ is necessarily bounded in this strip and the only way to satisfy all these conditions is if $j(s)$ is a constant independent of $s$. Expanding about $s=0$ one derives the algebra in \eqref{algincl2}. \footnote{One approach that we tried in order to prove the QNEC was to study $j(s)$ in the case where not all the conditions above are satisfied - in particular the nesting property of boosted regions \eqref{nest0} is generically going to fail for non-local modular Hamiltonians associated to a non vacuum state $\psi$. We did not get this approach to work and instead settled on the modular flow pattern in $f(s)$. }

Continuing on, this algebra  allows us to find an expression for the null deformed modular Hamiltonian in terms of an integral of the stress tensor:
\be
K^0\{ X_B^-\} = \int d^{d-2} y  \int_{-\infty}^{\infty} d x^- (x^- - X_B^-(y) ) T_{--}(x^-)
\ee
This was recently shown in \cite{Casini:2017roe} and we will give a slightly different proof of this in Appendix~\ref{app:modincl}.
The main ingredients in our proof are the algebra \eqref{algincl2}
as well as the recent computation of linearized shape deformations to the Rindler modular Hamiltonian \cite{Faulkner:2016mzt} which allows us to fix the modular Hamiltonian for small $X_B^-(y)$. This result proves the conjectured answer in \cite{Faulkner:2016mzt} that the higher order corrections in the $X_B^-$ expansion are essentially trivial. 

Note that these new modular Hamiltonians are not local in the sense that they do not generate local flows. With the result \eqref{algincl2} in hand one can then just go and calculate the algebra when $A$ and $B$ are both deformed null cuts (see Appendix~\ref{app:modincl} and \cite{Casini:2017roe}):
\be
\label{algboth}
\left[ K^0\{ X_B^-\}, K^0\{ X_A^-\}  \right] = i (K^0\{ X_A^-\} -  K^0\{ X^-_B \}) =  i P_-\{ X_B^- - X_A^- \}
\ee
such that:
\be
\label{newalg}
e^{- i K^0\{X_B^-\} s } e^{ i K^0\{ X_A^-\} s } = \exp\left( i(1 - e^{-s}) P_-\{ X_B^- - X_A^- \} \right)
\ee
where:
\be
P_- \{ X^-\} = \int d^{d-2} y  X^-(y) \int_{-\infty}^{\infty} d x^-  T_{--}(x^-)
\ee
While the action of these modular Hamiltonians is not local, it will become local when acting on operators in the lightcone limit (close to the Rindler horizon.) This should allow us to compute the action of the vacuum modular Hamiltonian perturbatively in the lightcone limit, which goes into computing $f(s)$. As we will see the details of this computation will not be important, excepting that they satisfy the modular inclusion algebra \eqref{algboth}. 

We make a final point returning to the simple uniform null cuts. The nested boosts relevant to $f(s)$ that we computed in \eqref{incl} tells that for large $s$ we simply have a null translation by a small amount $\delta x^-$. This means we can take $s$ large without the operators exploring too much of the spacetime and this will be important for us to claim the more general QNEC results. We can also understand what happens if we move the two entangling cuts away from each other slightly in the $v$ direction by an amount $\delta x^+$. The inclusion property $\mathcal{D}(B_0) \subset \mathcal{D}(A_0)$ is now only true if $\delta x^+ \leq 0$. These modular Hamiltonians are not constrained by the algebra of half sided modular Hamiltonians, but since here they are simple boosts we can just explicitly compute the modular flow. Consider the flow:
\be
e^{- i K_B^0 s } e^{ i K_A^0 s } \mathcal{O}(u_{\bar{A}},v_{\bar{A}}) \left| 0 \right>
= \mathcal{O}(u_{\bar{A}} + \delta x^- (1-e^{-s}), v_{\bar{A}} + \delta x^+ ( 1 - e^s) ) \left| 0 \right>
\ee
which for large $s$ still gives an operator shifted in the null $u = u_{\bar{A}} + \delta x^-$ direction, but the operator is now moving to large $ v\approx v_{\bar{A}} - \delta x^+ e^s$. If we plug this into the vacuum correlator in the denominator of $f(s)$ and expand for small $\delta x^+ e^s$ we have:
\be
\left< 0 \right| \mathcal{O}_B e^{- i K_B^0 s } e^{ i K_A^0 s } \mathcal{O}_{\bar{A}} \left| 0 \right>
= \frac{c_\Delta}{ \left( - \Delta v(\Delta u - \delta x^-) \right)^{\Delta_{\mathcal{O}}}} 
\left( 1  + e^s \frac{\Delta_{\mathcal{O}}\delta x^+}{(-\Delta v)} + \ldots \right)
\label{vshift2}
\ee
So unless we set $\delta x^+ =0$ we find a small but growing $e^s$ term which should be compared to \eqref{ex} and \eqref{vshift}. Without making this later $\delta x^+ e^s$ expansion the two operators will eventually become time-like separated from each other if $\delta x^+ >0$. Since in this case the two domains of dependence $\mathcal{D}(\bar{A}_0)$ and $\mathcal{D}(B_0)$ are not causally disconnected there is no issue with the necessary appearance of a branch cut in $s$ along $\Im s = 0$.  However this gives us some intuition for the growing $e^s$ QNEC term we are claiming for more general modular Hamiltonians and states. Consider a holographic theory. If the QNEC is violated then as one moves slightly inwards in the holographic $z$ direction the two bulk entanglement wedges for $\bar{A}$ and $B$ will come into causal contact. Since the JLMS \cite{Jafferis:2015del} result tells us that boundary modular flow equals bulk modular flow, a similar algebra for modular Hamiltonians should now apply except in the bulk and now determined via the relative position of the RT surface (elucidated further in the next subsection). Near the boundary we can approximate the cut with a bulk Rindler cut except slightly deformed due to the movement of the RT surface in the $v=x^+$ direction as we move inwards from the boundary.  From this consideration we expect to find the same $e^s$ growing term that we found
by shifting $\delta x^+$ on the boundary. In particular the wrong sign $\delta x^+ >0$ which applies when $Q_- < 0$ is an indication that the entanglement wedges are coming into causal contact. We now turn to a holographic calculation  demonstrating that indeed this bulk causality consideration is determined by the sign of the QNEC quantity $Q_-$. 

\subsection{Holographic proof and EWN}

Let us attempt to identify the gravitational time delay/advance directly in the bulk. In this section
we assume our CFT has a description in terms of a weakly coupled classical Einstein gravity theory. This is only true for a small class of theories, but these theories allow us to develop intuition for the general case. The results here are not new and were originally worked out in \cite{Koeller:2015qmn}, relating the QNEC in holographic theories to entanglement wedge nesting (EWN.) The EWN property states that if two boundary regions satisfy $\mathcal{D}(B) \subset \mathcal{D}(A)$ then the dual entanglement wedges must satisfy the same condition. The entanglement wedge of a region $A$ is the domain of dependence of the spacelike $A_b$ region located between $A$ on the boundary of AdS and the RT surface $\mathcal{E}_A$.  This requirement can be understood as being basic to the program of entanglement wedge reconstruction \cite{Akers:2016ugt}. 

We will work out a slight generalization for the integrated version of the QNEC in \eqref{introq}. For simplicity we will ignore many complications due to extrinsic curvature effects and effects arising due to a relevant deformation which takes us from the UV CFT to a more general QFT.  These more complicated effects were discussed carefully in \cite{Koeller:2015qmn}.

The metric solving Einstein's equations near the boundary of AdS has a Fefferman-Graham expansion:
\be
ds^2 = \frac{ - du dv + dy^2 + dz^2}{z^2} + z^{d-2} \tau_{\mu\nu} dx^\mu dx^\nu + \ldots
\ee
Similarly the two RT entangling surfaces parameterized via $v,u = X_{RT}^\pm(z,y)$ have an expansion:
\begin{align}
X_{RT,A}^-(z,y) &= X_A^-(y) + \mathcal{O}(z^2) \,, \qquad X_{RT,A}^+(z,y) =z^{d} p^A_-(y) + \ldots 
\\
X_{RT,B}^-(z,y) &= X_B^-(y) + \mathcal{O}(z^2) \,, \qquad  X_{RT,B}^+(z,y) =  z^{d} p^B_-(y) + \ldots
\end{align}
where we find this form by solving the extremal surface condition close to the boundary. Here $\tau_{\mu\nu},p^{A,B}_-$ are not fixed by the asymptotic boundary conditions. They are state ($\psi$) dependent and can be related to the CFT stress tensor and the shape variation of the holographic EE respectively:
\be
\label{toqft}
\tau_{\mu\nu} = \frac{16 \pi G_N}{d} \left< T_{\mu\nu}\right>_\psi \,, \qquad  p_{-}^{A} = \frac{8 G_N}{d} \frac{\delta S(A)}{\delta x^-(y)} \,,\qquad  p_{-}^{B} = \frac{8 G_N}{d} \frac{\delta S(B)}{\delta x^-(y)} 
\ee
The later relation may be less familiar to the reader, but can be thought of as the usual Hamilton-Jacobi relation between conjugate coordinates $(X^-, p_-)$ in the sense where $z$ is time and the area of the RT surface or $S_{EE}$ is like the action holding fixed the boundary value $x^-$: $p_- \sim  \partial_z X^+ \sim \delta S_{EE}/\delta x^- $. 
We will take $X^+_{A,B}(y)=0$ for simplicity to suppress additional leading terms that would arise in the $z$ expansion of $X^+_{RT}(z,y)$ multiplying various local extrinsic curvature invariants. 
We also remind the reader that $X_A^-(0) =0 $ and $X_B^-(0) = \delta x^-$. 

Now we consider a high energy particle moving near the boundary of AdS in the $\partial_u$ direction along a null geodesic with approximately fixed $v,z$ and paramaterized by the coordinate $u$. To leading order we only need
to track the small change $v \rightarrow v(u)$. This particle will be analogous to our $\mathcal{O}$ probe. To see if the two entangling wedges are causally disconnected we consider this null geodesic to pass through
the point $v(0) = X^+_{RT,A}(z,y=0)$ at $u=0, y=0$ and $z$ fixed. The particle then picks up a delay in the $v$ direction as it propagates to $u= X^-_{RT,B} \approx \delta x^-$: 
\be
v(\delta x^-) = X_{RT,A}^+(z,y=0) + v_{\rm delay} \,,  \qquad v_{\rm delay} = z^{d} \int_0^{\delta x^-} d u \tau_{--}(u, v=0,y=0) 
\ee
Comparing this new $v$ coordinate to the position of the $B$ RT surface we find this is determined by the QNEC quantity:
\begin{align}
v(\delta x^-) - X_{RT,B}^+(z,y=0) &= z^{d} \left( \int_0^{\delta x^-} d u \tau_{--}(u, v=0) 
+ p^A_{-} - p^B_{-} \right)_{y=0} \\
&= \frac{16\pi G_N}{d} z^d Q_-(A,B; y=0) \geq 0
\end{align} 
where we have used \eqref{toqft}. This result should then be compared to \eqref{vshift} and \eqref{vshift2} to find a consistent story between the bulk and the boundary. 

The lightcone limit allows us to study particles propagating near the boundary of AdS and weakly interacting via graviton exchange with the state $\psi$. This turns out to be a useful picture in any interacting CFT \cite{Fitzpatrick:2014vua,Fitzpatrick:2012yx,Komargodski:2012ek}, and this was the original motivation for studying the lightcone limit in this context. The delay one extracts from this picture is the total delay of the particle integrated over all boundary times $-\infty < u < \infty$ - causality then imposes the ANEC constraint. The boundary theory ANEC is in this way related to the Gao-Wald causality condition on the bulk \cite{Kelly:2014mra}, that the fastest path in the full spacetime between two null separated points on the boundary is a null line on the boundary. 
By studying causal curves that reach into the bulk and are sensitive to the boundary theory stress tensor $\tau_{--}$ term in the metric the authors \cite{Kelly:2014mra} used Gao-Wald to prove the ANEC.
This does not usefully constrain a more local version of the NEC because propogating the particle from the boundary into a fixed $z$ causes a large $v$ delay which swamps the delay/advance due to the $\tau_{--}$ term in the metric. This can only be removed by taking the two points on the boundary to be infinitely separated in the null $u$ direction. Entanglement wedge nesting is a much more fine grained version of causality that allows us to directly study the gravitational delay/advance at a fixed $z$ coordinate via the introduction of the entangling surfaces. And it turns out the way to extract this from the boundary theory is with the correlator in $f(s)$. 

\begin{figure}[h!]
\centering
\includegraphics[scale=.9]{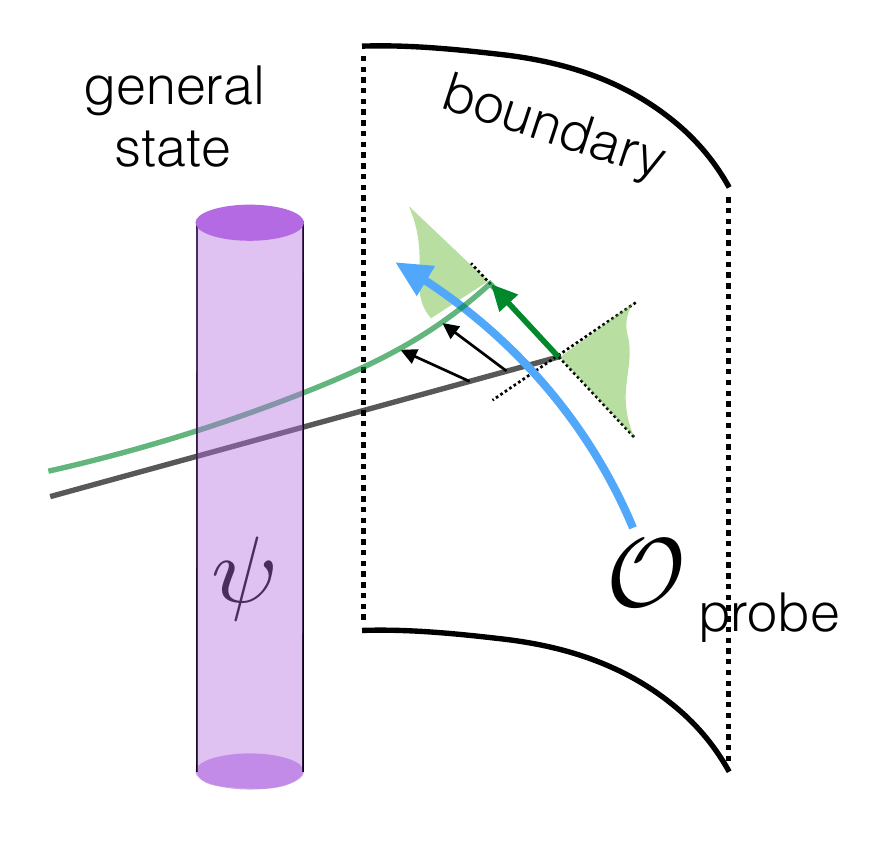}
\caption{A schematic of the AdS/CFT setup for understanding $f(s)$. The lightcone limit of a four point function $\left< \psi \mathcal{O} \mathcal{O} \psi \right>$ can be interpreted in terms of a dual holographic setup where the dual particle excitation to $\mathcal{O}$ and  $\psi$ stay far away from each other in AdS space by having a large relative angular momentum. We use this setup as inspiration for our computation, where we add into the mix two entangling surfaces which start null separated at the boundary and fall into the bulk.  EWN is the statement that these surfaces should be spacelike separated as one moves into the bulk. The $\mathcal{O}$ particle in the high energy/lightcone limit probes these entangling surfaces near the boundary after we act with modular flow on this particle. }
\end{figure}

\subsection{Tomita-Takesaki theory}

\label{sec:tt}

In order to prove various (non perturbative) properties of $f(s)$ we will need to have a better understanding of modular flow for a more general class of states than for the vacuum of a QFT. For now we will present (a brutalized version of) the abstract algebraic discussion of Tomita-Takesaki theory, see for example \cite{haag2012local}. This will pertain to the action of a single modular flow - the double modular flow will be discussed later in Section~\ref{sec:f}. 
The idea is to consider a von Neumann algebra of bounded operators $\mathcal{A}$ associated with some local region in spacetime say $\mathcal{D}(A)$. If we additionally have a state $\left| \psi \right>$ on the total Hilbert space that is cyclic and separating for $\mathcal{A}$ - meaning that $\mathcal{O}_A \left| \psi \right>$ is dense in the total Hilbert space for all operators $\mathcal{O}_A \in \mathcal{A}$  and that $\mathcal{O}_A$ cannot
annihilate $\left| \psi \right>$, then one can define the following modular operators:
\be
J_A \Delta^{1/2}_A \mathcal{O}_A \left| \psi \right> =  \mathcal{O}_A^\dagger \left| \psi \right>
\qquad J_A \Delta_{A}^{1/2} J_A = \Delta_A^{-1/2}  \qquad J_A \left| \psi \right> =  \left| \psi \right>
\qquad \Delta_A^{1/2}  \left| \psi \right> =  \left| \psi \right>
\ee
where $J$ is anti-unitary and $\Delta_A$ is positive and Hermitian, but generally unbounded. To make contact with the (full) modular Hamiltonian one writes $\Delta_A = e^{ - 2\pi K_A}$ where now $K_A$ will not be a positive operator. One can then show that:
\be
J_A \mathcal{A} J_A = \mathcal{A}' \qquad \Delta_A^{is} \mathcal{A} \Delta_A^{-is} = \mathcal{A}
\qquad s \in \mathbb{R}
\ee
where $\mathcal{A}'$ is the commutant which is then the bounded operators associated to the region $\mathcal{D}(\bar{A})$. 

Physically, cyclic and separating just means that the state has a large amount of entanglement between $\mathcal{D}(A)$ and $\mathcal{D}(\bar{A})$ and we expect that all reasonable QFT states  one might consider have this property. For the case of the vacuum the Reeh-Schlieder theorem \cite{schlieder1965some} rigorously establishes this fact. In a  quantum system with a finite dimensional Hilbert space this condition would be equivalent to the statement that the reduced density matrix $\rho_A$ (for a finite quantum system $\Delta_A = \rho_A \otimes \rho_{\bar A}^{-1}$) has full rank and so is invertible \cite{papadodimas2014state}, however it will be important to acknowledge the fact that in an infinite quantum system, since $\Delta_A$ is unbounded we have to carefully specify the domain on which it acts.  For example it is known that $\mathcal{A} \left| \psi \right>$ is generally in the domain of $\Delta_A^\alpha$ for $0 < \alpha < 1/2$ and $\mathcal{A}' \left|\psi \right>$ is generally in the domain of $\Delta_A^\alpha$ for $-1/2 < \alpha < 0$ \cite{haag2012local}. 

An important consequence of this structure is an abstract version of the KMS condition. To understand this we consider the correlator (which is a baby version of $f(s)$):
\be
h(s) \equiv \left< \psi \right| \mathcal{O}_{A} \Delta^{-\frac{i s}{2\pi}}  \mathcal{O}_{\bar A} \left| \psi \right> =  \left< \psi \right| \mathcal{O}_{A} e^{ i K_A s}  \mathcal{O}_{\bar A} \left| \psi \right>
\ee
which can be analytically continued into complex $s$ in the strip $ -\pi  < {\rm Im} s < \pi$. On the upper/lower edge we have: 
\be
h(t +  i\pi) = \left< \psi \right| \widetilde{\mathcal{O}}_A e^{ - i t K_A} \mathcal{O}_A \left| \psi \right> \qquad h(t -  i\pi) = \left< \psi \right|  \mathcal{O}_A e^{ i t K_A}  \widetilde{\mathcal{O}}_A \left| \psi \right> 
\ee
where $\widetilde{\mathcal{O}}_A =  J_A \mathcal{O}_{\bar A}^\dagger J_A \in \mathcal{A}$.  The difference across the cut, which arises in the $s$ strip after we identify $s \equiv s + 2\pi i$ at ${\rm Im } s = \pi$, is just the commutator $\left[ \widetilde{\mathcal{O}}_A, \mathcal{O}_A(t)\right]$
where $ \mathcal{O}_A(t)= e^{ - i t K_A} \mathcal{O}_A e^{ i t K_A}$.  Analyticity along $ {\rm Im} s =0$ is simply related to the fact that the original operators $\mathcal{O}_A$ and $\mathcal{O}_{\bar A}$ commute.  

We can give a less rigorous discussion of these results by appealing to the analogy with thermal systems. For example, if the subspace had a trace, we can then replace the correlators as:
\be
g(t +  i \sigma) = {\rm Tr}_A \rho_A^{ 1/2 +\frac{\sigma}{2\pi}  } \widetilde{O}_A \rho_A^{  1/2 -\frac{\sigma}{2\pi}}  \mathcal{O}_A (t)
\ee
where we have set $s = t + i \sigma$.
This expression demonstrates where the strip  $ -\pi  < \sigma < \pi$ comes from. In an infinite dimensional system the sum over intermediate eigenstates of $\rho_A$ is not guaranteed to converge outside of this range. 
In our case there is no trace, however we could regulate things around the entangling surface with a hard wall cutoff in order to introduce a trace. 

Moving forward we want to study the situation where there are now two algebras with a common cyclic and separating state and the inclusion property $\mathcal{A}_B \subset \mathcal{A}_A$. This is harder to study but we can use various results from the literature. We will explain these in a later section.

\section{Replica trick for the modular Hamiltonian}

\label{sec:replica}

Our computation of $f(s)$ will now begin in earnest. Our first task is to compute matrix elements of $K_A$ sandwiched between $\left| \psi \right>$ excited by the $\mathcal{O}$ operator insertions. To do this we will need to use the replica trick.

Previous discussion of using the replica trick to compute the modular energy of excited states has appeared in \cite{Lashkari:2015dia} (also \cite{Sarosi:2016oks,Ruggiero:2016khg}). This was then used to study the modular energy in 2d CFTs. While we will take a very similar approach there will be an important difference. We would like to write the answer in terms of twist operators in the orbifold theory $CFT^n/\mathbb{Z}_n$. It is not totally clear this is possible since, as noted in \cite{Lashkari:2015dia}, the replica trick in this case explicitly breaks the $\mathbb{Z}_n$ symmetry which cycles through the replicas. For this reason the results in \cite{Lashkari:2015dia} are left in the form of correlation functions on n-sheeted branched coverings without the $\mathbb{Z}_n$ symmetry. On the other hand the orbifold theory is much more under control since we can use standard results about defect CFTs \cite{Billo:2016cpy,Gliozzi:2015qsa,Gaiotto:2013nva,Billo:2013jda,Liendo:2012hy} in order to make progress with computations. This will be the main technical difficulty that we have to overcome here.

\subsection{Replica trick}

The replica trick is a way of computing properties of the operator $\ln \rho_A$ using the limit:
\be
\lim_{n \rightarrow 1} \partial_n \rho_A^{n-1} = \ln \rho_A
\ee
This is useful because it is sometimes possible to compute traces over $\rho_A^n$ for integer $n$ using a path integral. The limit is then only achievable once an analytic extension is found from integer $n$ to complex $n$. While this is usually subtle the replica trick has yielded many powerful results relating to entanglement entropy in QFT \cite{callan1994geometric,Calabrese:2004eu,Casini:2009sr,Lewkowycz:2013nqa}. 

We will firstly be interested in simply evaluating the half modular Hamiltonian: $2\pi H_A \equiv - \ln  \rho_A \otimes 1_{\bar{A}}$, thought of as an operator on the total Hilbert space, between matrix elements of the defining state $\left| \psi \right>$ excited by local operator insertions. This is not a totally well defined object in the continuum and
so will only be an intermediate step towards computing the full version: $K_A  = H_A - H_{\bar{A}}$ which is well defined. We will not be completely explicit about how we regulate $H_A$ to define it, but we will assume this regulator allows us to define a trace over the various tensor factors in the Hilbert space. 

Consider:
\be
\left< \psi \right| \mathcal{O}_{B} \left( \ln \rho_A \otimes 1_{\bar A}  \right)  \mathcal{O}_{\bar{A}}\left| \psi \right> = \lim_{n \rightarrow 1} \partial_n \left< \psi \right| \mathcal{O}_{B} \left( \rho_A^{n-1} \otimes 1_{\bar A}  \right)  \mathcal{O}_{\bar{A}}\left| \psi \right> \equiv  \lim_{n \rightarrow 1} \partial_n Z_n
\ee
which we can write as a trace over the Hilbert space $\mathcal{H}_A \otimes \mathcal{H}_{\bar A}$
\be
\label{tr}
Z_n = {\rm Tr}_A \rho_A^{n-1} {\rm Tr}_{\bar{A}} \left( \mathcal{O}_{\bar{A}} \left| \psi \right>
\left< \psi \right| \mathcal{O}_B \right)
\ee

The trace in \eqref{tr} can be computed using a path integral. We first write a path integral
representation of $\rho_A$ by integrating over Euclidean space with a branch cut running along $A$
and different boundary conditions above and below $A$ used to represent the density matrix. 
To be concrete let us take the state $\left| \psi \right>$ to be defined via local operator insertions which we will also denote as $\psi(x)$ (perhaps smeared appropriately). We place two operators
$\psi$ and $\psi^\dagger$ on the Euclidean section above and below the Cauchy slice $A \cup {\bar A}$ on each replica  The details of this state and the Euclidean path integral used to construct these states will not matter, except to  note that for now we take $\left| \psi \right>$ to be a pure state. We will extend the proof to the case of mixed states in Section~\ref{sec:loose}.

We now write a path integral representation for  ${\rm Tr}_{\bar{A}} \left( \mathcal{O}_{\bar{A}} \left| \psi \right> \left< \psi \right| \mathcal{O}_B \right)$ which differs from $\rho_A$ by the additional $\mathcal{O}$ operator
insertions within the path integral. We imagine slicing open the path integral along radial lines emanating outwards from $\partial A$ and integrating forward in a clockwise angular direction\footnote{We work clockwise because the entangling region $A$ starts on the left of the cut. This results in some funny minus signs, such as the Euclidean holomorphic coordinates close to the entangling surface satisfies $z = - \rho e^{ - i \theta} \rightarrow u = \rho e^s$ where $\theta$ increases in the clockwise direction and $\rho$ is the radius with $\rho >0, \theta=0$ specifying the $A$ region. We wick rotate as $\theta = i s$.  }  - so the ordering of operator insertions (including the operators $\psi(x)$ that create the state) in the $\mathcal{H}_A$ Hilbert space language is always angular ordering. 

Putting the various density matrices together and tracing  we can write the answer as a correlation function on a non-trivial manifold $\mathcal{M}_n(A)$ which consists of $n$ copies/replicas of the $d$ dimensional Euclidean space which are cut and joined cyclicly along $A$. Sometimes we will refer to this space as a branched manifold. The state operator insertions $\psi$ and $\psi^\dagger$ both arise on each replica and $\mathcal{O}_B, \mathcal{O}_{\bar{A}}$ live on the same single replica. Then:
\be
Z_n = \left<  \psi^{\otimes n}  \psi^{\dagger \otimes n} \mathcal{O}_B \mathcal{O}_{\bar A} \right>_{CFT \, {\rm on} \, \mathcal{M}_n}
\ee
where $\psi^{\otimes n}$ means insert the operator symmetrically on each replica.
This is not yet an orbifold correlation function.  The branched manifold can be alternatively represented by using a co-dimension 2 (non-local) twist defect operator living on $\partial A$:
\be
\label{twist}
Z_n \mathop{=}^? \left< \Sigma_n(\partial A)  \psi^{\otimes n}  \psi^{\dagger \otimes n} \mathcal{O}_B \mathcal{O}_{\bar A} \right>_{CFT^n/\mathbb{Z}_n \, {\rm on} \, \mathbb{R}^d}
\ee
where the orbifold/gauging of $CFT^n$ by the discrete cyclic permutation symmetry is necessary in order to remove the existence of $(n-1)$ extra conserved stress energy tensors from the new replicas - thus allowing us to apply standard CFT considerations to the orbifold theory on the original (unbranched) manifold $\mathbb{R}^d$ but now in the presence of a co-dimension $2$ twist operator.\footnote{Orbifolds of 2d CFTs are well studied \cite{polchinski1998string}. The higher dimensional versions have received less attention, see \cite{Belin:2016yll} for a recent discussion which however is complicated by non-trivial topology. We can literally view the resulting theory as a discrete gauging of the replica symmetry, by coupling the theory to a continuum version of a discrete gauge theory as reviewed in \cite{Banks:2010zn}. }
Indeed the state operator insertions are clearly symmetric under the $\mathbb{Z}_n$ symmetry so they are genuine orbifold operators. To unclutter the discussion moving forward we will often suppress the existence of these operators and consider them part of the definition of the twist operator $\Sigma_n(\partial A)   \psi^{\otimes n} \psi^{\dagger \otimes n}  \equiv \Sigma^\psi_n(\partial A) \equiv \Sigma_n$ where the later replacement is for further decluttering purposes.  

Unfortunately the operators $\mathcal{O}_B$ and $\mathcal{O}_{\bar A} $ are quite clearly not orbifold operators, so \eqref{twist} is not yet well defined.  We cannot simply symmetrize each individual $\mathcal{O}_B$ or $\mathcal{O}_{\bar{A}}$ operator over the action of the $\mathbb{Z}_n$ group since the two operators are necessarily inserted on the same replica. Thus we consider $\mathcal{O}_B \mathcal{O}_{\bar A}$ to be a bi-local operator with a non-local string attached whose sole job is to keep track of the relative position of the operator  on the different replicas, say when we move one of the operators around the twist defect relative to the other.
 We can then $\mathbb{Z}_n$-symmetrize this bi/non-local operator by summing this composite over the different replicas. We take this as our definition of \eqref{twist} which we rewrite as:
\be
\label{twist-new}
Z_n = \left< \Sigma_n^\psi(\partial A)  \overbracket{\mathcal{O}_B \mathcal{O}_{\bar A} }\right>_{CFT^n/\mathbb{Z}_n} \qquad  \overbracket{\mathcal{O}_B \mathcal{O}_{\bar A} }
\equiv \sum_{k=0}^{n-1} \mathcal{O}^{(k)}(x_B) \mathcal{O}^{(k)}(x_{\bar{A}})
\ee
where the superscript notation $\mathcal{O}^{(k)}(x_B)$ specifies which replica the operator descends from on $\mathcal{M}_n(A)$.
In Section~\ref{sec:loose} we will discuss a more precise definition of this bi-local operator where the string attached is actually a sum over Wilson lines for the orbifold gauge group $\mathbb{Z}_n$. For now we note that, due to the non-local nature of this operator, we must pick where we place the branch cuts in the definition of $\mathcal{M}_n$ in order to define which local operator $\mathcal{O}_B^{(k)}$ lives on which replica (and thus define what we mean by $k$). Excepting the effective string that remains attached between $x_B$ and $x_{\bar{A}}$ and moves past the twist operator interesecting the region $\bar{A}$, the choice of exactly where we place the branch cut goes away upon moving to the orbifold theory - as it must. 

Now we would like to compute  $K_A \equiv (- \ln \rho_A + \ln \rho_{\bar{A}})/2\pi$ by doing a similar replica trick to compute $\ln \rho_{\bar{A}}$. Notice that the difference here is the positioning of the branch cut. However in the orbifold theory, by definition, there is no knowledge of the position of the branch cut so one might conclude incorrectly that the answer, upon subtraction, is $0$. The reason we find a non-zero answer can be understood since moving the position of the branch cut from $\bar{A} \rightarrow A$ yields a different ordering for the bi-local operator  $ \overbracket{\mathcal{O}_B \mathcal{O}_{\bar A} }$. The conclusion is that we can compute the full modular Hamiltonian as:
\be
\label{sheet}
2\pi \left< \psi \right| \mathcal{O}_B \left( H_A^\psi - H_{ \bar{A}}^\psi \right) \mathcal{O}_{\bar A} \left| \psi \right>
= -  \lim_{n \rightarrow 1} \partial_n \left( \left< \Sigma_n  \overbracket{\mathcal{O}_B \mathcal{O}_{\bar A} }\right>  - \left< \Sigma_n \overbracket{\mathcal{O}_B \mathcal{O}_{\bar A} (\circlearrowleft) } \right>  \right)
\ee 
where for the later non/bi-local operator we have moved the $\mathcal{O}_{\bar{A}}$ relative to 
$\mathcal{O}_{B}$ around the twist operator once. That is:
\be
\overbracket{\mathcal{O}_B \mathcal{O}_{\bar A} (\circlearrowleft)}
\equiv \sum_{k=0}^{n-1} \mathcal{O}^{(k)}(x_B) \mathcal{O}^{(k-1)}(x_{\bar{A}})
\qquad k\equiv k+n
\ee

At this point we have now set up the problem. Computing any of these correlation functions seems difficult. We aim to make progress by bringing the $\mathcal{O}$ operators close to the twist defect $\Sigma$ and using a defect operator product expansion (dOPE.) We turn to this now.

\subsection{Defect OPE}

\label{sec:dope}

\begin{figure}[h!]
\centering
\includegraphics[scale=1]{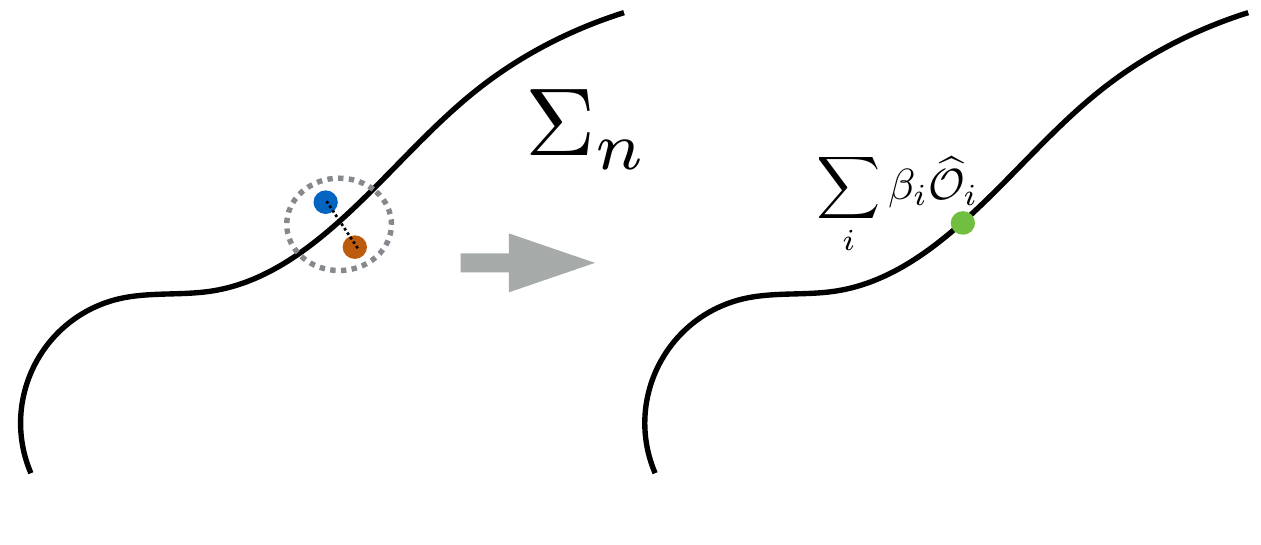}
\includegraphics[scale=.8]{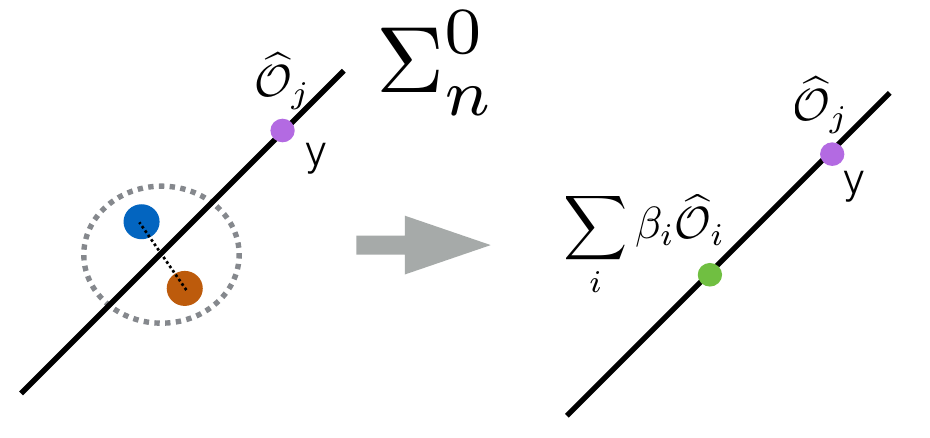}
\caption{(\emph{upper}) The defect OPE argument is based on bringing the two $\mathcal{O}$ operators close
to the defect and replacing these with a sum over defect operators. The dashed line between the operators represents the non-local string we need in order to study these operators in the orbifold theory. The dashed circle represents the radial quantization sphere in the defect theory on which we decompose the state in a basis of local operator insertions at the origin of this sphere. 
(\emph{lower}) The OPE coefficients $\beta^i$ can be computed by making the same replacement, but now on a defect which allows us to do the computation. That is on a flat defect in vacuum. The other operator $\widehat{\mathcal{O}}_j$ is inserted elsewhere on the defect so we can extract the various $\beta^i$. \label{fig:dope} }
\end{figure}

If we take the pair of operators $\mathcal{O}$ close to the defect, say at a point $y=0$ along the defect, we can imagine zooming out and
replacing these with a sum over local defect operators on $\Sigma_n$. That is $ \rightarrow \sum_i \beta^i \widehat{\mathcal{O}}_i(0)$ . Note that we might have done this in two steps, first
replacing the pair of operators by a sum of ambient local orbifold operators using the regular OPE, then bringing
these operators close to the defect. This would look roughly like:
\be
\label{bulkope}
\overbracket{\mathcal{O}_B \mathcal{O}_{\bar A}} \rightarrow \sum_{J} C_{B\bar{A}}^J \mathcal{O}_J \rightarrow \sum_i \sum_{J} Z_J^i C_{B\bar{A}}^J \widehat{\mathcal{O}}_i(0)
\ee
One might even think that there is another way to do this - first bring one of the operators $\mathcal{O}_{\bar{A}}$ close to the defect and expanding this in terms of defect operators. However this later method is not possible because individually $\mathcal{O}_{\bar{A}}$ is not an orbifold operator. Thus there is really only one channel we can evaluate this correlator in.\footnote{In a more familiar setting, if the later channel had been allowed, the equality between these two expansions would be akin to a bootstrap constraint on the defect CFT spectrum. See \cite{Liendo:2012hy,Gliozzi:2015qsa,Billo:2016cpy} for recent work on this bootstrap problem. } Additionally we will choose to ignore the intermediate step of the ambient OPE involving the sum over $J$ in \eqref{bulkope} - mostly because it turns out in the limit $n \rightarrow 1$ we can directly compute the dOPE coefficients $\beta^i$ for this full replacement  $\overbracket{\mathcal{O}_B \mathcal{O}_{\bar A}} \rightarrow \sum_i \beta^i \widehat{\mathcal{O}}_i(0)$ without having to sum over an infinite set of intermediate operators. 

We can compute the dOPE coefficients $\beta^i$ as follows. Firstly note that since the replacement is done locally we could have done the same replacement on a twist defect within a totally different setup but using the same replacement coefficient $\beta^i$.\footnote{We are lying a little here. It turns out that $\beta^i$ is sensitive to the local extrinsic curvature of the defect at $y=0$, in analogy to regular OPE
coefficients being sensitive to local curvature invariance of the metric if we make an OPE expansion of two local CFT operators in curved space \cite{Hollands:2008vx}. We will fix this lie in Section~\ref{sec:loose}. } Thus let us consider a flat/planar twist defect defined in the vacuum of a CFT and living along $\partial A_0$. To extract a particular OPE coefficient we must also insert some other defect operator $\widehat{O}_j$ far away from $0$ at a point $y$. All of this still in the presence of the bi-local $\overbracket{\mathcal{O}_B \mathcal{O}_{\bar A}}$. Now in this new setup take the $\mathcal{O}_{B,{\bar{A}}}$'s close to the defect simultaneously and make the same replacement we did above:
\be
\left< \Sigma_n^0 \widehat{O}_j(y) \overbracket{\mathcal{O}_B \mathcal{O}_{\bar A}} \right>
= \sum_{i} \beta^i G_{ij} \qquad G_{ij} =  \left< \Sigma_n^0  \widehat{\mathcal{O}}_i(0)  \widehat{\mathcal{O}}_j(y) \right> 
\ee
where in the notation we established above $\Sigma_n^0 = \Sigma_n^{\mathbb{1}}(\partial A_0)$ and $A_0$ is the uniform half space Rindler cut and the superscript $\mathbb{1}$ denotes the state operator insertions appropriate for the vacuum $\left| 0 \right>$. See Figure~\ref{fig:dope} for a schematic of this replacement.  This allows us then to extract $\beta^i$ after inverting the operator metric defined from the two point function of defect operators on the planar defect:
\be
\label{dope}
 \left< \Sigma_n  \overbracket{\mathcal{O}_B \mathcal{O}_{\bar A}} \right> 
 = \sum_{ij} \left< \Sigma_n \widehat{ \mathcal{O}}_i (0) \right> \left(G^{-1}\right)^{ij} \left< \Sigma_n^0 \widehat{O}_j(y) \overbracket{\mathcal{O}_B \mathcal{O}_{\bar A}} \right>
 \ee
The result is written as a sum over all local defect operators; and we are not being careful about the distinction between defect primaries and descendants, which is not really important as long
as we compute the operator metric $G_{ij}$ carefully. Actually the operators we will eventually be interested in will all be defect primaries such that $G$ will be diagonal. 

Up until now we have kept things general, and for integer $n$ all of the above should make sense.
However controlling the spectrum of defect operators and the resulting $\beta^i$ OPE coefficients is difficult. If we can now argue for an analytic continuation in $n$ then it turns out  there are several big simplifications that occur when taking the limit $n \rightarrow 1$. 

One of these simplifications is that we can move the $\partial_n$ so it only acts on the
last term in \eqref{dope} - the three point function term. This is because this is the only term that knows about the $\mathcal{O}_{B,\bar{A}}$ operator replica ordering which was discussed aobve. So when we compute $\ln \rho_A - \ln \rho_{\bar{A}}$ this term would vanish at $n=1$ since then there is only one replica   and the difference $\overbracket{\mathcal{O}_B \mathcal{O}_{\bar A} } -\overbracket{\mathcal{O}_B \mathcal{O}_{\bar A} (\circlearrowleft) }$ vanishes. So if the $\partial_n$ acts anywhere else the three point function term would give zero as we send $n \rightarrow 1$. 

The other simplification is that for small $n \approx 1$ the various correlators we need to compute in the presence of $\Sigma_n^0$ are fixed in terms of CFT correlators in flat space plus insertions of the modular Hamiltonian associated to the Rindler cut. Since this later insertion is a known integral over the CFT stress tensor \cite{Casini:2011kv} we can compute $\beta_i$ using the local data of the CFT. 

So we need to argue for an analytic continuation in $n$ of the defect operator spectrum as well as the  OPE coefficients $\beta^i$ computed in \eqref{dope}. Thus we turn to a discussion of the defect operator spectrum. 

\subsection{Local defect operators}

We start by reviewing what is known about local ambient space operators (away from the defect) for the replicated orbifold theory. They take the schematic form:
\be
\label{bulkorb}
\mathcal{O}_{ \{ \alpha_k\}} =  \bigotimes_{k=1}^n \mathcal{O}_{\alpha_k}^{(k)} + \mathbb{Z}_n {\rm- symm}
\ee
where we sum over $\mathbb{Z}_n$ cyclic permutations of the different replicas.
The conformal dimension of this operators is $\Delta_{ \{ \alpha_k\}} = \sum_k \Delta_{\alpha_k}$.
The operators are located at the same point on each replica.
In terms of including the effect of these ambient operators in certain entanglement computations the analytic continuation in $n$ has been successfully found for low dimension operators with a small number (fixed and independent of $n$)  of non unit operators inserted on each replica. These operators are important when replacing non-local twist operators with local operators when viewed from a distance, as in \cite{Headrick:2010zt,Calabrese:2010he,Cardy:2013nua,Agon:2015twa,Agon:2015ftl}, or when bringing a twist operator close to an anti-twist operator, as in \cite{Bousso:2014uxa,Casini:2017roe}. Note that our dOPE  should not be confused with the various OPE arguments used in these papers. 

Here we will only concern ourselves with single operator insertions on one replica symmeterized appropriately:
\be
\label{bulkone}
\mathcal{O}_{\alpha} \equiv \mathcal{O}_\alpha \otimes \mathbb{1} \ldots \otimes  \mathbb{1} 
+ \mathbb{Z}_n {\rm-symm} \equiv \sum_{k=0}^{n-1} \mathcal{O}_\alpha^{(k)}
\ee
where hopefully the notation does not cause confusion.
In order to discover the defect operator spectrum we need to take the local ambient orbifold operators \eqref{bulkorb} and bring them close to the defect. Any defect operator that is not discoverable in this way will not contribute to the answer in \eqref{dope}. 
In Appendix~\ref{app:multi} we argue that one can reproduce the full set of such defect operators by limiting oneself to single replica operators, \eqref{bulkone}, and bringing these close to the defect. Or in other words the more general ``multi-replica'' operators given in \eqref{bulkorb} do not add to the list of local defect operators when we bring \emph{these} close to the twist defect.  

Schematically we should find for the single replica bulk operators when we bring them close
to the defect we can rewrite these as a sum over defect local operators $\widehat{O}_j$:
\be
\label{dexp}
\lim_{ |w| \rightarrow 0} \sum_{k=0}^{n-1} \mathcal{O}_\alpha^{(k)} (w,\bar{w},0) \Sigma_n =
w^{-(\Delta_\alpha + \ell_\alpha)/2} \bar{w}^{-(\Delta_\alpha - \ell_\alpha)/2}\sum_{j} Z^j_\alpha w^{ (\widehat{\Delta}_j  +\ell_j )/2 }\bar{w}^{ (\widehat{\Delta}_j -\ell_j)/2 } \widehat{\mathcal{O}}_j(0) \Sigma_n
\ee
where we have written the expansion in Euclidean coordinates about the defect\footnote{With $w = -u$ and $\bar{w} = v$. The former minus sign is annoying and means we label operators in real times and imaginary times differently. Only the $B,\bar{A}$ operators, and later the one point functions $\left< T_{--}(u,v=0) \right>_\psi$ will be inserted in real
times and are labelled by $(u,v,y)$ while all other operators are labelled with $(w,\bar{w},y)$. Hopefully the distinction will be clear. }:
\be
ds^2 =  dw d\bar{w} + d y^2
\ee
Here $\widehat{\Delta}_j$ is the defect operator dimension and $\ell_j \in \mathbb{Z}$ is the angular momentum around the transverse plane to the defect, i.e. associated to the charge under $SO(2)$ rotations $w \rightarrow w e^{ - i \phi}$. 
Spinning ambient operators should be decomposed under the action of the $SO(2) \times SO(d-2)$ subgroup of the full Euclidean rotation group, and in this paper we will only need to consider scalar operators under $SO(d-2)$. 
If the resulting operator $\mathcal{O}_\alpha$ transforms nontrivially under $SO(2)$ rotations then $\ell_\alpha \neq 0$. 

In order to extract the operators in \eqref{dexp} generically we could draw a set of small radial quantization spheres $S_{d-1} \times \mathbb{R}^+$ around the point $y=0$ on the defect. The Hilbert space on the sphere $S_{d-1}$ is associated to the defect CFT (the defect lives on an $S_{d-3}\times \mathbb{R}^+$ subspace in radial quantization coordinates) with the symmetry group $SO(2)\times SO(d-1,1)$ of conformal transformations holding fixed the defect. We can then decompose operators into primaries and descendants and extract these operators by acting with the appropriate projection operators made out of Casimirs etc. This is a somewhat tedious procedure, especially for spinning operators. For the class of operators we will be interested in there is a quicker way. 

Let us consider the lowest dimension defect operator $\widehat{\Delta}_{\ell}$ of fixed spin $\ell$. Then we can extract this operator via a limit
\be
\label{dproj}
\widehat{\mathcal{O}}_{\ell}(0)  \Sigma_n
= \lim_{ |w| \rightarrow 0}\frac{ |w|^{ -\widehat{\tau}_\ell + \tau_\alpha } }{2\pi i}\oint \frac{d w}{w} w^{-\ell+\ell_\alpha} \sum_{k=0}^{n-1} \mathcal{O}_\alpha^{(k)} (w, |w|^2/w,0) \Sigma_n
\ee
where $\widehat{\tau}_\ell \equiv \widehat{\Delta}_\ell - \ell$ and $\tau_\alpha = \Delta_\alpha - \ell_\alpha$ define the twist
of the defect and ambient space operators respectively. 
It is natural to normalize these operators such that $Z_\alpha^{\ell} = 1$ which means the overall coefficient in their two point functions cannot be independently set to $1$. These operators will be the leading operators of interest to us when we take the lightcone limit since they have minimal twist at fixed $SO(2)$ spin. For this reason they are also necessarily defect primaries. 

We now argue for an analytic continuation in $n$. The most important thing that such a continuation should satisfy is locality of the bulk orbifold operators relative to the twist defect - that is, in Euclidean, moving a bulk operator by an angle $2\pi$ around the twist defect should return the original operator. 
 This quantizes $\ell_j \in \mathbb{Z}$ where $\mathbb{Z}/n$ is not allowed but would have been possible if we had not gauged the $\mathbb{Z}_n$ symmetry. 
This is also important to ensure  a consistent defect theory with well defined OPE coefficients
for all values of $n$. In particular this means that $\ell_j$ will stay an integer under $n$-analytic continuation, ruling out things like $ \ell_j \mathop{=}^? p n $ for $p \in \mathbb{Z}$.  All of these requirements are actually in line with the $n$ continuation advocated in \cite{Lewkowycz:2013nqa,Faulkner:2013yia} for computing holographic Renyi entropies.  With this in mind an appropriate continuation would define for real (and complex) values of $n$: the dimensions $\widehat{\Delta}_j(n)$ and the various defect OPE coeffcients $\beta^j(n)$ and $Z^j_J(n)$ which agree for integer $n$ are suitably well behaved for large $n$ and analytic in $n$.

How might we extract the defect operator dimensions as a function of $n$? 
We can think of two ways to study this. The first method only applies to holographic theories, however it is useful to give intuition into this problem and will give a method to extract the spectrum of defect operators for all $n$. Versions of this problem have been studied previously in holographic defect theories \cite{Aharony:2003qf} and it does not require much to adapt to the replica problem at hand \cite{twistdisplacement}. Essentially the idea is to study fields propagating on the Hyperbolic black hole, which is dual to the twist defect $\Sigma_n^0$ in a conformal frame where the defect lives at the boundary of $\mathbb{H}_{d-1} \times S^1$ . Then the spectrum of defect operator dimensions can be extracted by solving simple wave equations in this black hole subject to certain boundary conditions. Since this approach does not work for general CFTs we leave the details of this to Appendix~\ref{app:defect} where we apply it to the specific problem at hand. See Figure~\ref{fig:stressplot} below for the most illuminating picture of the defect spectrum that results. This then gives us a check on the second method that works only for $n \approx 1$ but now for general theories. 

The second way to extract the spectrum is to imagine we have on hand the ambient space two point function in the presence of the flat Rindler defect $\Sigma_n^0$. Actually it is rather simple to $n$-analytically continue this two point function following \cite{Faulkner:2014jva} where the answer can be
written in terms of a thermal correlator, at temperature $1/(2\pi n)$ for the CFT on $\mathbb{H}_{d-1} \times S^1$. While this thermal correlator is not known in general, it is however computable for $n=1$ and in an expansion about $n=1$.

For example $n=1$ gives back the CFT two point function on flat space to leading order. This two point function decomposes into correlators of defect operators living on a now imaginary defect lying along the Rindler cut. This decomposition can be understood as branching the operator representation of $SO(d+1,1)$ into representations of the subgroup $SO(2) \times SO(d-1,1)$, thus these defect operators have simple conformal dimensions related to the CFT operator dimension $\widehat{\Delta}(n=1) = \Delta + \mathbb{Z}_{\geq 0}$. Then it is interesting to understand the leading $(n-1)$ correction to these operators. We expect a correction to the conformal dimension $\widehat{\Delta}(n) = \widehat{\Delta}(1) + \mathcal{O}(n-1)$ and perhaps some mixing with other bulk operators, however in addition to this we will also find a new phenomonon can occur: completely new operators can arise that effectively decouple exactly at $n=1$ and which were not visible at the leading order. These only arise from spinning operators with spin $\geq 2$ and the displacement operator is an example which arises from the stress tensor. These new operators will play an essential role moving forward. 

\subsubsection{Example: scalar two point function}

\label{sec:scalar}

Let us go through a simple example using scalar orbifold operators built out of the CFT scalar operator $\phi$. Following \cite{Faulkner:2014jva}\footnote{The results in \cite{Faulkner:2014jva} were also derived without reference to the replica trick.  See also \cite{Faulkner:2017tkh,Faulkner:2015csl}. It is likely that our results here can also derived avoiding any mention of the replica trick, but at this time we have not worked this. } and the procedure outlined above  we consider the two point function and give an $n$ analytic continuation:
\be
\left< \Sigma_n^0 \phi   \phi \right>
= n \sum_{j=0}^{n-1} \left< \Sigma_n^0 \phi^{(j)} \phi^{(0)} \right> 
=\frac{n}{2\pi i}  \oint_C \frac{ d\lambda}{\lambda} \frac{1}{\lambda-1} \left< \phi ( z/\lambda, \bar{z} \lambda,y)
 \phi (w, \bar{w},0) \right>_{\mathcal{M}_n^0}
 \label{fform}
\ee
where the later correlator lives on the branched replica manifold (i.e. $n$ copies of the CFT with a $d-1$ dimensional branch cut/plane along the Rindler cut $\partial A_0$.) The $\lambda$ integral encircles $n$ poles on each replica at $\lambda =1$. The complex $\lambda$ plane reflects the branching structure of $\mathcal{M}_n$ along a fixed $y$ slice (i.e. there is a branch cut starting at $\lambda =0$ which we take to run along the negative real axis). The correlator additionally has branch cuts which for $y=0$ start at $\lambda = z/w$ and $\lambda = \bar{w}/\bar{z}$ due to lightcone singularities. These properties are totally general, and the simplest way to understand them is to conformally map $\mathcal{M}_n^0$
to $\mathbb{H}_{d-1} \times S_1$:
\be
ds^2 = \left( d\tau^2 + \frac{ d\rho^2 + d y^2}{\rho^2} \right) \qquad w = \rho e^{- i \theta}
\ee
the $S_1$ factor has length $ \theta \equiv \theta + 2\pi n$. The correlator maps to a thermal correlator for the CFT on the spatial manifold $\mathbb{H}_{d-1}$ and the $j$ sum over replicas is a sum over $\theta
= 2\pi j$ which we turn into a contour integral over the complex $s = - i \theta$ strip for $-2\pi n < {\rm Im} s < 0$ . The final form in \eqref{fform} follows from setting $\lambda = e^{-s}$ and passing back to the flat conformal frame. 

\begin{figure}[h!]
\centering
\includegraphics[scale=.75]{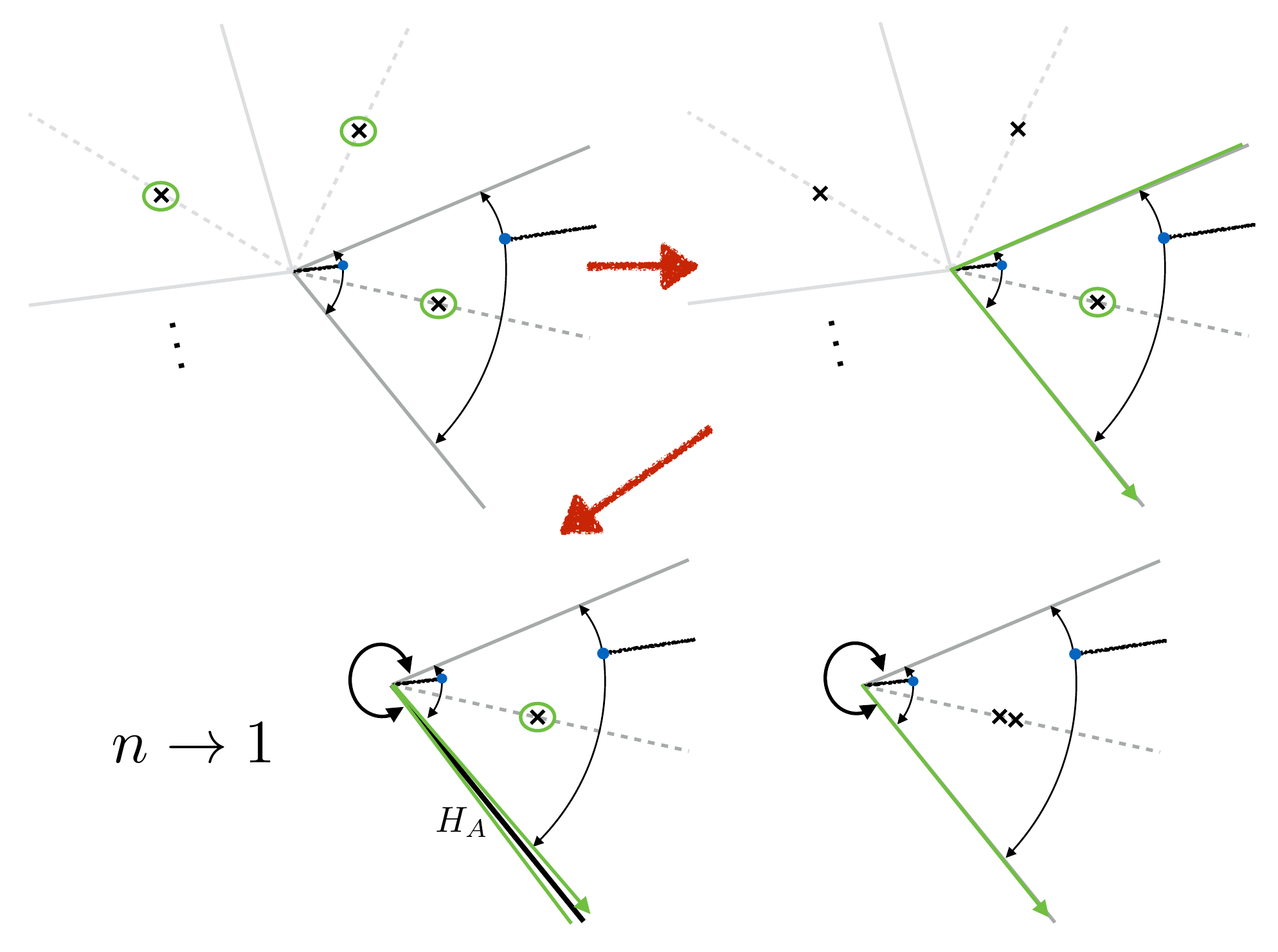}
\caption{\label{fig:cont-2pt} An illustration of the procedure to analytically continue the two point function in $n$ and extract the limit as $n \rightarrow 1$. We schematically plot the $\lambda$ plane with the slices of the pie representing different replicas and with the dashed lines lying along the positive real $\lambda$ axis on each replica. We integrate $\lambda$ over the green curves and the fuzzy black lines represent branch cuts coming from lightcone singularities. 
The cross represents a simple pole and the double cross is a double pole. The two top figures are at fixed integer $n$. The continuation proceeds from the second top picture and the leading correction as $ n \rightarrow 1$ comes from pulling down a factor of the modular Hamiltonian which is wedged between the straight line integration contour. This results in a commutator which gives rise to the double pole shown on the bottom right figure. The first term in \eqref{bothscalar} comes from the integral around the simple pole in the bottom left figure.  }
\end{figure}

Right now $n$ should still be an integer. However we can  pick the $C$ contour so that both the operators in \eqref{fform} stay on a single replica - that is $\lambda$ should itself remain on a single replica. The $n$ analytic continuation is then manifest since the two point function is well defined on   $\mathcal{M}_n$ for non integer $n$ due to its relation to a thermal correlator on $\mathbb{H}_{d-1}$
and the contour no longer depends on $n$ being integer. For example we can pick $C$ to wrap the replica branch cut around $\lambda =0 \rightarrow -\infty$ and the one remaining pole at $\lambda =1$ (see Figure~\ref{fig:cont-2pt}). A more detailed explanation for why this is the correct $n$ analytic continuation can be found in \cite{Faulkner:2014jva}.

With this $n$-continued correlator in hand we can in principle extract the spectrum of defect operators that couple to this operator $\phi$. Of course we do not in general know this correlator and so we must work with $n \approx 1$. The leading order answer at $n=1$ is then unsurprisingly just the flat space CFT correlator:
\be
\label{n0}
\left< \Sigma_1^0\phi \phi  \right>
=  \left< \phi( z,\bar{z}, y) \phi(w,\bar{w},0 ) \right> \ = \frac{c_\Delta}{( (w-z)(\bar{w}-\bar{z}) + y^2 )^{\Delta}} 
\ee
and the next correction is:
\begin{align}
\label{bothscalar}
(\partial_n-1) \left. \left< \Sigma_n^0\phi \phi \right> \right|_{n=1}
 = &\left( \vphantom{\int_0^\infty} - 2\pi \left< H_{A}^0 \phi( z,\bar{z},y) \phi(w,\bar{w},0 ) \right> \right. \\  & \qquad \left. - \int_0^{-\infty} \frac{d\lambda}{(\lambda-1)^2} \left< \phi( z/\lambda,\bar{z}\lambda,y) \phi(w,\bar{w},0 ) \right> \right)
 \nonumber
\end{align}
where at this order the $\partial_n$ pulls down the half modular Hamiltonian and 
the first term above comes from the one remaining pole at $\lambda=1$  (see the bottom part of Figure~\ref{fig:cont-2pt}). The shift $\partial_n \rightarrow
(\partial_n -1)$ can be achieved via an $(n-1)$ correction to the overall coefficient of the $n=1$ two point function which has a trivial effect on the defect spectrum. 

A very important property of \eqref{bothscalar} is that if we move the operator
$\phi(w,\bar{w},0)$ around the would be defect we get back to the original correlator - this was one of our requirements for an $n$ analytic continuation.  This happens here because
of a conspiracy between the two terms of \eqref{bothscalar}. In the second term we must deform the $\lambda$ contour as we move the operator around - this leaves a contribution from the double pole at $\lambda=1$ giving a commutator $-2\pi \left< \left[ H_A, \phi(z,\bar{z}) \right] \phi(w,\bar{w} ) \right>$ which cancels with the same commutator  coming from the first term in \eqref{bothscalar} which arrises as we move the operator past the modular Hamiltonian insertion. 

We can expand the expressions above at small $w,\bar{w},z,\bar{z}$ and extract from this the defect operator spectrum.  To directly extract the defect \emph{primary} operators we would need to study  the defect channel conformal blocks which can be found in \cite{McAvity:1995zd, Liendo:2012hy}. This is rather complicated and we are not interested in the complete spectrum of defect operators. Instead we will focus on the operators relevant for the lightcone limit where we set $ \bar{w} \rightarrow 0$ and $z \rightarrow 0$. We can smoothly take this limit on \eqref{n0} and this fact, along with the defect expansion given in \eqref{dexp}, tells us that there must be a set of defect operators with $SO(2)$ ``spin'' $\ell$ and $\widehat{\Delta}_\ell = \Delta + \ell$. Furthermore these operators are necessarily primaries. These are found by expanding the remaining correlator:
\be
\label{scalar0}
\left< \Sigma_1^0\phi \phi  \right>
= \frac{c_\Delta}{( - w \bar{z} + y^2 )^{\Delta}}  =  y^{-2\Delta}  \sum_{\ell=0}^\infty g_\ell^\phi \left( \frac{ w\bar{z} }{y^2} \right)^{\ell}
\qquad g_\ell^\phi =c_\Delta \frac{\Gamma(\Delta+\ell)}{\Gamma(\Delta) \ell! } 
\ee
thus $\ell \geq 0$ for these operators. We could have also extracted these operators using the projection \eqref{dproj} since they are the lowest dimension operators with fixed $\ell$. Also note that for $\ell < 0$ we would
take the limit $w \rightarrow 0$ and $\bar{z} \rightarrow 0$ and reproduce a similar set of operators
now with $\widehat{\Delta} = \Delta + | \ell|$. 

At the next order in the $(n-1)$ expansion attempting to directly set $\bar{w} \rightarrow 0$ and $z \rightarrow 0$ fails due to divergences that arise. This failure results in new $\log$ terms that can be understood as giving rise to anomalous defect dimensions. These $\log$'s are quite intricate and involve a delicate interplay between the two terms in \eqref{bothscalar}. 
The first term can be computed using the expression for the stress tensor OPE block of two auxiliary time like separated scalars, which turns out to be just the half modular Hamiltonian operator for the double cone region between the two operators \cite{Czech:2016xec,deBoer:2016pqk}. This region can be conformally mapped to the Rindler wedge.
Thus the first term is just the stress tensor conformal block for the two auxiliary operators and the two $\phi$ operators. The details of this computation and the lightcone expansion are left to the Appendix~\ref{app:defect}. 
We find:
\be
\label{n1}
\partial_n \left. \left< \Sigma_n^0 \phi \phi \right> \right|_{n=1}
= c_\Delta y^{-2\Delta} \left( P^\phi( w\bar{z}/y^2) + Q^\phi (w\bar{z}/y^2) \ln\left( \frac{z\bar{z} w\bar{w}}{y^4}\right) \right)
\ee
where $P(x)$ and $Q(x)$ have  power series expansions in their arguments and while $P$ is rather complicated we can write explicitly:
\be
\label{q1}
Q^\phi (x) =-  \frac{\Delta}{2 (d-1)} (1 - x)^{-\Delta +h-1}
= \sum_{\ell} q_\ell^\phi x^\ell
\ee
We note some consequences of this result. Most importantly there are no new operators that arise at order $(n-1)$. This is because $P^\phi$ and $Q^\phi$ have a regular Taylor series so, comparing to \eqref{scalar0}, $P^\phi$ only contributes to an order $(n-1)$ shift in the two point function of these operators ($\propto c_\Delta g_\ell^\phi+ \mathcal{O}(n-1)$) and $Q^\phi$ results in a shift in the scaling dimensions:
\be
\widehat{\Delta}_\ell = \Delta + \ell + (n-1) \frac{2 q_\ell^\phi}{g_\ell^\phi} + \ldots
\qquad \ell=0,1,2,\ldots
\ee
where explicitly the anomalous dimensions are:
\be
\left. \partial_n \widehat{\Delta}_\ell \right|_{n=1} = - \frac{\Delta}{(d-1)}
\frac{ (\Delta-h+1)_\ell }{(\Delta)_\ell }
< 0 
\ee
and $(x)_\ell \equiv \Gamma(\Delta+\ell)/\Gamma(\Delta)$ is the Pochhammer symbol. 
These shifts are always $(n-1) \times$ a negative number if $\Delta > h-1$ satisfies the unitarity bound.\footnote{This agrees with an independent calculation of these anomalous dimensions \cite{Lemos:2017vnx}. We thank the author's for pointing out a typo in their footnote 31. It is fixed in this version. }
While the details of this $(n-1)$ shift in the defect operator spectrum is not important for our final goal, we went through this exercise to make sure we have a good understanding of the important defect operator spectrum. We have also checked that this small $(n-1)$ correction is in agreement with the equivalent holographic technique for computing $\widehat{\Delta}(n)$, summarized in Figure~\ref{fig:scalaranalytic} of the Appendix.

\subsubsection{Stress tensor and the appearance of the displacement operator}

We will now analyze the defect operators that appear in the stress tensor channel. Again motivated by the lightcone limit we will limit ourselves to the correlator
of $T_{--}$ and $T_{++}$ where the $\pm$ are in the transverse directions to the defect.
These operators will give rise to defect operators with the lowest twist. The sum over replicas can be written:
\be
\label{tlam}
\left< \Sigma_n^0 T_{++} T_{--} \right> = \frac{1}{2\pi i} \int_C \frac{d\lambda}{\lambda} \frac{\lambda^2}{\lambda-1} \left< T_{++}( z/\lambda, \bar{z}\lambda,y) T_{--}(w,\bar{w},0) \right>_{\mathcal{M}_n}
\ee
Note the extra factor of $\lambda^2$ relative to \eqref{fform} which arrises from the rotation/boost applied to $T_{++}$. The contour is chosen as before to allow for an $n$ continuation. The $n=1$ answer is just the CFT correlator:
\be
\left< T_{++}(z, \bar{z},y) T_{--}(w,\bar{w},0) \right> = \frac{ c_T y^4}{4 \left( (w-z)(\bar{w}-\bar{z}) +y^2\right)^{d+2}}
\ee
from which we again find operators in the lightcone limit $\bar{w},z \rightarrow 0$ labelled by their spin and with:
\be
\label{stressops}
\widehat{\Delta}_\ell = d+ (\ell-2) \qquad \ell = 2,3,4, \ldots
\ee
as well as the conjugate operators with opposite charges $\ell \rightarrow -\ell$ but the same scaling dimensions. The shift in the spin arrises because $T_{--}$ already transforms with charge two under $SO(2)$ rotations so $\ell_\alpha = 2$ in \eqref{dexp}. If we had set $w,\bar{z} \rightarrow 0$ we would have found a set of operators with scaling dimensions $\widehat{\Delta}_\ell' = d- (\ell-2)$ for $\ell = 2,1, \ldots, -\infty$ which are now however not the same as the conjugate operators we found above - indeed they are not even operators with minimal twist $\widehat{\Delta}_\ell' - \ell = d+2$. This happens because of the asymmetry in the lightcone limit for spinning operators.

We will need to extract the two point function of these twist $d-2$ operators:
\be
\label{agl}
\left< T_{++}(0, \bar{z},y) T_{--}(w,0,0) \right> 
= y^{-d} \sum_{\ell=2}^\infty g_\ell \left( \frac{ w\bar{z} }{y^2} \right)^{\ell-2} \qquad
g_\ell  = c_T \frac{  \Gamma(d+\ell)}{4 \Gamma(d+2) (\ell-2)! }
\ee

One might have imagined that the $(n-1)$ corrections works as above for the scalar case - the
two point functions and scaling dimensions will shift by small amounts. There is however one new phenomenon that arrises from the $\lambda$ integral in \eqref{tlam}. The $(n-1)$ correction can be written as:
\be
(\partial_n-1) \left< \Sigma_n^0 T_{++} T_{--} \right>|_{\bar{w},z \rightarrow 0}
= -\int_0^\infty \frac{ d\lambda \lambda^2}{(\lambda-1)^2} \frac{ c_T y^4}{ 4 ( -w \bar{z} \lambda + y^2)^{d+2}} + H_A^0{\mathrm-term}
\label{hhterm}
\ee
We analyze the integral by rescaling $\lambda \rightarrow \lambda y^2 /(w\bar{z})$ and then expanding the $(\lambda- (w\bar{z}/y^2) )^{-2}$ term:
\be
\label{tterms}
= -c_T y^{-2d} \frac{y^2}{4 w\bar{z}}  \int_0^\infty d\lambda  \left( 1 + 2 \frac{ w\bar{z}}{y^2 \lambda} + 3 \left(\frac{ w\bar{z}}{y^2 \lambda}\right)^2 + \ldots  \right) \frac{ 1}{( -\lambda +1)^{d+2}}  
\ee
Note that the second term and higher in the expansion in brackets have divergent $\lambda$ integrals indicating an issue with the uniformity of this expansion as a function of $\lambda$ - a more careful limiting procedure will resolve this divergence into a $\ln(w\bar{z})$ term which then must be added to the $\ln$'s arising in the modular Hamiltonian term of \eqref{hhterm}. As with the scalar case we expect an overall $\ln (w\bar{w} z\bar{z})$ which results in a shift of the defect dimensions as in \eqref{n1}. However more interestingly the first term in  \eqref{tterms} is finite and can be seen to give rise to a new operator with $\ell=1$ and
\be
\widehat{\Delta}_1 = d-1 \qquad g_{1} = c_T \frac{(n-1)}{4 (d+1)}
\label{ag1}
\ee
This is not surprisingly the displacement operator \cite{Bianchi:2015liz,twistdisplacement}. We have then re-derived the results of \cite{Bianchi:2015liz,Faulkner:2015csl} for the coefficient of the two point function of the displacement operator as $n \rightarrow 1$:  $C_D = 4 g_{1} = c_T (n-1)/(d+1)$. The fact that $g_1 = \mathcal{O}(n-1)$ has interesting consequences for the defect OPE that we are interested in, and we will explore this in the following section. For now we simply note that this displacement operator has the same twist $d-2$ as the other operators we singled out \eqref{stressops}. 

The  $ \mathcal{O}(n-1)$ corrections to $g_\ell$ and $\widehat{\Delta}_\ell$ for $\ell \geq 2$ that we could attempt to find  using the correlators in \eqref{hhterm}
at this order are not important since they are only subleading effects to the defect operators we already found at $n=1$. Thus we will not bother to track down the anomalous dimensions etc.\footnote{The computation is even more involved than for the scalar operator considered above. For example since the modular Hamiltonian term in \eqref{hhterm} is now important the stress tensor three point function will play a role and this is rather involved to work with.}

\begin{figure}
\includegraphics[trim={-3cm 0cm 0cm 0cm},scale=0.8]{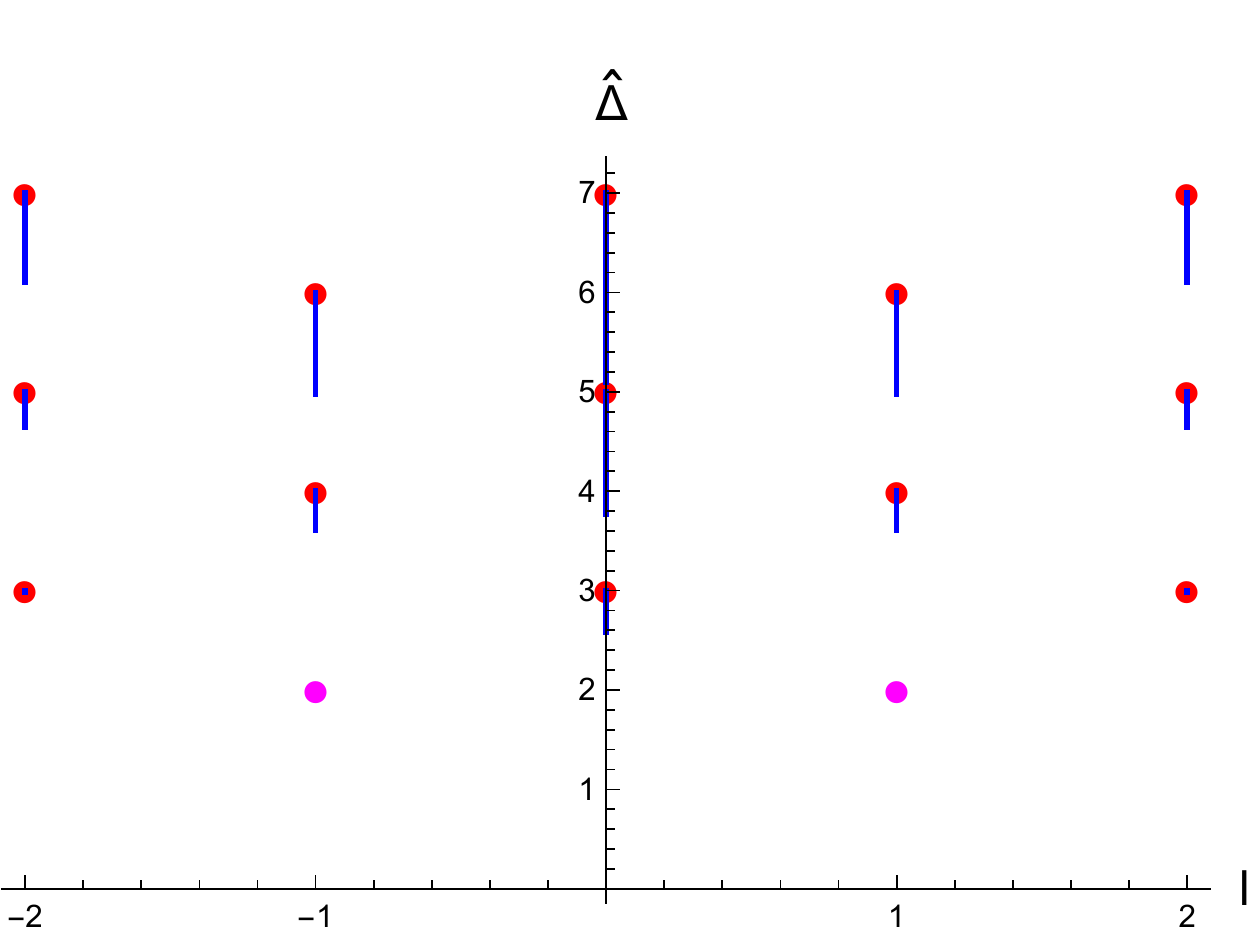}
\caption{Plot of $\widehat{\Delta}$ vs $l$, for the defect operators that appear in the dOPE of the stress tensor for a holographic model. The red dots are the values at $n=1$ and these results are universal to any CFT. The blue lines show the change in the values of the scaling dimension for $n \in (1,2)$. The displacement operator (shown in pink), which has a protected scaling dimension of $\widehat{\Delta} = d-1 = 2$, appears only for $n \neq 1$.}
\label{fig:stressplot}
\end{figure}

In the Appendix~\ref{app:defect} we computed the defect operator spectrum arising from the stress tensor directly using a holographic model. This gives a more complete picture of the spectrum of defect primaries that are scalars
under $SO(d-2)$ rotations as a function of $n$. Since our general CFT computations agree with the holographic ones close to $n=1$ we expect we have not missed any of the important defect operators. The picture we have of the defect spectrum is usefully summarized in Figure~\ref{fig:stressplot}.

\subsubsection{Higher spin versions of the displacement operator}

Using the same techniques, we can also compute the spectrum of displacement operators coming from a higher-spin operator $\mathcal{J}_{\mu_1 \mu_2 ... \mu_J}$ with conformal dimension $\Delta_J$ and spin $J > 2$ (we consider only symmetric traceless representation of $SO(d)$ rotations.) 
The minimal twist operators arise from $\mathcal{J}_{+...+}$ and $\mathcal{J}_{-...-}$ which we focus on. 
Note that these are not  conserved currents and $\Delta_J > d+J-2$.

Analogous to the case of stress tensor, the associated displacement operators emerge at order $\mathcal{O}(n-1)$ and can be computed by extracting the power-law terms $y^{-2\widehat{\Delta}_\ell}$ at order $\mathcal{O}(n-1)$ in the higher-spin two-point function:
\begin{eqnarray}
&&\langle \Sigma_n^0 \mathcal{J}_{+ \ldots +}\left(0,\bar{z},y\right)\mathcal{J}_{- \ldots -}\left(w,0,0\right)\rangle_{CFT^n/\mathbb{Z}_n} \mathop{=}_{n \approx 1} \sum^{J-1}_{\ell = 1} g_\ell y^{-2\widehat{\Delta}_\ell}\left(\bar{z}w\right)^{-J+\ell} + ...\nonumber\\
&& \widehat{\Delta}_\ell=\Delta_J-J+\ell, \qquad \;\;g_\ell =-\left(n-1\right)(-1)^{\ell-J}c_J \Gamma\left(J-\ell\right)\ell \frac{\Gamma\left(\Delta_J+\ell\right)}{\Gamma\left(\Delta_J+J\right)},\;
\label{eq:hs2PF}
\end{eqnarray}
where the $ \ldots$ in the first line include terms that correspond to ``normal" defect operators that survive at $n=1$, as well as their corrections at higher orders of $\mathcal{O}\left(n-1\right)$; $c_J$ is the coefficient of the CFT $\langle \mathcal{J}\mathcal{J}\rangle$ two-point function. We conclude that when we extend to the case of higher-spin operator with spin $J$, there emerges a family of $J-1$ displacement operators $\widehat{\mathcal{D}}^\ell$ labelled by their $SO(2)$ spin with $1\leq\ell\leq J-1$.

We are not completely sure how to interpret these new defect operators. One might conjecture that in holographic theories they represent new very massive fields living on the RT surface. However in the holographic case one works in a large $N$ theory and there will be double trace operators of higher spin that should equally well give rise to double trace displacement operators. We leave speculation about these to the discussion section. Either way the particular displacement operator $\widehat{\mathcal{D}}^{J-1}$ coming from an ambient operator $\mathcal{J}_{- \ldots -}$ of lowest twist at fixed \emph{even} $J$ play a definitive role when we examine a higher spin version of the QNEC as we will discuss in Section~\ref{sec:loose}.

\subsection{Summary}

\label{sec:summary}

In summary we will concentrate on the following defect operators coming from a limit of the bulk stress tensor close to the defect:
\begin{align}
\nonumber
\widehat{T}_\ell \equiv \frac{1}{2\pi i }\sum_{k=0}^{n-1} \oint \frac{d w}{w} w^{-\ell+2} T_{--}^{(k)}(w,0,0)  \qquad \ell =1,2, \ldots \\
\widehat{T}_{-\ell}  \equiv - \frac{1}{2\pi i } \sum_{k=0}^{n-1} \oint \frac{d \bar{z}}{\bar{z}} \bar{z}^{-\ell+2} T_{++}^{(k)}(0,\bar{z},y)
\qquad \ell =1,2, \ldots 
\label{dominant}
\end{align}
where all the operators we are interested in are primaries and have the lowest dimension/twist
with a fixed $\ell$. We have used \eqref{dproj} and taken the limit $|z|,|w| \rightarrow 0$ inside the contour integral which is allowed for twist $d-2$ operators. One can check that \eqref{dominant} reproduces the results in this section as long as one considers only the leading term that arises in the $(n-1)$ expansion for a fixed $\ell$. Since all these operators have the same twist they will contribute equally in the lightcone limit of interest. The defect operators coming from ambient operators with the lowest twist will dominate in this limit. While this could be a scalar operator with $(d-2)/2<\tau<d-2$ we 
will see how this contribution exactly cancels in the sum rule that we derive in Section~\ref{ssec:sum}.


\section{Evaluation of the modular Hamiltonian}

\label{sec:modham}

We would like to use what we just learnt about the defect operator spectrum as $n \rightarrow 1$
in order to evaluate \eqref{dope} which we reproduce here:
\be
\label{dope2}
 \left< \Sigma_n  \overbracket{\mathcal{O}_A \mathcal{O}_{\bar A}} \right> 
 = \sum_{ij} \left< \Sigma_n \widehat{ \mathcal{O}}_i (0) \right> \left(G^{-1}\right)^{ij} \left< \Sigma_n^0 \widehat{O}_j(y) \overbracket{\mathcal{O}_A \mathcal{O}_{\bar A}} \right>
 \ee
The operators that dominate in the lightcone limit can be found via the residue projections in \eqref{dominant}. Since these operators are primary $G_{ij}$ is diagonal and can be inverted easily. 
We take $i \rightarrow \widehat{T}_\ell$ and $j \rightarrow \widehat{T}_{-\ell}$
such that we can set $G^{-1}_{\ell \ell'} = (\delta_{\ell,-\ell'}) g^{-1}_\ell (y^2)^{-d-\ell+2} $ for $\ell =1,2, \ldots$ where $g_\ell$ was given in \eqref{ag1} and \eqref{agl} which we also reproduce here:
\be
g_1 = c_T \frac{(n-1)}{4(d+1)} \,, \qquad g_\ell = c_T \frac{\Gamma(d+\ell)}{4 \Gamma(d+2) (\ell-2)! }\,, \qquad \qquad  \ell \geq 2
\ee
To evaluated \eqref{dope2} we just need to compute the ``one point functions''  $\left< \Sigma_n \widehat{ \mathcal{O}}_i (0) \right> $ and the ``three point function'' terms $\left< \Sigma_n^0 \widehat{O}_j(y) \overbracket{\mathcal{O}_A \mathcal{O}_{\bar A}} \right>$.

\subsection{One point functions}

The one point functions are the only piece of data in the defect OPE that know about the state $\psi$ and the details of the entangling cut $A$. We will relate the $n\rightarrow 1$ limit to the entanglement entropy of $\psi$ reduced to $A$ as well as the one point functions of $T_{\mu\nu}$ in the state $\psi$. These are the two ingredients that go into the QNEC. 

Let's begin with $\ell \geq 2$. In this case the limit $n \rightarrow 1$ is trivial and we simply get local operators without the defect:
\be
\lim_{n \rightarrow 1} \left< \Sigma_n \widehat{T}_\ell \right>
= \frac{(-1)^\ell}{(\ell-2)!} \left< \left(\partial_u\right)^{\ell-2}  T_{--} (0)\right>_\psi
\ee
where the $(-1)^\ell$ comes from the awkward minus sign relating the holomorphic coordinates to lightcone coordinates $w = - u$.

We now move to $\ell=1$ where we find the physics of the displacement operator. See \cite{Dong:2017xht} for a similar discussion based on \cite{Allais:2014ata,Rosenhaus:2014woa}. 
We would like to evaluate:
\be
\left< \Sigma_n \widehat{T}_1 \right> = \frac{1}{2\pi i} \oint d w \left< \Sigma_n T_{--}(w,0,0) \right>_{CFT/\mathbb{Z}_n }
\ee
where $T$ here is the orbifold stress tensor, which includes the sum over replicated stress tensors. 
Note that the analytic continuation in $n$ is trivial since there is only a single $T$ operator
insertion on each replica. 
Since the $n=1$ limit gives the stress tensor in the state $\psi$, we do not find any
$1/w$ pole that gives a non-zero answer.  Hence this one point function must be
$\propto (n-1)$. At this next order we pull down the half modular Hamiltonian and it is certainly possible that $T_{--}$ is singular in the presence of $H^\psi_A$.
We extract an answer if there is a pole $H^\psi_A T_{--} \propto 1/w$ then:
\be
\label{onet1}
\lim_{n \rightarrow 1} \left< \Sigma_n \widehat{T}_{1} \right>
 = i (n-1) \oint d w \left< H^\psi_A T_{--}(w,0,0) \right>_\psi
\ee

Let us back up a little now and rather compute the answer at finite $n$. We can find the $1/w$ pole in this case as follows. The displacement operator is related to the non-conservation of the stress tensor in the presence of the twist defect. We have:
\be
\label{dfun2}
\nabla^\mu \left< \Sigma_n  T_{\mu \alpha}(w,\bar{w},y) \right>  = \delta_{\partial A}(w,\bar{w}) \left<\Sigma_n D_\alpha(y)  \right>
\ee
where, recall that $y$ are the $d-2$ the coordinates along the defect the transverse coordinates to the defect are $w,\bar{w}$ (at least close to $y=0$).
And the displacement operator in turn is related to the shape deformation of the orbifold
partition function with the twist defect. That is the shape deformation of the Renyi entropy:\footnote{
Recall the definitions $S_n = (1-n)^{-1} \ln {\rm Tr} \rho_A^n = (1-n)^{-1} \ln( Z_n/(Z_1)^n)$.}
\be
\delta_{\rm shape} S_{n}(A) = \frac{1}{1-n} \delta_{\rm shape} \ln \left< \Sigma_n \right>
= - \frac{1}{1-n} \int d y \sqrt{h} \,\, \delta x^-(y) \frac{ \left< \Sigma_n D_-(y) \right>}{\left< \Sigma_n \right> }
\ee
Now the Ward identity in \eqref{dfun2} implies the necessary $1/w$ pole. That is we must have:
\be
\label{twistsing}
\left< \Sigma_n T_{- - }(w,\bar{w},y)\right> = -\frac{ \left<\Sigma_n D_-(y)  \right>}{2 \pi w} + {\rm non \,\, singular}
\ee
such that $\partial_+ w^{-1}= \partial_{\bar{w}}  w^ {-1} = \pi \delta_{\partial A}(w,\bar{w})$ gives the desired delta function. Thus we have:
\be
\left< \Sigma_n  \widehat{T}_{1} \right>
=-  \frac{1}{2\pi} \left< \Sigma_n D_-(0) \right> =-  \frac{(n-1)\left< \Sigma_n \right> }{2\pi} \frac{ \delta   S_n(A)}{\delta x^-(0) }
\ee
where we define the functional derivative to absorb the measure factor $\sqrt{h}$. The limit
$n \rightarrow 1$ of the Renyi entropies give the EE, and so we find the desired behavior:
\be
\label{ee}
\lim_{n \rightarrow 1} \frac{1}{n-1} \left< \Sigma_n \widehat{T}_{1} \right> = 
  \left< D_{-}(0) H^\psi_A \right>_\psi
= 
- \frac{1}{2\pi} \frac{ \delta S_{EE}(A)}{ \delta x^-(0) }
 \equiv  - \frac{1}{2\pi} \mathcal{P}_-^A
 \ee
where we could have extracted the displacement operator directly from the $n=1$ limit from the pole 
of $T_{--}$ in the presence of the modular Hamiltonian:
\be
\left< H_A T_{--}(w,0,0) \right> = - \frac{ \left<  D_-(0) H_A^\psi \right>}{2 \pi w} + {\rm non\,\, singular}
\ee
taking the limit $n \rightarrow 1$ on \eqref{twistsing}.

Note that $\mathcal{P}_-$ depends non-locally on the shape of the entangling surface.
So for example functional derivatives of this with respect to  $x^-(y)$ are non-zero also when $y \neq 0$. It is also non-linear in the state $\left| \psi \right> \left< \psi \right|$. 
This should be contrasted with the $T_{--}$ contribution which is localized to the null generator at $y=0$ and is the expectation value of a linear operator on the Hilbert space. 

There is a potential issue here relating to divergences that might naturally appear in $S_{EE}$. These  would show up in the language of the displacement operator as divergences in the one point function
$\left< \Sigma_n \widehat{T}_{1} \right>$. Thus we need to specify a regulator to define $S_{EE}$ as well as $\widehat{T}_1$. The regulator will depend on a small parameter $\epsilon$ with units of length. Actually many natural regulators (the brick wall for example) can be designed to preserve the boost invariance around the undeformed cut $\partial A_0$
such that $\delta S_{EE}^{(\epsilon)}(A_0)/\delta x^- = 0$ for the vacuum, so one might not expect additional divergences for excited states.\footnote{ \label{statedep} Entanglement entropy has potentially state dependent divergences which would invalidate this argument \cite{Marolf:2016dob}. However these state dependent divergences also afflict the definition of the stress tensor, via improvement terms, in such a way that the state dependence cancels between the various terms we find in the lightcone expansion. This should be clear in the final answer, at least for null cuts of the Rinlder horizon since then the QNEC quantity is related to the relative entropy which does not suffer from state dependent divergences. } Of course additionally divergences might show up since $A$ is not locally flat - an effect we will ignore until Section~\ref{sec:locgeo}. If we do not find a regulator where this divergence goes away for the flat cut, then actually the dOPE method forces a resolution. In this case we would find that the one point function of the displacement operator on the flat cut is non-zero:
\be
\left< \Sigma_n^0 \widehat{T}_{1}^{(\epsilon)} \right> \neq 0
\ee
This means that $ \widehat{T}_{1}^{(\epsilon)}$ is no longer a primary operator. 
This fact would then lead to some non-trivial mixing when we make the defect OPE argument. For example the displacement operator would now mix with the defect unit operator:
\be
G_{\widehat{T}_1 \widehat{\mathbb{1}} }  \equiv \left< \Sigma_n^0 \widehat{T}_{1}^{(\epsilon)} \widehat{\mathbb{1}} \right>
\neq 0 
\ee
We can deal with this mixing simply by removing the one point function:
\be
\widehat{T}_{1}^{(reg)} \equiv  \widehat{T}_{1}^{(\epsilon)} - \left< \Sigma_n^0 \widehat{T}_{1}^{(\epsilon)} \right> \widehat{\mathbb{1}} 
\ee
The end result is that the displacement operator one point function must involve a vacuum subtraction of the entropy for the flat cut $\partial A_0$:
\be
\label{subtraction}
\mathcal{\mathcal{P}}_-^A  \rightarrow \frac{\delta}{\delta x^-(0)} \left( S^{(\epsilon)}_{EE}(A; \left|\psi \right>) - S^{(\epsilon)}_{EE}(A_0; \left| \Omega \right>) \right)
\ee
where we have made explicit the state dependence. 
This vacuum subtraction will be very important later when we deal with the local geometric terms that we are ignoring for now.

\subsection{Three point functions}

It turns out the three point function term in \eqref{dope2} can be treated in a similar way to the ambient space two point correlator which was used above to extract the defect spectrum and two point function coefficients $g_\ell$. 
The operators $ \mathcal{O}_B \mathcal{O}_{\bar A}$ are either always on the same replica or displaced by one replica. We need to then sum over the different replicas. We treat the case where the two operators $\mathcal{O}_A$  are on the same replica: 
\be
\overbracket{\mathcal{O}_B \mathcal{O}_{\bar A} }
\equiv \sum_{k=0}^{n-1} \mathcal{O}_B^{(k)} \mathcal{O}_{\bar A}^{(k)}
\ee
and then do a simple continuation, as discussed in Section~\ref{sec:replica} to find the other case. 

We must additionally insert $\widehat{T}_{-\ell}$ which also involves a symmeterization $\sum_{k'=0}^{n-1} T_{++}^{(k')}$ over each replica. As a result of the $\mathbb{Z}_n$ symmetry around the defect there is a non-trivial sum left over labeled by $j =k-k'$ the separation in replicas:
\be
 \left< \Sigma_n^0 \widehat{T}_{-\ell}(y) \overbracket{\mathcal{O}_B \mathcal{O}_{\bar A}} \right>
 = - \frac{n}{2\pi i }  \oint \frac{d \bar{z}}{\bar{z}} \bar{z}^{-\ell+2} \left( \sum_{j=0}^{n-1}  \left<  T_{++}^{(j)} (z,\bar{z},y) \mathcal{O}_B^{(0)}  \mathcal{O}_{\bar A}^{(0)} \right>_{\mathcal{M}_n} \right)_{z \rightarrow 0}
\ee
Actually, as we discovered before, the limit $z \rightarrow 0$ is not well defined because of the appearance of log's. These will not afflict the final answer for the full modular Hamiltonian but for now we work with a finite but small $z$. The analytic continuation of the term in brackets  can be treated as we did in Section~\ref{sec:scalar} turning the sum into a contour integral $\int d\lambda$. Since the $n=1$ term vanishes in the full modular Hamiltonian we can concentrate on the $(n-1)$ piece. For the term in brackets we have:
\begin{align}
\label{applyto}
\lim_{n \rightarrow 1} (\partial_n - 1) \sum_{j=0}^{n-1}  \left<  T_{++}^{(j)} (z,\bar{z},y) \mathcal{O}_B^{(0)}  \mathcal{O}_{\bar A}^{(0)} \right>_{\mathcal{M}_n}
 &=   -2\pi \left< H_A^0 \mathcal{O}_B  T_{++}(z,\bar{z},y)  \mathcal{O}_{\bar A}  \right>  \\  & \qquad  -\int_0^{-\infty} \frac{d\lambda \lambda^2}{\bar{z} (\lambda-\bar{z})^2} \left<T_{++}(z \bar{z}/\lambda , \lambda,y) \mathcal{O}_B  \mathcal{O}_{\bar A}\right>  \nonumber
\end{align}
where in the first term we insert $H_A^0$ along the region $A_0$ just before the operator $\mathcal{O}_B$ insertion in the Euclidean clockwise ordering sense. In the second term we have rescaled $\lambda \rightarrow \lambda/\bar{z}$ relative to the equivalent expression in \eqref{bothscalar}. The specified $\lambda$ contour is such  that $T_{++}$ naturally stays away from the $\mathcal{O}$ operator insertions with $T_{++}$ again inserted just before $\mathcal{O}_B$ in the clockwise ordering. 
Note that the second term depend on the CFT 3 point function $\left<T\mathcal{O}\mathcal{O}\right>$ which is fixed by conformal invariance but the first modular Hamiltonian term is now determined by a four point function $\left<TT\mathcal{O}\mathcal{O}\right>$ and is not fixed by conformal symmetries.  
 
Fortunately when we compute the full modular Hamiltonian this term turns into $\left<K_A^0 \mathcal{O} T \mathcal{O}\right>$ and since $K_A^0$ is a  conserved charge this correlator is now related via a Ward identity to the universal 3 point function. More specifically we may apply the continuation $\mathcal{O}_{\bar{A}}  (\circlearrowleft)$ directly to \eqref{applyto} and subtract. The resulting expression has a finite $z \rightarrow 0$ limit, allowing us to make the projection onto the defect operators:
\be
\label{twoc}
 \lim_{n \rightarrow 1} (- \partial_n) \left< \Sigma_n^0 \widehat{T}_{-\ell}(y) \left( \overbracket{\mathcal{O}_B \mathcal{O}_{\bar A}}
 - \overbracket{\mathcal{O}_B \mathcal{O}_{\bar A} (\circlearrowleft)} \right) \right>
 \equiv 2\pi \left(  C^{(1)}_\ell + C^{(2)}_\ell   \right)
\ee
where the first term in \eqref{applyto} becomes:
 \be
C_\ell^{(1)} = -  \oint \frac{d \bar{z}}{ 2\pi i \bar{z}} \bar{z}^{-\ell+2} \left< T_{++}(0,\bar{z},y) \mathcal{O}_B \left[ K_A^0,\mathcal{O}_{\bar{A}} \right] \right> 
\label{acl1}
 \ee
In fact this term will only be non-zero for $\ell \geq 2$ due to the lack of a $1/\bar{z}$ that is necessary for a non-zero answer for $\ell=1$. Thus the displacement operator gets no contribution from this modular Hamiltonian term. For $\ell \geq 2$ the commutator $\left[ K_A^0, \mathcal{O}_{\bar{A}}\right]$ simply moves the local operator around a little and so summing over $C_\ell^{(1)}$ in the defect OPE gives the same contribution as the stress tensor exchange $\mathcal{O}\mathcal{O} \rightarrow T \rightarrow \psi \psi$ of:
\be
\label{unitstress}
2\pi \left< \psi \right| \mathcal{O}_B \left[ K_A^0, \mathcal{O}_{\bar{A}} \right] \left| \psi \right>
\ee
evaluated in the lightcone limit. This is the subject of \cite{Hartman:2016lgu}. Indeed this allows us to include another term we have thus far ignored - the \emph{unit operator} contribution which will exactly reproduce the unit operator exchange in the correlator of \eqref{unitstress}.

\begin{figure}[h!]
\centering
\includegraphics[scale=.65]{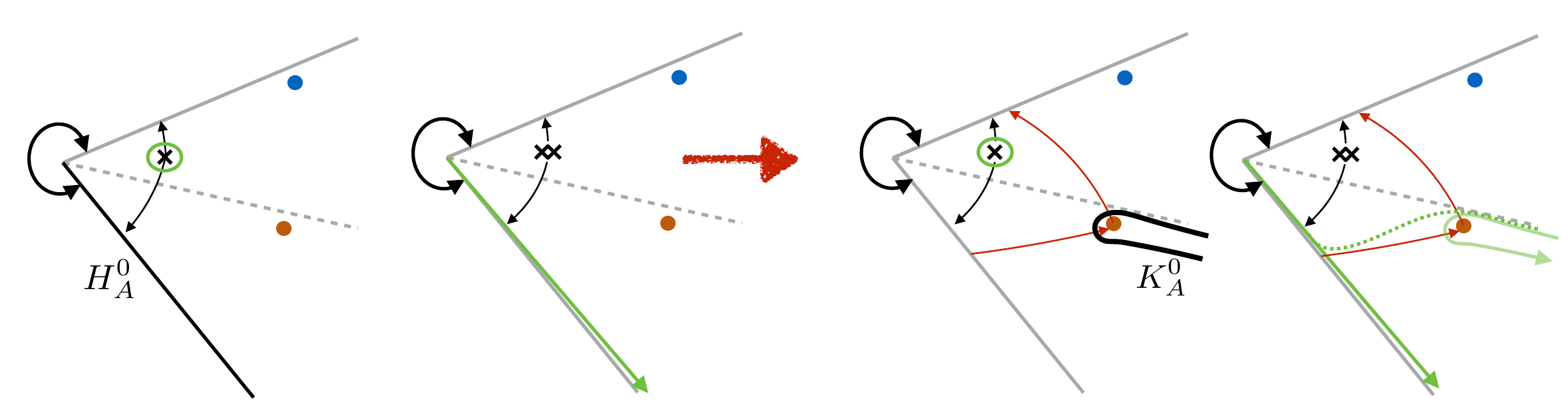}
\caption{\label{fig:contour} The complex $\lambda$ plane represented as a slice of pie with the edges identified. Note that $\lambda$ is an anti-holomorphic coordinate so ordering is now anti-clockwise compared to the discussion in the text. The green curves are the contours we integrate over. The two figures on the left represent the two terms in \eqref{applyto} and on the right we have the two terms $C_\ell^{(1,2)}$ in \eqref{acl1} and \eqref{acl2} which arises after computing the full modular Hamiltonian using the operator deformation of \eqref{sheet}. }
\end{figure}

So at this point we reassess our goal. We can remove the $C_\ell^{(1)}$ term in the three point function by computing the following vacuum subtracted modular Hamiltonian instead:
\be
\label{prer}
 \left< \psi \right| \mathcal{O}_B \left[ K_A- K_A^0, \mathcal{O}_{\bar A}\right] \left| \psi \right>
= \sum_{\ell=1}^{\infty} \lim_{n \rightarrow 1} \left( \left< \Sigma_n \widehat{T}_{\ell} \right>
g_{\ell}^{-1} \right) C_\ell^{(2)} + \ldots
\ee 
where we have used $K_A \left| \psi \right> = 0$. The second term $C_\ell^{(2)}$ that remains from \eqref{twoc} is
\be
C_{\ell}^{(2)} = - \frac{1}{2\pi} \oint \frac{d \bar{z}}{2\pi i \bar{z} } \bar{z}^{-\ell+2} \int_{\mathcal{C}(\bar{A})} \frac{d\lambda \lambda^2}{(\lambda-\bar{z})^2 \bar{z}} \left<T_{++}(0,\lambda,y) \mathcal{O}_B  \mathcal{O}_{\bar A}\right>
\label{acl2}
\ee
The contour $\mathcal{C}(\bar{A})$ encircles the operator $\mathcal{O}_{\bar{A}}$ in the clockwise direction, straddling the branch cut that runs out to $\lambda \rightarrow \infty$. This particular contour arrises because we are computing the full modular Hamiltonian. Tracking the motion of $\mathcal{O}_{\bar{A}}  (\circlearrowleft)$ we find we have to deform the $\lambda$ contour in \eqref{applyto} so that $T_{++}$ avoids $\mathcal{O}_{\bar{A}}$. This results in $\mathcal{C}(\bar{A})$ when we subtract the two terms.\footnote{Note that $\lambda$ is an anti-holomorphic coordinate so the ordering prescriptions are reversed and the deformation $\mathcal{O}_{\bar{A}}(\circlearrowleft)$ involves a clockwise rotation in the $\lambda$ plane.} See Figure~\ref{fig:contour} for an illustration of this.

If we switch the order of integration, which is justified because $\lambda$ stays well away from the origin, we can now do the $\bar{z}$ integral:
\be
C_\ell^{(2)} = \frac{\ell}{2\pi}  \int_{C_{\bar{A}} } d \lambda \lambda^{1-\ell} \left<T_{++}(0,\lambda,y) \mathcal{O}_B  \mathcal{O}_{\bar A}\right>
\label{int1}
\ee
We finally need the following CFT 3 point function:
\begin{align}
\label{cft3}
\left< T_{++}(0,\lambda,y) \mathcal{O}_B \mathcal{O}_{\bar A} \right>_{n=1}
&= \frac{ c_{\mathcal{O}}^T (\Delta u)^2 y^4 (-\Delta v \Delta u)^{-\Delta + h-1}  }{4 (\lambda u_B + y^2)^{h+1}  (\lambda u_{\bar A} + y^2)^{h+1} }  
\end{align}
Here $\Delta u = u_B - u_{\bar A}$
and $ \Delta v = v_B - v_{\bar A}$ and we have dropped various terms like $u_B v_B \ll 1$ which are small in the lightcone limit. The $\lambda$ integral over this 3 point function can be done and written in terms of a Hypergeometric function. We find that for $\ell \geq 2$ we can then rewrite $C_\ell^{(2)}/g_{\ell}$ as another Hypergeometric integral:
\be
\frac{C_\ell^{(2)}}{g_\ell (y^2)^{-d-\ell+2} } = - i c_{\mathcal{O}}  (- \Delta u  \Delta v)^{-\Delta_{\mathcal{O}}}  \mathcal{N} 
\int_{u_{\bar{A}}}^0 du \ell (-u)^{\ell-2}\left( \frac{(u-u_{\bar{A}})(u_{B}-u)}{ (u_B-u_{\bar{A}})^2 }\right)^h \qquad \ell \geq 2
\label{int2}
\ee
with
\be
\mathcal{N} = \frac{16 \pi G_N \Delta_{\mathcal{O}}}{d} \Delta u (- \Delta u  \Delta v)^{h-1}
\ee
That is, one can explicitly check that the two integrals \eqref{int2} and \eqref{int1} agree after dividing by $g_\ell$ and using the definition of $G_N^{-1} \propto c_T$  that appears in the normalization $\mathcal{N}$ via:
\be
\label{gnct}
c_T = \frac{\Gamma(d+2) }{8\pi G_N \pi^h \Gamma(h)(d-1)} \qquad c_{\mathcal{O}}^T =- c_{\mathcal{O}} \frac{ d\Delta_{\mathcal{O}} \Gamma(h)}{2 \pi^h (d-1)}
\ee
Here $c_\Delta$ sets the overall normalization of the $\mathcal{O}\mathcal{O}$ two point function
and is related to $c_\Delta^T$ via the Ward identity.  The relation between $c_T$ and $G_N$ is by definition the usual relation in holographic theories  with $R_{AdS}=1$. We use the definition here for convenience, however we stress that we need not have $G_N$ small. 
Additionally for $\ell=1$ we find:
\be
\label{ac12}
\frac{C_1^{(2)}}{g_1 (y^2)^{-d+1}}  = - \frac{i c_{\mathcal{O}} (- \Delta u  \Delta v)^{-\Delta_{\mathcal{O}}}  \mathcal{N}}{(n-1)} \left( - \frac{ u_{\bar{A}} u_B}{(u_B-u_{\bar{A}})^2} \right)^h
\ee
where the $1/(n-1)$ compensates the displacement operator one point function which is $\mathcal{O}(n-1)$. 

\subsection{Putting it together}

We can now put everything together. The term with $\ell \geq 2$ can be easily summed using the integral representation \eqref{int2} and:
\be
\sum_{\ell \geq 2} \frac{\ell u^{\ell-2}}{(\ell-2)!} \left< \left(\partial_u\right)^{\ell-2} T_{--}(0) \right>_\psi= u^{-1} \partial_u\left( u^2  \left<  T_{--}(u) \right>_\psi \right)
= i  \left< \left[K_A^0, T_{--}(u) \right] \right>_\psi
\ee
Adding in the displacement operator we find for \eqref{prer}:
\begin{align}
\label{commkk0}
\mathcal{R} &\equiv  \frac{\left< \psi \right| \mathcal{O}_B \left[ K_A- K_A^0, \mathcal{O}_{\bar A}\right] \left| \psi \right>}{ \left< \Omega \right| \mathcal{O}_B  \mathcal{O}_{\bar A} \left| \Omega \right>}  \\
&= \mathcal{N}  \left(  \int_{u_{\bar A}}^0 du  \left< \left[K_A^0, T_{--}(u) \right] \right>_\psi \left( \frac{(u-u_{\bar{A}})(u_{B}-u)}{ (u_B-u_{\bar{A}})^2 }\right)^h 
+ i \left( \frac{u_B(- u_{\bar{A}})}{(u_B - u_{\bar{A}})^2} \right)^h \frac{ \mathcal{P}_-^A}{2\pi} \right) + \ldots
\nonumber
\end{align}

While this result is only an intermediate step towards our final goal it is instructive to work with this a little. Imagine now taking a limit where $u_B,-u_{\bar{A}} \rightarrow \infty$. 
Of course we should always keep $u_B \ll 1/| v_B|$ etc. so we remain in the lightcone limit. After integrating by parts we discover:
\be
\mathcal{R} = i \frac{ 16\pi G_N \Delta_{\mathcal{O}}}{d} \frac{1}{\Delta v} \left( \frac{ u_B u_{\bar{A}} \Delta v}{ \Delta u} \right)^{h}  \left( \int_{-\infty}^0 du
 \left< T_{--}(u) \right>_\psi  - \frac{ \mathcal{P}_-^A}{2\pi}  \right) + \ldots
 \label{aR}
\ee
Note that the object in brackets can be interpreted as the first order change in the CFT relative entropy for the region $\bar{A}$ (with the reference state being the vacuum) under a null deformation in the positive $u$ direction.

At this stage we could try to constrain the sign of the shape deformation of $\mathcal{R}$ which is in turn related to matrix elements of the shape deformation of the full modular Hamiltonian for the state $\psi$. If we consider changes of the subregion $A$ under inclusion then since in this case $\delta_{\rm shape} K$ is a negative semi-definite operator then one might argue that
 $\delta_{\rm shape} \mathcal{R}$ is negative.\footnote{This would require the $\mathcal{O}$ operators to be inserted such that they are Hermitian conjugates of each other, which can be achieved via a  $\pi/2$ Euclidean rotation of both operators around the entangling surface.}  Indeed such deformations applied to \eqref{aR} will lead to something proportional to the QNEC quantity, however we cannot claim  $\delta_{\rm shape} \mathcal{R}$ has a definite sign since we are subtracting the action of $K_0$ and this subtraction will change under the deformation. 
We note in passing that our original goal was to derive the QNEC in this way, however this method ultimately failed for the reasons explained. Although once we had the result \eqref{aR} in hand it was not hard to come up with an argument that does work involving modular flow. We turn to this now. 

\section{Modular flow}

\label{sec:modflow}

\subsection{Warm up}

\label{sec:singmodflow}

In this section we warm up with a simple example of modular flow, using the results of the previous section. Consider:
\be
\label{modflwu}
\left< \psi \right| \mathcal{O}_B e^{ i s K_A} \mathcal{O}_{\bar{A}} \left| \psi \right>=  \left< \psi \right| \mathcal{O}_B e^{ i s K_A} \mathcal{O}_{\bar{A}} e^{- is K_A} \left| \psi \right>
\ee
we would like to evaluate this in the lightcone limit.\footnote{Note we continue to label $\mathcal{O}_B$ as if it is in the region $\mathcal{D}(B)$, despite the $B$ region playing no role here. If the reader  likes take $B=A$ in this warm up section.} We will do this using perturbation theory, where the small paramater is $|u v| \ll 1$ for both $u = \{ u_B, u_{\bar{A}} \}$ and $v = \{v_B,v_{\bar{A}} \}$. We expect we can apply this here
because we know that the action of $K_A$ within correlators of operators which are close to light like separated (i.e. in the lightcone limit) is well approximated by $K_A^0$ plus corrections in powers of $uv$. This is the content of \eqref{commkk0} where the leading correction is suppressed $\sim (u v)^{\tau/2}$ for $\tau= d-2$ the twist of the stress tensor. Since the modular flow generated by  $K_A^0$ at zeroth order does not take us outside of the lightcone limit we then might expect we can continue to apply \eqref{commkk0} to higher orders in the actin of $K_A$ which is necessary to generate modular flow. In this section we will use this method to compute \eqref{modflwu}.

The method we will use here is actually not completely justified since in reality \eqref{commkk0} is a statement about matrix elements  of $K_A$ in a subset of states $\mathcal{O} \left| \psi \right>$ for operators light like separated from the defect.\footnote{There is a useful analogy here to the quantum error correction language of bulk reconstruction advocated in \cite{Almheiri:2014lwa,Dong:2016eik,Jafferis:2015del}. These ``lightcone'' states are like the code subspace where the action of modular flow is well approximated by $K_A^0$. This is the analogous content of the JLMS \cite{Jafferis:2015del} result that the bulk modular Hamiltonian is the boundary modular Hamiltonian when acting on low energy states in the gravitational dual. However more work is required to show that this bulk/boundary equivalence applies to modular flow; see footnote 5 for a justification \cite{Faulkner:2017vdd}. For example this does not work for the quantum error correction codes discussed in \cite{Harlow:2016vwg} where the relation between bulk and boundary modular Hamiltonians requires a double sided projection $\Pi_c$ onto the code subspace: $K_{\rm bulk} = \Pi_c K_{\rm bdry} \Pi_c$ . Roughly speaking in our case we want a version involving only a half sided projection $K_{\rm bulk} = \Pi_c K_{\rm bdry} =  K_{\rm bdry} \Pi_c $ although a weaker version should work too. We thank Aitor Lewkowycz, Don Marolf  and Xi Dong for discussion on this. \label{foot}} 
While to compute higher order powers $(K_A)^m$ we might transition outside of the light cone limit and find matrix elements not well approximated by $K_A^0$. To address this issue we need a method to directly compute higher order powers of $K_A$ in these states. There is a very efficient method to compute these higher order terms, by essentially directly computing \eqref{modflwu} using the replica trick. Roughly speaking we can compute correlators involving $\rho_A^p \mathcal{O}_B \rho_A^{-p}$ for integer $p$ and analytically continue this $p \rightarrow is/2\pi$.\footnote{We thank Aitor Lewkowycz for suggesting this to us.} This method syncs well with the defect OPE computation. We present this  in Appendix~\ref{sec:appmodflow} and find \emph{exactly} the same result as the perturbative method that we present now, thus providing a complete justification of these results (and also likely a justification of the method.)

The method here starts by using the second expression in \eqref{modflwu}, writing $K_A = K_A^0 + (K_A-K_A^0)$ and then expand in $(K_A-K_A^0)$. We work with the second expression in \eqref{modflwu} rather than the first because the perturbative series arranges itself differently for these two expressions - the end result is the same, but we choose the most efficient route.  Expanding using time dependent perturbation theory:
\be
\label{modpert0}
\left< \psi \right| \mathcal{O}_B e^{ i s K_A} \mathcal{O}_{\bar A} e^{-is K_A} \left| \psi \right>
\approx \left< \psi \right| \mathcal{O}_B  \mathcal{O}_{\bar A} (s) \left| \psi \right>
+ i \int_0^s d t \left< \psi \right| \mathcal{O}_B(t-s)
\left[ K_A-K_A^0 ,  \mathcal{O}_{\bar{A}}(t) \right] \left| \psi \right>
\ee 
where we define:
\be
\mathcal{O}_{\bar{A}}(s) = e^{ i s K_A^0} \mathcal{O}_{\bar{A}} e^{- i s K_A^0}  \qquad {\rm etc.}
\ee
which is still a local operator just at a different point in spacetime. We write this point as:
\be
x_{\bar{A}}(s) \equiv ( u_{\bar A}(s),v_{\bar A}(s) ) \equiv ( u_{\bar A} e^s, v_{\bar A} e^{-s} )
\ee
and similarly for the modular flow of $x_B(t-s)$ etc. 

In order to arrive at this result, for terms already at the order $\mathcal{O}(K_A-K_A^0)$, we are free
to make the following manipulations: $ e^{ i K_A^0 t} \left| \psi \right> 
\approx e^{ i K_A t} \left| \psi \right> = \left| \psi \right>$. We have made such a replacement several times in the second term of \eqref{modpert0} with the goal of writing the answer in terms
of the commutator in $\mathcal{R}$ defined \eqref{commkk0}.  We will need this commutator
for operators that are no longer at the location of $\mathcal{O}_B,\mathcal{O}_{\bar{A}}$ so let us define instead:
\be
\label{defR}
\mathcal{R}(x_2,x_1;A) \equiv \frac{ \left< \psi \right| \mathcal{O}_2 \left[ K_A-K_A^0, \mathcal{O}_1 \right] \left| \psi \right>}{ \left< \Omega \right| \mathcal{O}_2 \mathcal{O}_1 \left|  \Omega \right> }
\ee
where the $A$ in $R(.. ;A)$ refers to the fact that we are using the $A$ modular Hamiltonian. In the next section we will need the $B$ version of $\mathcal{R}$.  The lightcone limit of $\mathcal{R}$ can be succinctly written in terms of the following function:
\be
F(x_2,x_1;u) \equiv \frac{16\pi G_N \Delta_{\mathcal{O}}}{d} \Delta u \left( -\Delta u \Delta v \right)^{h-1} \left(\frac{(u_2 - u)(u-u_1)}{(u_2-u_1)^2} \right)^h
\label{Fx2x1u}
\ee
such that:
\be
\mathcal{R}(x_2, x_1;A) = i \left(  \int_{u_1}^0 du \left<T_{--}(u)\right>_\psi u^2 \partial_u \left( u^{-1} F(x_2, x_1; u) \right)
+ \frac{F(x_2, x_1; 0)}{2\pi} \mathcal{P}_-^A \right) + \ldots
\label{Ranswer}
\ee
We will also need the expansion of $\left< \psi \right| \mathcal{O}_2 \mathcal{O}_1 \left| \psi \right>$
in the lightcone limit - which was computed in \cite{Hartman:2016lgu}. This can also be written simply in terms of $F$:
\be
\label{hartp}
\mathcal{P}(x_2,x_1) \equiv  \frac{ \left< \psi \right| \mathcal{O}_2  \mathcal{O}_1 \left| \psi \right>}{ \left< \Omega \right| \mathcal{O}_2 \mathcal{O}_1 \left|  \Omega \right> }-1
=  - \int_{u_1}^{u_2} du \left< T_{--} (u) \right>_\psi F(x_2,x_1; u) + \ldots
\ee
Note that for this computation we will always take $ u_1 < 0 < u_2$. Now that everything is written in terms of $F$, we note that it satisfies the following properties:
\be
F(x_2(t), x_1(t); u)  = e^t F(x_2, x_1; e^{-t} u) 
\qquad F(x_2, x_1; u_{1,2}) =0 
\ee
where $x_2(t)$ is the modular flow/boost of the coordinate $x_2$ with respect to the action of $K_0$ etc. These identities will be useful for the various manipulations we make below.

Putting all the above definitions together we can now write the correlator we are interested in \eqref{modflwu} as:
\begin{align} 
\frac{\left< \psi \right| \mathcal{O}_B e^{ i s K_A} \mathcal{O}_{\bar A} \left| \psi \right>} { \left< \Omega \right| \mathcal{O}_B  \mathcal{O}_{\bar A} (s) \left| \Omega \right>}
\approx
\left( \vphantom{\int_0^s} 1 + \mathcal{P}(x_B, x_{\bar A}(s) )
+ i \int_0^s d t \mathcal{R}\left(x_B(t-s), x_{\bar{A}}(t) ; A \right)  + \ldots \right)
\label{PR}
\end{align}
It turns out we can simplify a lot this later $t$ integral:
\begin{align} \nonumber
- \int_1^{e^s} d\lambda \left( \int_{u_{\bar A}(t)}^0 du \left< T_{--}(u) \right>_\psi u^2 \partial_{u} \left( u^{-1} F(x_B(-s), x_{\bar A}(0); u/\lambda) \right) + \frac{F(x_B(-s), x_{\bar{A}}(0); 0)}{2\pi} \mathcal{P}_-^A \right)
\label{bigeq}
\end{align}
where $\lambda = e^t$ and $u_{\bar{A}}(t) = \lambda u_{\bar{A}} $. Now we would like to exchange the $\lambda$ and $u$ integrals. In order to do this we have to make the lower limit of integration
$\lambda$ independent. So we insert a step function $H(u- \lambda u_{\bar{A}})$ (where keep in mind that $u_{\bar{A}} <0$) so we can extend the lower range of integration to $u_{\bar{A}}(s) = e^s u_{\bar{A}}$. 
Additionally using the fact that:
\be
\partial_u ( u^{-1} F(x_B(-s), x_{\bar{A}};u/\lambda) ) 
= - u^{-2} \partial_\lambda (  \lambda F(x_B(-s) x_{\bar{A}};u/\lambda))
\ee
we can then easily do the $\lambda$ integral:
\begin{align}
\int_{u_{\bar{A}}e^s}^0 d u \left< T_{--}(u)\right>_\psi \left[ \lambda F(x_B(-s), x_{\bar{A}}; u/\lambda) \right]_{\lambda = {\rm max}(1, u/u_{\bar{A}} )}^{\lambda = e^s} - (e^{s}-1) F(x_B(-s), x_{\bar{A}}; 0) \mathcal{P}_-^A
\end{align}
Adding this term to the $\mathcal{P}$ term in \eqref{PR} we see there are various cancelations and we have:
\begin{align}
\nonumber
&= 1 - \int_{u_{\bar{A}}}^{u_B} du \left<T_{--}(w)\right>_\psi \left(\vphantom{\int} H(u) F(x_B, x_{\bar{A}}(s); u) +H(-u)  F(x_B(-s), x_{\bar{A}}; u) \right)  \\ &\qquad  \qquad \qquad -  (1-e^{-s}) F( x_B, x_{\bar{A}}(s);0)   \mathcal{P}^A_- 
\end{align}
which can also be compactly written as:
\be
\label{togen}
\frac{\left< \psi \right| \mathcal{O}_B e^{ i s K_A} \mathcal{O}_{\bar A} \left| \psi \right>} { \left< \Omega \right| \mathcal{O}_B  \mathcal{O}_{\bar A} (s) \left| \Omega \right>} = 1 - \int_{u_{\bar{A}}}^{u_B}du \left< T_{--}(u)\right>_\psi A_s(u) - \frac{\mathcal{J}A_s(0)}{2\pi} \mathcal{P}^A_- + \ldots
\ee
where
\begin{align}
\nonumber
A_s(u) &= F(x_B, x_{\bar{A}}(s); u) \qquad 0 < u< u_B \\
&= F(x_B(-s), x_{\bar{A}}; u) \qquad u_{\bar{A}} < u< 0
\label{region}
\end{align}
and $\mathcal{J} A_s(0) \equiv (A_s(0+ \epsilon) - A_s(0-\epsilon))$ is the jump discontinuity in the function $A_s(u)$ at $u=0$. There are several checks on this result. Firstly when $s=0$ we trivially go back to the result of \cite{Hartman:2016lgu} given in \eqref{hartp}. Secondly we know via Tomita-Takesaki theory that this expression should be analytic in the strip $- \pi/2 <  \Im s < \pi/2$. One can check this is indeed the case, and relies intricately on the region of $u$ integration in \eqref{region} staying on the opposite side to the action of modular flow in the coordinates of $F$. 

It is interesting that the integral is contained within $u_{\bar{A}} < u < u_A$ despite the fact that zeroth order modular flow moves the operators further away (for example at intermediate steps above this is not the case). This is linked to the fact that $K_A \left| \psi \right> =0$ which implies that we can do the computation above in several different ways (say starting from the first expression in \eqref{modflwu} or
$\left< \psi \right| e^{ - i s K_A} \mathcal{O}_B e^{ i s K_A } \mathcal{O}_{\bar{A}} \left| \psi \right>$) and consistency of these different computations tells us that the final $u$ integral region must be contained within $u_{\bar{A}} < u < u_A$. We do not have a deeper understanding of this. 

Finally later when we bound the function $f(s)$ we will need to consider single modular flow but for the case where the two operators are initially inserted in the same region. This is a mild extension of the above computation, although now we must be more careful with operator ordering and branch cuts. The answer is the naive generalization of \eqref{togen}
\begin{align}
\nonumber
& \frac{\left< \psi \right| \mathcal{O}_{\bar{A}}' e^{ i s K_A} \mathcal{O}_{\bar{A}} \left| \psi \right>} { \left< \Omega \right| \mathcal{O}_{\bar{A}}'  \mathcal{O}_{\bar{A}} (s) \left| \Omega \right>}  = 1 - \int_{u_{\bar{A}} }^{0}du \left< T_{--}(u)\right>_\psi  \widetilde{F}(x_{\bar{A}}'(-s),x_{\bar{A}}; u)  \\ 
& \qquad \qquad + \int_{u_{\bar{A}}'}^{0} du \left< T_{--}(u)\right>_\psi  \widetilde{F}(x_{\bar{A}}',x_{\bar{A}}(s); u)   
 - (1-e^{-s}) \frac{\widetilde{F}(x_{\bar{A}}',x_{\bar{A}}(s);0)}{2\pi} \mathcal{P}^A_- + \ldots
 \label{sameside}
\end{align}
where for the operator ordering shown we should take $-\pi < \Im s <0$ in order to pick the correct branch cut prescription when doing the $u$ integral. Here:
\be
\widetilde{F}(x_{\bar{A}'},x_{\bar{A}};u) \equiv \frac{16\pi G_N \Delta_{\mathcal{O}}}{d} e^{ - i \pi h}  \Delta u \left( -\Delta u \Delta v \right)^{h-1} \left(\frac{(u-u_{\bar{A}}' )(u-u_{\bar{A}})}{(u_{\bar{A}}'-u_{\bar{A}})^2} \right)^h
\ee
and we take the case $u_{\bar{A}}, u_{\bar{A}}' <0$, $u_{\bar{A}}' - u_{\bar{A}} >0$ and  $v_{\bar{A}}' - v_{\bar{A}} <0$ (so the two operators are initially space-like separated before the action of modular flow). Other cases can be arranged via analytic continuation of this answer.

\subsection{Double modular flow}

\label{sec:dmodflow}

\begin{figure}[h!]
\centering
\includegraphics[scale=0.27]{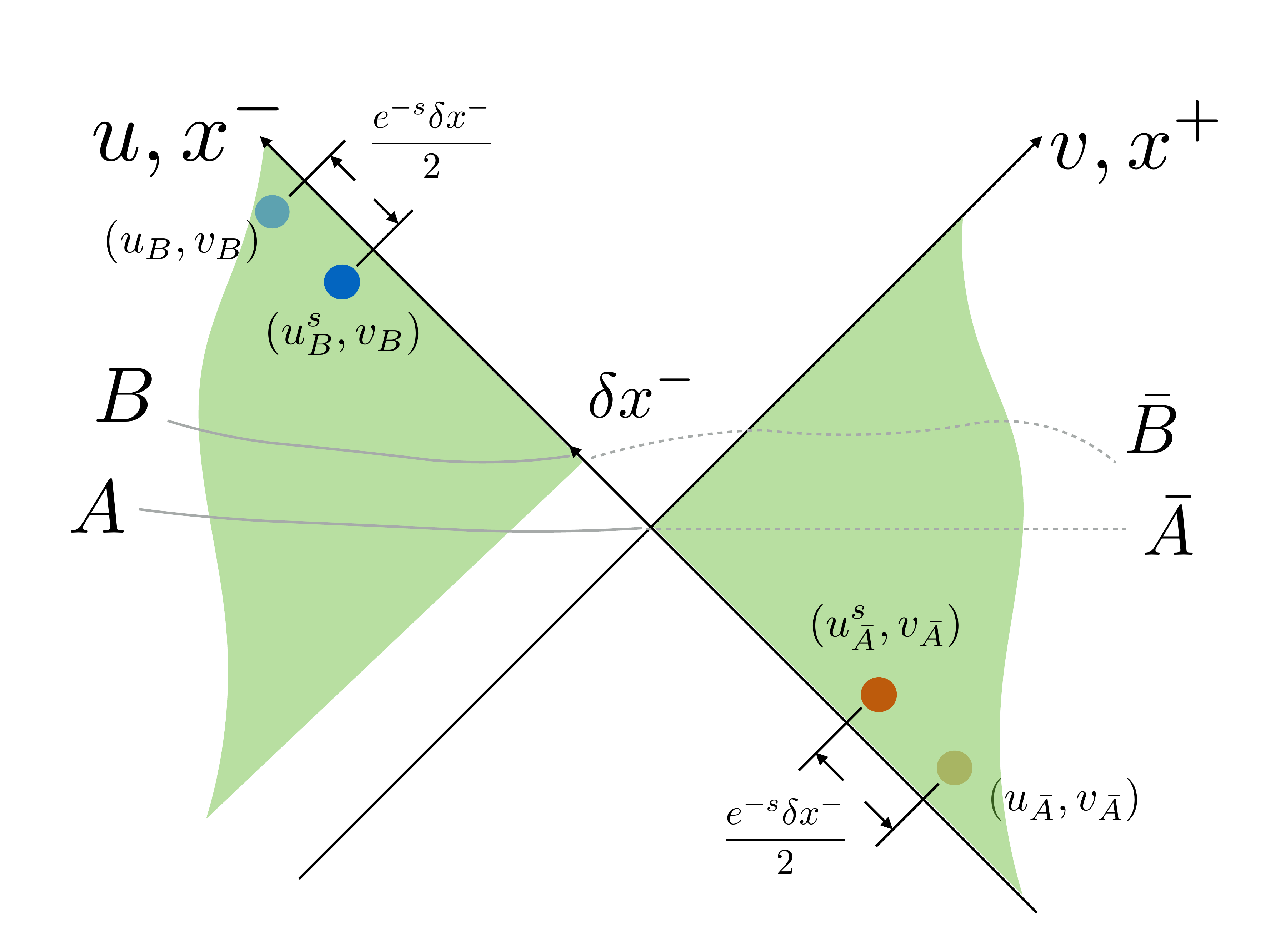}
\caption{Configuration for computing $f(s)$, the double modular flow $e^{-is K_B}e^{is K_A}$ are defined w.r.t entangling regions $B$ and $A$, whose boundaries $\partial B$ and $\partial A$ are located at $(u=\delta x^-,v=0)$ and $(u=0,v=0)$, at the transverse point of interest $\vec{y}=0$. The $s$-dependent operators $\mathcal{O}^s_{B,\bar{A}}$ (solid dots) are inserted at $(u^s_{B,\bar{A}},v_{B,\bar{A}})$, defined relative to some original points of insertions (fade dots) at $(u_{B,\bar{A}},v_{B,\bar{A}})$.}
\label{fig:double_mod_flow}
\end{figure}

In this section we would like to use the above method to compute the more complicated double flow correlator (see Figure \ref{fig:double_mod_flow}):
\be
\label{affs}
f(s) = \frac{ \left< \psi \right| \mathcal{O}_B^s e^{  - i s K_B} e^{ i s K_A} \mathcal{O}_{\bar{A}}^s \left| \psi \right>}{\left< \Omega \right| \mathcal{O}_B^s e^{  - i s K_B^0} e^{ i s K_A^0}  \mathcal{O}_{\bar A}^s \left| \Omega \right> }
\ee
The operators $\mathcal{O}^s_{B,\bar{A}}$ are inserted at: 
\bea
\mathcal{O}^s_B &=&\mathcal{O}(x^s_B),\;\mathcal{O}^s_{\bar{A}}=\mathcal{O}(x^s_{\bar{A}}),\;\;x^s_B=\left(u^s_B,v_B\right),\;x^s_{\bar{A}}=\left(u^s_{\bar{A}},v_{\bar{A}}\right)\nonumber\\
\;\;u_B^s &=& u_B-\frac{\delta x^-}{2}e^{-s},\;u^s_{\bar{A}}=u_{\bar{A}}+\frac{\delta x^-}{2}e^{-s}
\label{shifts}
\eea
The superscript on the coordinates is to emphasize that we have shifted the operator insertions by a small $s$-dependent amount, as illustrated in Figure \ref{fig:double_mod_flow}. We have done this for future convenience where this will lead to several cancelations that are important for the final bound that we derive. These were found by trial and error, however a complete understanding of these can be found in later work \cite{Ceyhan:2018zfg} where they naturally arise from relative modular flow. See for example (3.25) of that paper - where the shifts are encoded in the $V$ factors. 

In the next section we will discuss general properties of $f(s)$ in the strip $ - \pi/2 < {\rm Im} s < \pi/2$.
Here we would like to compute $f(s)$ in the lightcone limit where we take $s$ large but not too large.
For now we keep $s$ fixed and $\mathcal{O}(1)$. 
Either way taking $v_{B,\bar{A}}$ small, the dominant contribution will be the lowest twist defect operators which for now we take to be the stress tensor and the displacement operator. We will use the same method as in the simple example above. We now have the complication of two modular flows to keep track of, however we can use the same perturbative method: writing $K_A = K_A^0 + (K_A- K_A^0)$ and $K_B = K_B^0 + (K_B-K_B^0)$ and expanding. Just as for single modular flow this method needs proper justification. Again we go through an alternative method in Appendix~\ref{sec:appmodflowdouble} that gives \emph{exactly} the same answer as the perturbative approach. The method involves using a Cauchy-Schwarz bound to write \eqref{affs} at the order we wish to compute it, but now only using single modular flow. We have an independent (not perturbative) argument that our computations of singular modular flow are under control and so the results of this section are justified. 

In the perturbative approach we expand starting from:
\be
\left< \psi \right| \mathcal{O}^s_B  e^{  - i s K_B}  \left(e^{ i s K_A}  \mathcal{O}^s_{\bar{A}}  e^{ -i s K_A}\right) e^{   i s K_B} \left| \psi \right>
\ee
The leading order term (after replacing $K_A \rightarrow K_A^0$ etc.) is constrained by the algebra of modular inclusions.
We will use the following notation to denote the zeroth order modular flow  $\mathcal{O}^s_{\bar{A}}(s_A,-s_{B}) \equiv e^{  - i s K_B^0}  \left(e^{ i s K_A^0}  \mathcal{O}^s_{\bar{A}}  e^{ -i s K_A^0}\right) e^{   i s K_B^0}$ and  this is a local operator at:\footnote{Note the ordering of the argument in  $\mathcal{O}_{\bar{A}}(s_A,-s_{B})$  is important and the subscript tells us which modular Hamiltonian to evolve with, $K_A^0$ or $K_B^0$. } 
\be
x^s_{\bar{A}}(s_A,-s_{B}) \equiv (u^s_{\bar{A}}(s_A,-s_{B}) ,v_{\bar{A}}(s_A,-s_{B}) )
\equiv ( u^s_{\bar{A}} + (1 - e^{-s}) \delta x^-, v_{\bar{A}} )
\ee
We will use similar notation to describe other zeroth order modular flows below. 
Expanding we find:
\begin{align}
\nonumber
& \left< \psi \right| \mathcal{O}^s_B  e^{  - i s K_B}  \left(e^{ i s K_A}  \mathcal{O}^s_{\bar{A}}  e^{ -i s K_A}\right) e^{   i s K_B} \left| \psi \right>
 = \left< \psi \right| \mathcal{O}^s_B   \mathcal{O}^s_{\bar{A}} (s_A,-s_B) \left| \psi \right> +  \\
& \qquad \qquad \qquad \qquad \qquad+  i \int_0^s dt \left( \vphantom{\int_0^s} \left< \psi \right| \mathcal{O}^s_B(s_{B},(t-s)_A) \left[ K_A-K_A^0,
\mathcal{O}^s_{\bar{A}}(t_A) \right] \left| \psi \right>  \right. \\& \qquad \qquad \qquad \qquad \qquad \qquad \qquad
\left.  \vphantom{\int_0^s}  - \left< \psi \right| \mathcal{O}^s_B(t_{B}) \left[ K_B-K_B^0,
\mathcal{O}^s_{\bar{A}}(s_A,(t-s)_{B}) \right] \left| \psi \right> \right) + \ldots \nonumber
\end{align}
We can write this  in terms of the correlators used in the warm up section $\mathcal{R}(x_2,x_1 ;A), \mathcal{R}(x_2,x_1 ;B)$ and $\mathcal{P}(x_2,x_1)$ which are in turn related to the block function $F$. In particular note that all denominators
in $\mathcal{P},\mathcal{R}$ are the same for all the terms above: $=\left< \Omega \right|
\mathcal{O}^s_B(s_B) \mathcal{O}^s_{\bar{A}}(s_A) \left| \Omega \right>$. Factorizing out this denominator we find:
\begin{align}
f(s) = 1 &+ \mathcal{P}(x^s_B,x^s_{\bar{A}}(s_A,-s_{B} ) ) \\ &+  i \int_0^s d t \left( \mathcal{R}(x^s_B(s_B,(t-s)_A),x^s_{\bar{A}}(t_A);A) - \mathcal{R}(x^s_B(t_B),x^s_{\bar{A}}(s_A,(t-s)_B);B) \right)   + \ldots \nonumber
\end{align}
In a slight generalization to the previous section we have defined:
\be
\mathcal{R}(x_2, x_1;B) = i \left(  \int_{u_1}^{\delta x^-} du \left<T_{--}(u)\right>_\psi (u-\delta x^-)^2 \partial_u \left(\frac{ F(x_2, x_1; u) }{u-\delta x^-}\right)
+ \frac{F(x_2, x_1; \delta x^-)}{2\pi} \mathcal{P}_-^{B} \right) + \ldots
\ee
relevant for the action of $\left[ K_B-K_B^0, \,\,\cdot\,\, \right]$. Very similar manipulations to the previous section follow, in particular we switch the order of integration from $d\lambda \leftrightarrow d u$
where $\lambda = e^t$. We also need some slightly more general identities for $F$ relating
to boosts around the $u = \delta x^-$ surface:
\be
F(x^s_B(t_{B}), x^s_{\bar{A}}(s_A,(t-s)_{B} ) ) = e^t F(x^s_B, x^s_{\bar{A}}(s_A,(-s)_{B} );
(u-\delta x^-)/\lambda+ \delta x^- )
\ee
Turning the crank we find the answer:
\be
\label{fcrank}
f(s) = 1 - \int_{u^s_{\bar{A}} }^{u^s_B} d u \left< T_{--}(u) \right>_\psi B_s(u)
- \frac{\mathcal{J} B_s(0)}{2\pi} \mathcal{P}_-^A -  \frac{\mathcal{J} B_s(\delta x^-)}{2\pi} \mathcal{P}_-^{B}
\ee
where we have the piecewise defined function:
\begin{align}
B_s(u) &= F(x^s_B(s_{B},-s_A),x^s_{\bar{A}}; u)  \qquad  u^s_{\bar{A}} < u < 0 \\
&= F(x^s_B(s_{B}),x^s_{\bar{A}}(s_A); u)  \qquad  0 < u < \delta x^- \\
&= F(x^s_B,x^s_{\bar{A}}(s_A,-s_{B}); u)  \qquad  \delta x^- < u < u^s_B
\label{Bsu}
\end{align}
This is quite a remarkable result. Again the stress tensor is only integrated between the positions of the original operator insertions, and cancellations occur in the computation to ensure this.
Although our intermediate steps involve $T_{--}$ inserted in a larger $u$ window, we note that since we are re-summing local operators at the location of the defects to find this integral, the computation is actually never sensitive to the expectation of $T_{--}$ outside of this window.

Because of all the definitions that go into the above result, the reader might get lost in these expressions so we write these out more fully:
\begin{align} \nonumber
f(s) & = 1 + \frac{16 \pi G_N \Delta_{\mathcal{O}} }{d} \frac{1}{\Delta v} \left( -\frac{\Delta v}{ \Delta u } \right)^h  \left( \int_{u^s_{\bar{A}} }^0 du \left<T_{--}(u)\right>_\psi b_L(u) +
 \int_{\delta x^- }^{u^s_B} du \left<T_{--}(u)\right>_\psi b_U(u)  \right. \\ & \left. \qquad \qquad +    \int_0^{\delta x^-} du \left<T_{--}(u)\right>_\psi b_M(u)  
 +  \frac{ 1-e^{-s}}{2\pi} \left( b_M(0)\mathcal{P}_-^A - b_M(\delta x^-) \mathcal{P}_-^{B} \right) \right)
 \label{complicated}
\end{align}
where $\Delta u = u_B - u_{\bar{A}}  - \delta x^-$,\footnote{The $s$-dependence we introduced in the insertion positions $u^s_{B,\bar{A}}$ has the desired effect of eliminating the would-be $s$-dependence in the term $(-\Delta u)^h$. This $s$ dependence was in v1 of this paper, and leads to issues later. } and the functions are:
\begin{align}
b_L(u) & = \left( - u^s_{\bar{A}}+u\right)^h \left( u^s_B - \delta x^- ( 1- e^{-s}) - u\right)^h  \qquad & u^s_{\bar{A}} < u< 0 \\
b_M(u) & = e^s \left( - u^s_{\bar{A}} +e^{-s} u\right)^h \left( u^s_B - \delta x^- (1 - e^{-s}) - e^{-s} u\right)^h  \qquad & 0 < u<  \delta x^- \\
b_U(u) & = \left( - u^s_{\bar{A}} - \delta x^- (1 - e^{-s}) +u\right)^h \left( u^s_B - u\right)^h  \qquad & \delta x^- < u< u^s_A
\end{align}
Note that $b_M(0) = e^s b_L(0)$ and $b_M(\delta x^-) = e^s b_U(\delta x^-)$. We can check what happens in the large $s$ limit. Only the middle part of the piecewise function has a large $e^s$ growing term $b_M \rightarrow (- u_{\bar{A}} (u_B -\delta x^-) )^h e^s$. So we find:
\be
f(s) \approx 1 + \frac{16\pi G_N \Delta_{\mathcal{O}}}{d} \frac{e^s}{\Delta v} \left( \frac{\Delta v (u_B -\delta x^-)  u_{\bar{A}} }{(u_B- \delta x^-) - u_{\bar{A}} } \right)^h \left( \int_0^{\delta x^-} du \left< T_{--}(u)\right>_\psi
+ \frac{ \mathcal{P}_-^A -  \mathcal{P}_-^{B} }{2\pi} \right) 
\ee
We thus have the promised $e^s$ growing term multiplying the QNEC quantity. Later we will find it necessary to place the operators symmetrically about the two entanglement cuts with $u_{\bar{A}}   = - u_B +\delta x^-$, then:
\be
f(s) \approx 1 + \frac{ 16\pi G_N \Delta_{\mathcal{O}} }{d} \frac{e^s}{\Delta v} \left(- \frac{\Delta v (\Delta u - \delta x^-)}{4} \right)^h Q_{-}(A,B) + \ldots
\ee
where $\Delta u = u_B - u_{\bar{A}} > \delta x^-$ and $\Delta v = v_B - v_{\bar{A}} < 0$ .  This was the main goal of this section. In the following section we will study more general properties of $f(s)$ in the complex strip and use this to show that the quantity $Q_-$ above is positive. 

\section{General properties of $f(s)$}

\label{sec:f}

\subsection{Analyticity in the $s$-strip}

\label{ssec:a}

We start by noting that in this section in order to apply various theorems from the algebraic approach to QFT we need to take the operators $\mathcal{O}^s_{B,\bar{A}}$ to be non-distributional bounded operators. This can be achieved by firstly smearing the local operators over a small neighborhood $\delta^d$ in spacetime keeping this region within the respective domains of dependence $\mathcal{D}(B)$ or $\mathcal{D}(\bar{A})$. Secondly to make the operators bounded we could apply a projection from the spectral decomposition of the operator or we could also simply use fermionic operators. As long as the lengths scales that these procedures introduce (e.g. $\delta$) are much smaller than the various length scales in our final setup we expect the details of this procedure will not matter and in our final results we can replace the operators again with local bosonic operators. 

Let us first define two functions which are un-normalized versions of $f$:
\begin{align}
\label{defgminus}
g_-(s) \equiv \left< \psi \right| \mathcal{O}^s_B e^{ - i s K_B} e^{ i s K_A} \mathcal{O}^s_{\bar{A}} \left| \psi \right>  \\
g_+(s)  \equiv \left< \psi \right| \mathcal{O}^s_{\bar{A}} e^{ - i s K_A} e^{ i s K_B} \mathcal{O}^s_B \left| \psi \right>
\label{defgplus}
\end{align}
We can write the functions as inner product on two states:
\be
\label{gpm}
g_-(s) = \left( e^{ i s^\star K_B} \left(\mathcal{O}^s_B\right)^{\dagger} \left| \psi \right> , e^{ i s K_A} \mathcal{O}^s_{\bar A}  \left| \psi \right> \right) \qquad g_+(s) = \left(e^{ i s^\star K_A} \left(\mathcal{O}^s_{\bar A}\right)^\dagger   \left| \psi \right> , e^{ i s K_B} \mathcal{O}^s_B \left| \psi \right> \right)
\ee
The two functions $g_{\pm}$ are well defined for $\left\lbrace g_- : -\pi \leq {\rm Im } s \leq 0\right\rbrace$ and $\left\lbrace g_+ : 0 \leq {\rm Im } s \leq \pi\right\rbrace$ as long as $x^s_B\in \mathcal{D}(B), x^s_{\bar{A}}\in\mathcal{D}(\bar{A})$, this is satisfied for $\text{Re}s>s_0=\text{max}\left\lbrace \ln{\left(\frac{\delta x^-}{2|u_B-\delta x|}\right)},\ln{\left(\frac{\delta x^-}{2|u_{\bar{A}}|}\right)}\right\rbrace$. These are semi-strips which we simply refer to as strips, inside of which the functions are bounded using the Cauchy-Schwarz inequality by the norms of these states
\begin{align}
& | g_-(s) | < \| e^{ {\rm Im} (s) K_B} \left(\mathcal{O}^s_B\right)^\dagger \left| \psi \right>  \|  \| e^{ - {\rm Im} (s) K_A} \mathcal{O}^s_{\bar A} \left| \psi \right>  \| \\
& | g_+(s) | < \| e^{ {\rm Im} (s) K_A} \left(\mathcal{O}^s_{\bar A}\right)^\dagger \left| \psi \right>  \|  \| e^{ - {\rm Im} (s) K_B} \mathcal{O}^s_{B} \left| \psi \right>  \|
\end{align}
The bound is finite in their respective $s$-strips thanks to Tomita-Takasaki theory (see Section~\ref{sec:tt}). 
This also establishes analyticity since Tomita-Takasaki theory tells us that the maps taking
$\mathbb{R} \rightarrow \mathcal{H}_{CFT}$ or $\mathcal{H}_{CFT}^\star$ :
\be
s \rightarrow e^{ is K_A} \mathcal{O}_{\bar{A}} \left| \psi \right>
\qquad s \rightarrow \left< \psi \right| \mathcal{O}_B e^{ - i s K_B}
\ee
have analytic continuations into the strip $\left\lbrace -\pi \leq \Im s \leq 0,\; \Re s > s_0\right\rbrace$ for the vector and dual vector in $g_-(s)$ of \eqref{gpm}. This is similarly true for the vectors and dual vectors in $g_+(s)$ but now for the strip $\left\lbrace 0 \leq \Im s \leq \pi,\;\Re s>s_0\right\rbrace$.  

The fact that the operators are inserted in an $s$ dependent manner leads to a slight subtlety here since for complex $s$ the operators (or smeared versions thereof) are no longer clearly part of the algebra. This is not a problem since at somewhat large $s$ the shifts are small and can be achieved with a simple Taylor series expansion in $\delta x^- e^{-s}$. We can then make the argument for each term in this expansion. Another justification of this should be achievable with relative modular flow, since this gives rise to these operator shifts \cite{Ceyhan:2018zfg}, and has a controlled analytic structure. See also Section~\ref{sec:locgeo} for further discussion of this point. 

Now notice that $g_+(s) = g_-(s)$ for $\left\lbrace {\rm Im} s =0, \;\Re s>s_0\right\rbrace$ since the modular flowed operators $e^{isK_A}\mathcal{O}^s_{\bar{A}}e^{-isK_A}$ and $e^{isK_B}\mathcal{O}^s_{B}e^{-isK_B}$ commute with each other. This is due to the inclusion property $B \subset A$ and $\bar{A} \subset \bar{B}$. Note that for real $s$ the small $s$-dependent shifts are real so indeed we can associated the operators to their respective algebra.
 
Therefore we have $g_{\pm}(s)$ as holomorphic functions agreeing along the real $s$ axis where they are continuous functions, by the edge of the wedge theorem, they must be complex analytic continuations of each other. Thus we can define $g(s)$ holomorphic in the full strip $\left\lbrace -\pi \leq {\rm Im s} \leq \pi,\;\Re s > s_0\right\rbrace$. We will then use $g(s)$
in our definition of $f(s)$ after normalizing appropriately. 

To make contact with some of the theorem's used when studying half sided modular inclusions
we can prove analyticity in a slightly different way. Consider the structure theorem proven in various ways \cite{wiesbrock1993half, borchers1995use,araki2005extension, buchholz1990nuclear} which states that 
\be
V_-(s) \equiv e^{ - i s K_B} e^{ i s K_A} \qquad \bar{A} \subset \bar{B}
\ee
has an analytic extension to the complex strip $\left\lbrace 0 \leq {\rm Im} s \leq \pi\right\rbrace$ as a holomorphic function with values in the space of bounded operators on the QFT Hilbert space. 
Furthermore in this strip the operator norm is bounded $\| V_-(s) \|  \leq 1$. Similar statements hold for the opreator:
\be
V_+(s) \equiv e^{ - is K_A} e^{ is K_B} \qquad B \subset A
\ee
however now $\| V_+(s) \| \leq 1$ in the strip $\left\lbrace - \pi \leq {\rm Im} s \leq 0\right\rbrace$. 
We will not go over the details of the proof except to note that $B \subset A$ implies that
$V_+(s - i\pi) = J_A V_+(s) J_B$ so the operator $V_+$ is unitary on the top and bottom of the strip
$\Im s =0, - \pi$, and the bound on the norm in the interior follows roughly from the maximum modulus principle.

Note that in our definitions of $V_{\pm}$ the $s$-strips are reversed compared to where we defined $g_\pm$. We can thus use the results above to extend the definition of $g_{\pm}(s)$ to the full region $\left\lbrace -\pi \leq \Im s \leq \pi, \;\Re s>s_0\right\rbrace$ which then must satisfy $g_+(s) = g_-(s) \equiv g(s)$ everywhere in this strip again via the edge of the wedge theorem. For example now we can set:
\begin{align}
g(s) &= \left< \psi \right| \mathcal{O}^s_{\bar{A}} V_+(s) \mathcal{O}^s_B \left| \psi \right> 
\qquad - \pi \leq \Im s \leq 0,\;\Re s>s_0 \\
g(s) &= \left< \psi \right| \mathcal{O}^s_B V_-(s) \mathcal{O}^s_{\bar{A}} \left| \psi \right> 
\qquad 0 \leq \Im s \leq \pi,\;\Re s>s_0
\end{align}

We can see, in a specialized example, how the bound on the norm of $V_\pm$ arises. Take the  modular flow for Rindler cuts in vacuum, i.e. those relevant for defining the denominator in $f(s)$. In this case the modular Hamiltonians $K_{A,B} \rightarrow K_{A,B}^0$ satisfy the Poincare algebra and we can simply compute $V_\pm^0$:
\be
V^0_+(s) = U( -\delta x^- (1 - e^{-s} ) ) \qquad V^0_-(s) = U(\delta x^- (1 - e^{-s} ) )
\ee
where $U$ is a null translation in the $x^-$ direction. This translation is generated by a positive null momentum operator $P_-$ with $U(a) = e^{ i a P_-}$. Such that $a = \mp \delta x^-(1- e^{-s})$ with $\Im a >0$ in the appropriate $s$ strip. This situation will generalize to cases that satisfy the contraint of half-sided modular inclusions $e^{ i s K_A^0} \mathcal{A}_B e^{ -i s K_A^0} 
\subset \mathcal{A}_B$ for $s>0$ (see Section~\ref{modinclmain} and Appendix~\ref{app:modincl}). This only applies to vacuum states for special cuts, and we emphasize this is not a general result for the $\psi$ modular Hamiltonians.  In fact later we will give evidence that the growing $Q_-$ term only arises in the case where the algebra does \emph{not} apply. However the case for half-sided modular inclusions will always be used to normalize $g$. So we give a complete definition of $f$:
\begin{align}
g_0(s) &= \left< \Omega \right| \mathcal{O}^s_{\bar{A}} V^0_+(s) \mathcal{O}^s_B \left| \Omega \right> 
\qquad - \pi \leq \Im s \leq 0,\;\Re s>s_0 \\
g_0(s) &= \left< \Omega \right| \mathcal{O}^s_B V^0_-(s) \mathcal{O}^s_{\bar{A}} \left| \Omega \right> 
\qquad 0 \leq \Im s \leq \pi,\;\Re s>s_0
\end{align}
with $f(s) = g(s)/g_0(s)$. Here the region of analyticity for $g_0(s)$ is at least that of $g(s)$, although we now have to consider the possibility that $g_0(s)$ has a zero. For Rindler modular Hamiltonians in a CFT we can just compute this quantity:
\be\label{eq:normalization}
g_0(s) = c_{\mathcal{O}} \left(- ( v_B- v_{\bar{A}})( u_B - u_{\bar{A}}  - \delta x^- ) \right)^{-\Delta_{\mathcal{O}}}
\ee
where this function is $s$-independent and has no zeros. So we conclude that $f(s)$ is analytic in the complex $s$-strip of interest. 

For example the small corrections we computed in Section~\ref{sec:dmodflow} for the lightcone limit of $f(s)$, which were in general complicated functions of $s$ \eqref{complicated}, demonstrate this analyticity in a non-trivial way. Although it should be noted that it is important for the QNEC proof that we can make the analyticity argument for general $s$ including very large $s$ where we move outside of the lightcone limit and where we do not have an explicit expression for $f(s)$. 

We end by noting that the operator bound on $V$ is not useful for us since the norm of the state $\mathcal{O}_A \left| \psi \right>$ is dependent on the details of the smearing of the operator, so this will be a very weak bound scaling with some inverse power of the small distance scale $\delta$ over which we smear the operator. We now move onto proving a more refined bound on the boundaries of the reduced strip $\left\lbrace -\pi/2 \leq \Im s \leq \pi/2,\;\Re s>s_0\right\rbrace$. 

\subsection{Cauchy-Schwarz bound}
 We now study constraints on $f(s)$ along the lines $ \left\lbrace \Im s = \pm i \pi/2,\;\Re s>s_0\right\rbrace$. 
Our considerations here are in line with, and thus motivated by, the derivation of the chaos \cite{Maldacena:2015waa} and causality \cite{Hartman:2015lfa} bounds. Starting with $\Im s = \pi/2$ we use $g_+$ defined in 
\eqref{defgplus}.
We use the Cauchy-Schwarz inequality to show:
\be
\label{cs}
{\rm Re} f(s) < |f(s)| < \frac{ \|e^{  \pi K_A/2}  \left(\mathcal{O}^{t+i\pi/2}_{\bar{A} }\right)^\dagger \left| \psi \right> \|
\| e^{ - \pi K_B/2}  \mathcal{O}^{t+i\pi/2}_{B }\left| \psi \right> \| }{|\left< \Omega \right| \mathcal{O}^{t+i\pi/2}_{\bar{A}} e^{ - i s K_A^0} e^{ i s K_B^0} \mathcal{O}^{t+i\pi/2}_B \left| \Omega \right>|} 
,\;\; s = t + i\pi/2,\;t\in\mathbb{R},\;t>s_0
\ee

The denominator is simply given by (\ref{eq:normalization}), which is $s$-independent. The numerator terms in \eqref{cs} are determined by two point functions, for example:
\bea
\label{normcorr}
\| e^{  \pi K_A/2}  \left(\mathcal{O}^{t+i\pi/2}_{\bar{A} }\right)^\dagger \left| \psi \right> \|  &=& \sqrt{\left< \psi \right| \mathcal{O}^{t+i\pi/2}_{\bar{A}} e^{\pi K_A}  \mathcal{O}^{t-i\pi/2}_{\bar{A}}  \left| \psi \right>}\nonumber\\
\| e^{  -\pi K_B/2}  \mathcal{O}^{t+i\pi/2}_B \left| \psi \right> \|  &=& \sqrt{\left< \psi \right| \mathcal{O}^{t-i\pi/2}_B e^{-\pi K_B}  \mathcal{O}^{t+i\pi/2}_B \left| \psi \right>}
\eea
where we have taken $\mathcal{O}_{\bar{A},B}$ to be a Hermitian operator, and thus $\left(\mathcal{O}_{\bar{A},B}^s\right)^\dagger=\mathcal{O}_{\bar{A},B}^{s^*}$. By examining the norms in \eqref{normcorr} let us introduce two normalized single modular flowed correlators: 
\bea
h_{\bar{A}}(s)&=&\frac{\left< \psi \right| \mathcal{O}^{s}_{\bar{A}} e^{\pi K_A}  \mathcal{O}^{s-i\pi}_{\bar{A}} \left| \psi \right>}{\left< \Omega \right| \mathcal{O}^{s}_{\bar{A}} e^{\pi K^0_A}  \mathcal{O}^{s-i\pi}_{\bar{A}}  \left| \Omega \right>}\nonumber\\
h_B(s)&=&\frac{\left< \psi \right| \mathcal{O}^{s-i\pi}_B e^{-\pi K_B}  \mathcal{O}^{s}_B \left| \psi \right>}{\left< \Omega \right| \mathcal{O}^{s-i\pi}_B e^{-\pi K^0_B}  \mathcal{O}^{s}_B  \left| \Omega \right>}
\label{hsubtract}
\eea 
These are designed to be analytic in $s$ as well as to reproduce the norms in \eqref{normcorr} for $s = t + i \pi/2$. 
The denominators in $h_{\bar{A},B}(s)$ are also $s$-independent, and agree with one another as well as with that of $f(s)$ if we choose: 
\be
u_B= \frac{\Delta u + \delta x^-}{2},v_B = -\frac{\Delta v}{2},\;\;u_{\bar{A}}= -\frac{\Delta u - \delta x^-}{2},v_{\bar{A}} = \frac{\Delta v}{2}
\label{positions}
\ee
So we pick $\lbrace u_{\bar{A},B}, v_{\bar{A},B}\rbrace$ to be these values from now on. It is very important that the denominators match so that the leading terms for the Cauchy-Schwarz bound below in the light-cone limit all cancel.

The functions $h_{\bar{A},B}(s)$ can be computed using the single modular flowed correlators discussed in Section \ref{sec:singmodflow}, and have the same analyticity properties as $f(s)$; they are real and positive along $\left\lbrace\Im s =\pi/2, \;\Re s>s_0\right\rbrace$ because they are proportional to norms of states there. Now define 
\be 
F(s)\equiv f(s)-\frac{1}{2}h_{\bar{A}}(s)-\frac{1}{2}h_B(s)
\label{Fs}
\ee
The Cauchy-Schwarz bound therefore translates into: 
\bea
&&\Re f(s)\; \leq\; |f(s)|\; \leq\; h_{\bar{A}}(s)^{1/2}h_B(s)^{1/2}\;\leq\; \frac{1}{2}h_{\bar{A}}(s)+\frac{1}{2}h_B(s)\nonumber\\
\rightarrow && \Re F(s) = \Re\left(f(s)-\frac{1}{2}h_{\bar{A}}(s)-\frac{1}{2}h_B(s)\right)\;\leq\; 0,\;\;s=t+i\pi/2,\;t>s_0
\label{cscs}
\eea
By starting with the definition of $g(s)$ using $g_-(s)$ and that $h_{\bar{A},B}(t-i/2)=h_{\bar{A},B}(t+i/2)$ for $t\in \mathbb{R}, t>s_0$, one can show that the same is also true for $\Im s=-\pi/2, \;\Re s>s_0$. We thus conclude that 
\be
\Re F(s) < 0\;  , \;\;\;\Im s = \pm i \pi /2,\;\Re s>s_0
\ee

\subsection{The sum rule}

\label{ssec:sum}

In order to make contact with the ANEC sum rule derived in \cite{Hartman:2016lgu} let's map the strip $-\pi/2 < \Im s < \pi/2$ to the upper half plane via:
\be
\label{oursig}
\sigma = \frac{i} {\Delta u - \delta x^-} e^{-s} \,, \qquad (\Delta u - \delta x^-)  > 0
\ee
and define 
\be
\label{ourz}
z^2 = - \Delta v (\Delta u - \delta x^-)/4 \geq 0
\ee
note that $z^2 \rightarrow \eta$ was used in \cite{Hartman:2016lgu}. 

\begin{figure}[h!]
\centering
\includegraphics[scale=0.22]{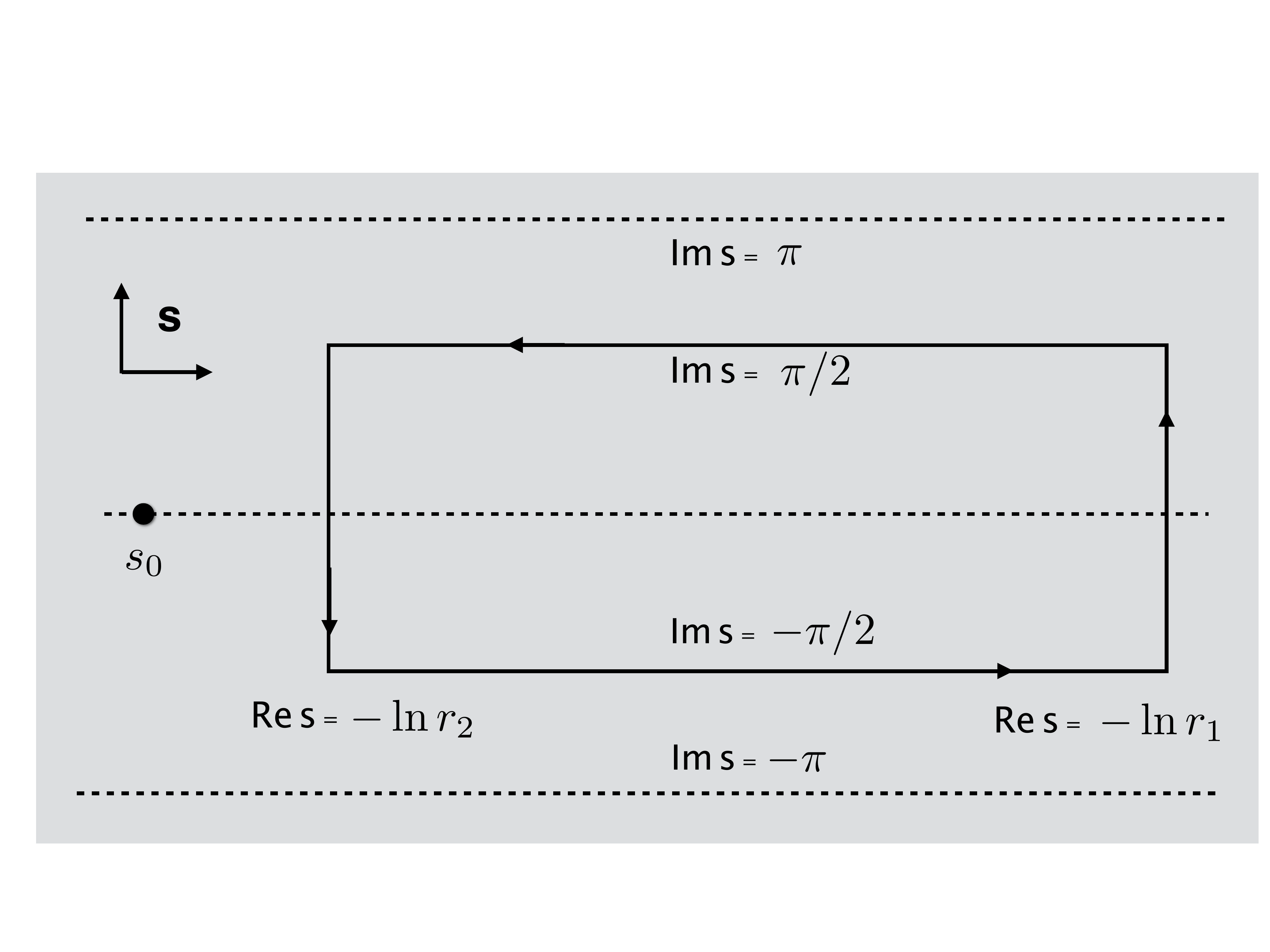}
\includegraphics[scale=0.22]{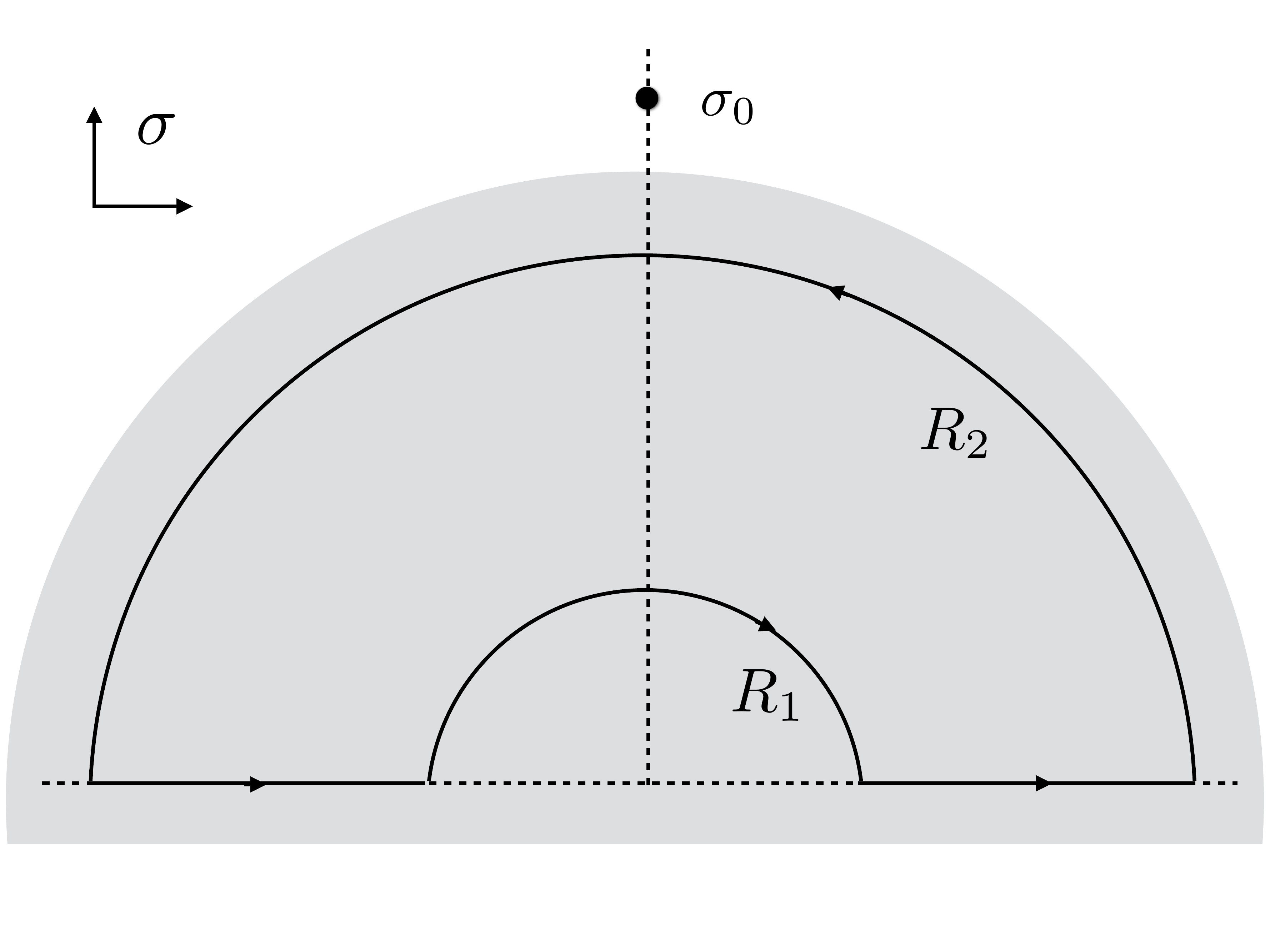}
\caption{Left: the contour $\Gamma$ in the $s$-stripe; right: the image of $\Gamma$ in the $\sigma$-plane. $\Gamma$ is chosen to be away from the possible branch-cut singularity $s_0$, so that $F(s)$ is analytic in the region bounded by $\Gamma$.}
\label{fig:sum_rule}
\end{figure}

To proceed, let us consider a contour $\Gamma$ in the $s$-stripe where $F(s)$ is analytic, consisting of $\Im s = \pm \pi/2$ and $\Re s =-\ln{r_1},\;\Re s=-\ln{r_2}$, where $s_0 < -\ln{r_1}<-\ln{r_2}$. This is then mapped to semi-circles of radii $R_{1,2}=\frac{r_{1,2}}{\Delta u -\delta x^-}$ connected by straight line segments in the $\sigma$-plane (Figure \ref{fig:sum_rule}).  

By analyticity $\oint_\Gamma d\sigma F(\sigma) = 0$, we have that: 
\bea\label{eq: sum_rule_contour}
&&-\int^{R_2}_{\text{semi-circle}} d\sigma F(\sigma) + \int^{R_1}_{\text{semi-circle}} d\sigma F(\sigma) = \int^{-R_1}_{-R_2} d\sigma F(\sigma) + \int^{R_2}_{R_1} d\sigma F(\sigma) \nonumber\\
&=&\frac{1}{\Delta u -\delta x^-}\int^{-\ln{r_1}}_{-\ln{r_2}} dt\; e^{-t} \left\lbrace F\left(t+i\pi/2\right)+F\left(t-i\pi/2\right)\right\rbrace
\eea
where $\int^R_{\text{semi-circle}}$ denotes an anti-clockwise integral over the semi-circle of radius $R$. In general, $F(s)$ is some complicated function we know nothing about other than its analytic properties. However, when the light-cone expansion is valid (i.e. moderate $t$), we can approximate the double-modular flowed term $f(s)$ by (\ref{complicated}), and the same-side single modular flowed terms $h_{\bar{A},B}(s)$ using (\ref{sameside}). Let us write the leading order light-cone approximation thus obtained as $\tilde{F}(s)$. After some tedious algebra one can show that : 
\be
\tilde{F}\left(t+i\pi/2\right)+\tilde{F}\left(t-i\pi/2\right)=0,\;\;t\in\mathbb{R} 
\label{Fstrip}
\ee
This fact is somewhat analogous to the statement that the double discontinuity defined for a 4 point CFT correlation function in \cite{Caron-Huot:2017vep} vanishes when applied to a single conformal block in the s-channel.  The expressions in (\ref{complicated}) are analogous to the light cone limit of these blocks. We emphasize that \eqref{Fstrip} is a non trivial result for which it is important that we made the appropriate small $s$ dependent shifts on the operator locations. This fact will also be essential to our proof of the QNEC bound as we will expand upon below. 

In light of (\ref{eq: sum_rule_contour}), this implies that for any $R_1, R_2$ where the light cone expansion is valid, we have:
\be\label{eq:radii_indepedendence}
 \int^{R_2}_{\text{semi-circle}} d\sigma \tilde{F}(\sigma)=\int^{R_1}_{\text{semi-circle}} d\sigma \tilde{F}(\sigma)
\ee
independent of the radii $R_1, R_2$. Suppose we expand: 
\be 
\tilde{F}(\sigma) = \sum_{\ell\in \mathbb{Z}} \mathcal{Q}_\ell \sigma^\ell 
\ee
and recall that 
\be
\int^{R}_{\text{semi-circle}} d\sigma  \sigma^{\ell} =\begin{cases}
i\pi,\;\;\; \ell = -1\\
-\frac{2}{1+\ell}R^{\ell+1},\;\;\ell \in \text{even}\\
0,\;\;\ell \neq -1\text{ and }\ell \in \text{odd}
\end{cases}
\ee
The $R$ independence of (\ref{eq:radii_indepedendence}) implies that we must have $\mathcal{Q}_{2n}=0$ for $n\in \mathbb{Z}$, thus $\tilde{F}(\sigma)$ only contains terms of odd integer power, of which only the simple pole $\mathcal{Q}_{-1}\sigma^{-1}$ survives the semi-circle integral:
\be\label{eq:sum_rule} 
\int^{R}_{\text{semi-circle}} d\sigma  \tilde{F}(\sigma) = i\pi\mathcal{Q}_{-1}
\ee 
The residue of the $\sigma$ pole in the light cone limit is:
 \be
\tilde{F}(\sigma) = - i \frac{4\pi G_N \Delta_{\mathcal{O}}}{d} z^{d-2} \sigma^{-1} Q_-(A,B)  + \ldots
\ee  

At this point let us also comment on the reason for the $s$-dependence in the operator insertions $x^s_{B,\bar{A}}$. Had we chosen to proceed with $s$-independent insertions, say simply placing $\mathcal{O}_{B,\bar{A}}$ at $\left(u_{B,\bar{A}},v_{B,\bar{A}}\right)$ (see Figure \ref{fig:double_mod_flow}), which was how we initially attempted the proof, then the corresponding $\tilde{F}(\sigma)$ thus computed will be contaminated by even powers of $\sigma$. Doing the semi-circle integral will yield corrections to (\ref{eq:sum_rule}) which are supressed by powers of $R$ but might be leading in $z$. In order to suppress these corrections we would need to be in the large $t$ (small $R$) limit compared to the other small paramater $z^\#$ controlling the light cone limit, at which point the light-cone approximation $F(s)\approx \tilde{F}(s)$ can no longer be trusted. In particular the order of limits is very important: we must take $z$ small before we take $s$ large.  We can only achieve this, while projecting onto the term of interest, once \eqref{Fstrip} is satisfied.
The $s$-dependence in the positions of the operator insertions $x^s_{\bar{A},B}$ is introduced for this purpose.

As an example of the power of this statement we can now give the reason that low twist scalar contributions can be ignored. While such contributions might dominate the light cone limit they are killed by the above contour integrals. More explicitly this is because they don't have a $\sigma^{-1}$ term but they will still satisfy the constraint \eqref{Fstrip}. So for example if we had not proceeded with the $s$-dependent shifts of the operator locations then indeed such low twist contributions would have given the leading answer. 

Now consider the contour $\mathcal{C}$ consisting of the semi-circle $\lbrace |\sigma|=R, \Im \sigma\leq 0\rbrace$ and the straight-line segment $\left\lbrace -R\leq \sigma\leq R, \sigma \in \mathbb{R}\right\rbrace$, inside a region where $F(\sigma)$ is analytic and the light-cone approximation $F(s)\approx \tilde{F}(s)$ is valid. Taking the real part of 
\be 
\oint_{\mathcal{C}} d\sigma F(\sigma)=0
\ee
we get
\bea
-\int^R_{-R}\;d\sigma\; \Re F(\sigma)=\Re \int^R_{\text{semi-circle}}F(\sigma) \approx \frac{4\pi^2 G_N \Delta_{\mathcal{O}}}{d} z^{d-2} Q_-(A,B) 
\eea
Applying the Cauchy-Schwarz bound we obtained before, which in the $\sigma$-plane says: 
\be 
\Re F(\sigma)\leq 0,\; \sigma \in \mathbb{R}
\ee
we can extract the QNEC quantity as a positive sum rule:
\be
Q_-(A,B) = -\frac{d }{4 \pi^2 G_N \Delta_{\mathcal{O}}} \lim_{z \rightarrow 0}  z^{2-d} \int_{-R}^R d \sigma \Re F(\sigma) \geq 0
\ee
This ends our proof of the QNEC.  


\section{Loose ends}

\label{sec:loose}

\subsection{Mixed states}

Here we address the issue of mixed states. Until now we have only considered pure states $\left| \psi \right>$, and since entanglement entropy is non-linear in the state we cannot simply extrapolate the QNEC for mixed states from the QNEC for pure states (this should be contrasted with the ANEC where this was possible.) Indeed we will see an interesting effect relating to mixed states when interpreting our results through the lens of a putative gravitational dual theory: the entangling surface splits into two surfaces, one for $A$ and one for $\bar{A}$. This is also true holographic theories if we use the quantum extremal surface \cite{engelhardt2015quantum} in our computation of the quantum corrected entanglement entropy. So our results to follow give further, theory independent, evidence for the importance of the quantum extremal surface (see \cite{Dong:2017xht} for a derivation in holographic models.)

Our methods rely on the existence of a modular operator $K_A$ associated to a sub region. While we could define $- \ln \rho_A + \ln \rho_{\bar{A}}$ for any mixed state $\rho_\psi$, this is not the correct generalization of $K_A$.
For example it has very different properties than what we might hope, most notably there is no sense in which this guess annihilates the state. It is also, up to a sign, the same operator for $A$ and $\bar{A}$ which would mean our subsequent arguments, if they had been possible, would not distinguish 
$S_A$ from $S_{\bar{A}}$.  The correct thing to do, it turns out, is to instead look for a purification of $\rho_\psi$ in an enlarged Hilbert space $\mathcal{H}_A \otimes \mathcal{H}_{\bar{A}} \otimes \mathcal{H}_C$. Then the correct modular operator is
\be
2\pi K_A = (- \ln \rho_A  + \ln \rho_{\overline{A}+C})
\ee
Which now manifestly differs from the minus of:
\be
2\pi K_{\bar{A}} = (- \ln \rho_{\bar{A}}  + \ln \rho_{A+C })
\ee
Finding a purification is always possible and at worst we must take this new Hilbert space to be a double of the original CFT Hilbert space. Let us embrace this ``worst case'' scenario and map $\rho_\psi$ to a pure state in the thermofield double Hilbert space $\mathcal{H}_{CFT} \otimes \mathcal{H}_{CFT'}$. To illustrate this we consider a mixed sum over primary states:
\be
\rho_\psi = \sum_{\alpha} \lambda_\alpha  \left| \alpha \right> \left< \alpha \right| 
\ee
where $\left| \alpha \right> = \psi_\alpha (0) \left| \Omega \right>$ is a real scalar primary operator insertion working in radial quantization about the origin and  $\left< \alpha \right| =\lim_{x \rightarrow \infty} |x|^{2\Delta_\alpha} \left< \Omega \right| \psi_\alpha (x) $. This density matrix is a state on the $S_{d-1}$ CFT Hilbert space and we would like to consider tracing over some spatial sub-region $\bar{A} \subset S_{d-1}$. We can take the purification to be:
\be
\left| \psi \right> = \sum_{\alpha} \sqrt{\lambda_{\alpha}} \left| \alpha \right> \left| \alpha \right>
 \in \mathcal{H}_{CFT} \otimes \mathcal{H}_{CFT'}
\ee
We now aim to compute matrix elements of the modular operator $K_A$ using the replica trick, relating the answer to the $n$ analytic continuation of a genuine correlator in the $\mathbb{Z}_n$ orbifold theory.  For example matrix elements of $\ln \rho_A$ can be computed via the $n \rightarrow 1$ limit of:
\be
Z_n(A) = \sum_{ \{\alpha_k;k=0,\ldots n-1\}} \left(\prod_{k=0}^{n-1} \lambda_{\alpha_k} \right) \left< \Sigma_n(\partial A) \overbracket{ \psi_{\{ \alpha_k\}}(0) \psi_{\{ \alpha_k\}}(\infty)}  \overbracket{\mathcal{O}_B \mathcal{O}_{\bar A} } \right>_{CFT^n/\mathbb{Z}_n}
\ee
where:
\be
\label{bilocmix}
\overbracket{ \psi_{\{ \alpha_i\}}(0) \psi_{\{ \alpha_i\}}(\infty)} 
= \bigotimes_{k=0}^{n-1} \psi_{\alpha_k}^{(k)}(0)   \psi_{\alpha_k}^{(k)} (\infty) + \mathbb{Z}_n \, {\rm symmeterization}
\ee
and the symmeterization only acts on the pairwise operators. Note that the operators that create the state, in the mixed case, are inserted in a correlated  way at $0$ and $\infty$ on a fixed replica.
This is just like the bi-local $\mathcal{O}_B \mathcal{O}_{\bar{A}}$ operator insertion that we had to deal with previously. We can understand how these specific bi-local operators arise from the purifcation perspective since tracing out $CFT'$ gives:
\be
\rho_A^n = \left({\rm Tr}_{\bar{A}+\rm CFT'} \left| \psi \right> \left< \psi \right| \right)^n =
\sum_{\{\alpha_k\}, \{\beta_k\} } \left( \prod_{k=0}^{n-1} \sqrt{ \lambda_{\alpha_k} \lambda_{\beta_k} }  \delta_{\alpha_k, \beta_k} {\rm Tr}_{\bar{A}} \left| \alpha_k\right> \left< \beta_k \right|
\right)
\ee
Thus correlating operators on the same replica. 

When computing matrix elements of 
$\ln \rho_{\bar{A} + CFT'}$ for the state $\left| \psi \right>$ it turns out the replica trick gives exactly the same prescription we have been working with for pure states:
\be
\label{hence}
Z_n(\bar{A}+CFT') =  \sum_{ \{\alpha_k;k=0,\ldots n-1\}} \left(\prod_{k=0}^{n-1} \lambda_{\alpha_k} \right) \left< \Sigma_n(\partial A) \overbracket{ \psi_{\{ \alpha_k\}}(0) \psi_{\{ \alpha_k\}}(\infty)}  \overbracket{\mathcal{O}_B \mathcal{O}_{\bar A}(\circlearrowleft) } \right>_{CFT^n/\mathbb{Z}_n}
\ee
that is all we have to do is take $\mathcal{O}_{\bar{A}}$ and move it in a clockwise direction around the $\Sigma_n$ defect. This follows after two steps. Firstly the trace over the purification now results in:
\be
{\rm Tr}_{CFT'} \left( \rho_{\bar{A}+CFT'}^n \right) =  {\rm Tr}_{\rm CFT'} \left( {\rm Tr}_{A} \left| \psi \right> \left< \psi \right|\right)^n
 = \sum_{\{\alpha_k\}, \{\beta_k\} } \left( \prod_{k=0}^{n-1} \sqrt{ \lambda_{\alpha_k} \lambda_{\beta_k} }  \delta_{\alpha_k, \beta_{k+1}}  {\rm Tr}_{A} \left| \alpha_k \right> \left< \beta_k \right| \right)
 \ee
which correlates operators $\delta_{\alpha_k, \beta_{k+1}}$ on \emph{shifted} replicas. The second step comes after we trace over $A$ in the above expression and write the answer
as a correlation function on a branched manifold $\mathcal{M}_n'$. In order to produce the same branch cut structure as the replicated manifold
$\mathcal{M}_n$ used for the computation of $Z_n(A)$, we must deform the cuts, which for $\mathcal{M}_n'$ initially lie along $\bar{A}$, back to $A$. Since the branch cut is now moved across the state operator insertions (say at $0$) this effectively undoes the shift in the correlation of $\psi_\alpha$ operators, thus giving back the same bilocal operators \eqref{bilocmix} as appearing in $Z_n(A)$. The only difference coming when we include the $\mathcal{O}_{B,\bar{A}}$ operator insertions one of which shifts replicas as we deform the branch cut. Hence \eqref{hence}.

The end result is satisfying and gives the same prescription for deformation the $\mathcal{O}_{B,\bar{A}}$ as we worked with before. So our proof continues for mixed states also. Note that the different correlation between state operator insertions means that we will find a different answer for $K_A$ versus $K_{\bar{A}}$.

There is another formalism for dealing with the various bi-local operators and oddities that we have defined in this paper. We sketch the picture here, and leave details for future discussion. Again the idea is that we want to write the replica trick computation using objects which are intrinsic to the orbifold theory. The oribold theory is a discrete gauging of the $\mathbb{Z}_n$ cyclic permutation symmetry for the $CFT^n$ theory. Gauging this symmetry results in new non-local operators that live naturally in this theory. For example the entanglement region $A$ where one identifies the different replicas, can originally be thought of as a co-dimension $1$ operator inserted along $A$ and ending on $\partial A$. After gauging we remove the co-dimension $1$ operator leaving a co-dimension $2$ twist operator $\Sigma_n$ living at $\partial A$. In particular the position of branch cut on the replicated manifold becomes irrelevant. 

This twist operator carries a $\widetilde{\mathbb{Z}}_n$ charge, corresponding to a $d-2$ form generalized global symmetry \cite{Gaiotto:2014kfa}, which arrises when we gauge the original ($0$-form) replica symmetry. The twist operators are somewhat analogous to flux tubes and we can measure the charge of the flux tube by encircling the twist operator with a Wilson loop for the discrete gauge symmetry: $W^q(\mathcal{C})$. These are labeled by an integer $q=0,\ldots n-1$ such that:
\be
\left< W^q(\mathcal{C} )  \Sigma_n(\partial A) \right>_{CFT^n/\mathbb{Z}_n}
= e^{ i q 2\pi/n} \left<  \Sigma_n(\partial A) \right>_{CFT^n/\mathbb{Z}_n}
\ee
Here $\mathcal{C}$ circles $\partial A$ (on the un-replicated/un-branched space.) We can define local operators that are charged under the gauge symmetry via:
\be
\psi_{ \{\alpha_k\}}^{q}(x) \equiv \sum_{l = 0}^{n-1} e^{ i q l 2\pi/n} \bigotimes_{k=0}^{n-1} \psi_{\alpha_{k+l}}^{(k)}(x)
\ee
and $k,l,q$ etc are defined modulo $n$. These will only make sense in the orbifold theory (for $q\neq 0$) if we attach them to Wilson lines. Thus we propose to define our bi-local operators as:
\be
\overbracket{ \psi_{\{ \alpha_k\}}(x) \psi_{\{ \alpha_k\}}(x')}^{\mathcal{C}(x,x')} \equiv \sum_{q=0}^{n-1} \psi^{-q}_{\{\alpha_k\}} (x) W^q(\mathcal{C}(x,x')) \psi^q_{\{\alpha_k\}}(x')
\ee
where $\mathcal{C}(x,x')$ is some open curve between the points $x,x'$. The curve $\mathcal{C}$ was previously implicitly defined by our convention of which point is on which replica for the branched covering (i.e. where we choose our branch cuts on $\mathcal{M}_n$ before gauging.)  

This definition also works for the bi-local probe operators:
\be
\overbracket{\mathcal{O}_B(x_B) \mathcal{O}_{\bar{A}}(x_{\bar{A}}) }^{\mathcal{C}(x_B,x_{\bar{A}})}
\ee
The two different prescriptions for $\ln \rho_A$ and $\ln \rho_{\bar{A}+ {\rm CFT'}}$ correspond to different choices of curves the $\mathcal{C}(x_B,x_{\bar{A}})$, passing to the left or right of the twist operator at $\partial A$. 

Let us illustrate this with the following example.
Consider a more standard replica trick computation - that of computing thermal Renyi entropies $S_n(A)$.
Actually the well known methods for computing this are not obviously intrinsic to the orbifold theory for the same reasons discussed above.  The computation of $S_n(A)$ would be expected to be governed by a twist correlator on a space $\mathbb{S}^1 \times \mathbb{R}_{d-1}$. However since the twist operator is co-dimension $2$ in the orbifold theory and does not have a branch cut or a co-dimension $1$ object ending on it, this correlator is not sensitive to the difference between $A$ and $\bar{A}$. Since $S_{n}(A) \neq S_{n}(\bar{A})$ for mixed states this would give the wrong answer. The issue arrises because the sums over the states in the thermal density matrix on each replica are independent, and in particular \emph{do not} involve a sum over the action $g$ of the replica symmetry $\sum_{k=0}^{n-1}{\rm Tr}_{CFT^n} g^k e^{ - \beta H_{CFT^n}}$, necessary to project to the symmetric states after gauging $\mathbb{Z}_n$. It turns out we can remove the the projection onto symmetric states by introducing a sum over Wilson lines $W^q(\mathbb{S}^1)$ wrapping the thermal circle. That is:
\be
e^{S_n(A)} \propto \sum_{q=0}^{n-1} \left< W^q (\mathcal{C}) \Sigma_n(\partial A)  \right>_{CFT^n/\mathbb{Z}_n}
\ee
where $\mathcal{C}$ wraps $\mathbb{S}^1$.
The difference between the $A$ and $\bar{A}$ Renyi entropies corresponds to picking  
the curve $\mathcal{C}$ to either intersect $A$ or $\bar{A}$ respectively.  
In our computation, where now $\rho_\psi$ corresponding to the thermal state, we should pick both the curve $\mathcal{C}$ as well as an open curve between the local operators $\mathcal{O}_{B,\bar{A}}$ on which to place these Wilson lines. 

\subsection{Local geometric terms}
\label{sec:locgeo}

We now address the existence of local geometric terms that appear in the expansion of $f(s)$ in the lightcone limit. These terms might actually contaminate the expression we derived so far. Such that
the leading correction to $1$ will not be given by the QNEC quantity. The holographic QNEC proof \cite{Koeller:2015qmn} suggests that these new terms do not appear if we specify that locally the entangling surface is stationary under deformations in the $x^-$ direction, that is we should require that the extrinsic curvature $\mathcal{K}^+_{ab}(y=0) = 0$ and enough $y$ derivatives thereof about the point of interest ($y=0$) should kill these terms. Of course it would be nice to derive this condition in our general proof of the QNEC without resorting to holography. We aim to do this here. 

Recall that we are picking a coordinate system close to $y\approx 0$ such that the entangling surface $\partial A$ is defined by the equation:
\be
\partial A \,: \qquad \left( v = X_A^+(y) \,,\, u = X_A^-(y) \right) \qquad X_A^\pm(y=0) = 0
\ee
The extrinsic curvatures are $\mathcal{K}^{A\pm}_{ab}(y) = \partial_{y^a} \partial_{y^b} X_A^\pm(y)$.
In our final setup for the QNEC we also have the other region $B$ which we can take to be defined close to $y=0$ via:
\be
\partial B \,: \qquad \left( v = X_B^+(y) \,,\, u =  X_B^-(y) \right) \qquad X_B^-(0) = \delta x^- 
\,, X_B^+(0)=0
\ee
Without loss of generality we can set $\partial_y X_A^\pm(0) =0$ which implies that $\partial_y X_B^+(0) =0$ otherwise we violate the nesting condition $\mathcal{D}(B) \subset \mathcal{D}(A)$. For full generality we will leave $\partial_y X_B^-(0)$ non-zero. At second order in the $y$ expansion the nesting condition means that:
\be
\label{matpos}
\left(\mathcal{K}_{ab}^{A}(0)-\mathcal{K}_{ab}^{B}(0) \right) \geq 0
\ee
is a positive semi-definite matrix. This then implies that $\partial_y^2 X_A^+(0) - \partial_y^2 X_B^+(0) \geq 0$ since the Laplacian $\partial_y^2$ is the sum of the eigenvalues of the matrix \eqref{matpos}.  

In our language the origin of the sought after divergent terms is from a more careful study of the dOPE replacement when computing the modular Hamiltonian via the replica trick:
\be
\label{onthis}
 \left< \Sigma_n^\psi (\partial A) \overbracket{ \mathcal{O}_B \mathcal{O}_{\bar{A}} } \right> 
 = \sum_i \beta^i   \left< \Sigma_n^\psi(\partial A) \widehat{O}_i \right> 
\ee
where previously we assumed $\beta^i$ can be calculated by making this same replacement on a flat defect in vacuum $\partial A_0$ in the presence of a defect operator $\widehat{\mathcal{O}}_j$. One might worry that this does not correctly capture the shape dependence of the surface $\partial A$, since the flat defect is a rather brutal replacement. We initially did not worry about this since shape deformations away from the flat defect are achieved via the displacement operator and its derivatives which would appear in this expansion via the defect operator insertions $\widehat{\mathcal{O}}_j$.  However we
now show that this replacement does not capture correctly the local geometric terms defined at the point $y=0$ - that is the extrinsic curvature and a finite number of $y$ derivatives thereof. So while it does capture the non-local shape dependence - i.e. the behavior of $\partial A$ far from $y=0$ we need to work harder to account for the local shape dependence. 

The way to fix this problem is to realize that $\beta^i$ is still sensitive to local geometric quantities at the point $y=0$. Thus we propose to think of the $ \beta^i \equiv \beta^i( \mathcal{K}^\pm(0), \partial \mathcal{K}^\pm(0), \ldots)$ as a function of $\mathcal{K}^\pm$ and its derivatives at that point. Note the extrinsic curvature is the only geometric quantity of interest in Minkowski space, however in curved space we would have dependence on the various local curvature invariants. We leave a complete study of this to future work.

Computing the $\beta^i$ is now more involved.
One must  do the replacement on a curved defect that has the same extrinsic curvature and derivatives thereof locally at $y=0$. Note that the replacement is still made in vacuum which then will make it possible to compute $\beta^i$.

\subsubsection*{The case with $\mathcal{K}^+=0$ and  $\mathcal{K}^- \neq 0$}

Let us address the simplest case where $X_{A}^+ \approx 0$\footnote{We take the notation
$\approx 0 $ in this section to denote that $X^+_A$ and a finite number of derivatives at $y=0$ vanishes. The exact number we would require depends on the spacetime dimensionality since we only must ensure these terms are sub-leading compared to the QNEC term appearing in $f(s$). }
around the point $y=0$ so we are free to replace the entangling cut with one lying along a null cut of the Rindler horizon: $u= \widetilde{X}_A^-(y)$  where we choose this function such that $\widetilde{X}_A^- \approx X_A^-$ around $y=0$. Actually in this case the replacement yields a surface where we know the exact modular Hamiltonian:
\be
K^0\{\widetilde{X}_A^-\} =\int d^{d-2} y \int_{-\infty}^{\infty} d u \left(u-\widetilde{X}_A^-(y) \right) T_{--}(0,u,y) 
\ee
We discuss this in Appendix~\ref{app:modincl}. This was also proven in \cite{Casini:2017roe} using similar methods. At the order of interest the only effect of $X^-_A(y)$ that we care about is on the OPE coefficient for the defect unit operator $\beta^{\widehat{\mathbb{1}}}$. After taking the $n \rightarrow 1$ limit 
on \eqref{onthis}, this gives rise to a shift in the leading order term for the modular Hamiltonian matrix elements:\footnote{Note the stress tensor exchange in this expression still agrees with the brutal flat defect replacement at the order of interest.}
\be
 \left< \psi \right| \mathcal{O}_B K_A\mathcal{O}_{\bar A} \left| \psi \right> = \left< \psi \right| \mathcal{O}_B \left[ K^0\{\widetilde{X}_A^-\}, \mathcal{O}_{\bar{A}} \right] \left| \psi \right>
 + \{ T_{--}, \mathcal{P}_- \}-{\rm terms} + \ldots
\ee
Thus when we compute the function $\mathcal{R}$ defined in \eqref{commkk0}, modular flow and all the related quantities we must replace the leading terms with modular flow using this new zeroth order modular Hamiltonian: $K^0\{ \widetilde{X}_A^- \}$
and also $K^0\{ \widetilde{X}^-_B \}$ appropriate for the other surface $\partial B$. 

This necessitates two changes to our functions $f(s)=g(s)/g_0(s)$ and the subtraction terms $h_{A,B}(s)$ in \eqref{hsubtract}. Firstly, since the new leading order term is different it makes sense to redfine $f(s)$ where we pick the denominator $g_0(s)$ to involve vacuum modular flow for the new null deformed cuts $\widetilde{X}^-_{A,B}$. We should similarly do this for the $h_{A,B}(s)$ denominators. This procedure then removes the potentially offending terms due to the extrinsic curvature $\mathcal{K}^-$. It means that in the light cone expansion the leading contribution to each of these functions is $1$ and the next contribution is $\mathcal{O}(z^{d-2})$. 

For example the new denominator of $f(s)$ is:
\begin{align}
\label{newden}
g_0(s) \equiv \left< \Omega \right| \mathcal{O}_B^s e^{  - i s K^0\{\widetilde{X}^-_B\}} e^{ i s K^0\{\widetilde{X}^-_A\}}   \mathcal{O}_{\bar A}^s \left| \Omega \right> 
&= \left< \Omega \right| \mathcal{O}_B^s \left( U_{ (1- e^{-s})/2} \right)^2  \mathcal{O}_{\bar A}^s \left| \Omega \right>  \\
U_a \equiv  U\left\{ a \left( \widetilde{X}^-_B -\widetilde{X}^-_A\right) \right\} 
\label{usus}
\end{align}
where we have used the more general algebra that these modular operators \eqref{newalg} satisfy. Here $U$ is a generalized null translation operator, which roughly speaking, translates each null generator along the Rindler horizon by a non-uniform amount. It can be written as an exponentiation of the ANEC operator:
\be
U\left\{X^-\right\} \equiv \exp\left( i \int d^{d-2} y X^-(y) \int_{-\infty}^{\infty} du T_{--}(u) \right) 
\equiv U_1 \equiv e^{i P}
\label{defpp}
\ee

After we do this replacement with the null deformed modular Hamiltonians we encounter a new issue. We have now potentially contaminated the Cauchy-Schwarz (CS) bound in \eqref{cscs} in the light cone limit at some lower order in the $z$ expansion. It turns out to fix this we need the leading contributions to the denominators of \eqref{hsubtract} to match the denominators in \eqref{cs}. Note that the numerators were designed to match after computing the CS bound.  The denominators matched in the previous calculation after imposing the relations in \eqref{positions} for the location of the operator insertions. Now that these denominator terms involve non-geometric modular flow (albeit still for the vacuum state) this is seemingly much harder to arrange. 

So the second change to $f(s)$ we must make is as follows.
In fact we can fix the above issue by applying the small $s$-dependent shifts to the operator insertion by using vacuum modular flow itself, rather than just moving the operators by hand. That is consider
\begin{equation}
\mathcal{O}_B^s \equiv U_a \mathcal{O}_B U_{-a}
\qquad \mathcal{O}_{\bar{A}}^s \equiv  U_{-a}  \mathcal{O}_{\bar{A}} U_{a} \qquad a =  \frac{1}{2} (1- e^{-s})
\label{newdef1}
\end{equation}
where $U_a$ was defined just above and we allow for $a$ complex.  These are exactly the shifts predicted in \cite{Ceyhan:2018zfg}. Note that for uniform entangling cuts $\partial A, \partial B$ these flows are geometric and the shifts given above are the same as the shifts in \eqref{shifts} (up to simple translations by $\delta x^-/2$ that we could absorb into $u_B, u_{\bar{A}}$.)  For non-uniform cuts these new operators become non-local, although they are approximately local which is sufficient for our computations in $f(s)$ etc.

One can then check, using the algebra of half-sided modular inclusions discussed in Appendix~\ref{app:modincl}, that with this replacement the denominators in \eqref{hsubtract} and \eqref{cs} are all $s$ independent (as they were previously.) We finally need them to all be equal. This was achieved previously via \eqref{positions} but here we must again use vacuum modular flows:
\begin{equation}
\mathcal{O}_{B} \equiv J_B^0 U_{-1} \mathcal{O}_{\bar{A}} U_1 J_B^0 
\label{newdef2}
\end{equation}
Note that $U_1= U_{ (1- e^{-s})/2}|_{s=i\pi}$ is a generalized translation that provably sends $\mathcal{O}_{\bar{A}}$ to an operator in the algebra $\mathcal{A}'_B$ and then the conjugation operators $J_B^0$ sends this to an operator $\mathcal{O}_B$ in $\mathcal{A}_B$.  
This relation then replaces \eqref{positions} for this more general case, and it is possible to check that our new choice reduces to \eqref{positions} for uniform cuts (again up to some simple $\delta x^-/2$ shifts.) One can also now check that the denominators in $f(s)$ and $h_{A,B}(s)$ agree and are $s$ independent. 

We can run the QNEC proof using the new definitions of these operators (\ref{newdef1}-\ref{newdef2}) and with the appropriate replacement of the vacuum modular flow by the vacuum modular flow for the deformed null cuts.
Note that the issues we deal with above only afflicts the leading terms in the light-cone limit. They give sub-leading correctons to the $z^{d-2}$ terms where the QNEC lives and so these terms will agree with previous considerations.

We finally need to revisit analyticity of $f(s)$ and $F(s)$ in \eqref{Fs}. We have already seen that the denominators, appropriately defined, are $s$ independent so we may drop them. 
We are left to worry about the dependence on $s$ in the numerator, especially through the operators $U_a$. Consider the followinng numerator for $0 \leq {\rm Im} s \leq \pi$:
\begin{equation}
\label{newg}
g(s) = \left( e^{ i s^\star K_A} U_{-a^\star} \mathcal{O}_{\bar{A}} U_{a^\star} \left| \psi \right> , e^{ i s K_B} U_{a}  \mathcal{O}_{B}  U_{-a} \left| \psi \right> \right) \,, \qquad a = (1-e^{-s})/2
\end{equation}
The basic question is: can we give an analytic continuation of
\begin{equation}
e^{i s K_B} \mathcal{O}_{B}^s \left| \psi \right>
= e^{i s K_B} U_{a} \mathcal{O}_{B} U_{-a} \left| \psi \right>
\end{equation}
into the strip $0\leq {\rm Im} s \leq \pi$ and similarly for the bra in \eqref{newg}. If we can, then the resulting function has the desired analyticity. A similar discussion is necessary for $-\pi \leq {\rm Im} s \leq 0$ with a different function, however we will not spell this out.
We addressed the issues of analyticity previously in Section~\ref{ssec:a}, where we had to worry about the $s$ dependence in the operator locations spoiling analyticity. We can give a similar discussion here. At large $s$ we can expand:
\begin{equation}
\label{expexp}
e^{i s K_B} \mathcal{O}_{B}^s \left| \psi \right> \approx \sum_m \frac{(-2i)^{-m}}{m!} e^{-s m} e^{i s K_B} \left[ P, \left[ P, \ldots \left[ P, U_{1}\mathcal{O}_{B}  U_{-1} \right] \right] \right] \left| \psi \right>
\end{equation}
where $P$ was defined in \eqref{defpp}
and for large $s$ this expansion should converge. For sufficiently local operators these commutators are well defined and result in an operator that is still inside the algebra $\mathcal{A}_A'$. Thus we can analytically continue these terms into the strip using Tomita-Takesaki theory applied to the state dependent flow $e^{i s K_B}$. Likely there is a more rigorous proof of this using ideas similar to \cite{Ceyhan:2018zfg}. 

We also need to consider the numerator of $h_A(s)$ and $h_B(s)$. There is a similar discussion for these. For example the numerator of $h_B(s)$ can now be written as:
\begin{equation}
\left< \psi \right| U_{-a+1} \mathcal{O}_B U_{a-1} e^{-\pi K_B} U_a \mathcal{O}_B U_{-a} \left| \psi \right> 
\end{equation}
and we can apply a similar expansion as in \eqref{expexp} to convince ourselves of analyticity for large $s$. 


In conclusion all elements of the QNEC proof (computability, CS bound and analyticity) work for two entangling cuts that become null cuts of the same Rindler horizon close to $y =0$.  This seems to be a general lesson - when we can replace the defect locally with one where we know the modular Hamiltonian as a local integral over the stress tensor, then the QNEC result applies. 
 
Note that the appropriate quantities in the QNEC $\mathcal{P}_-$ should now be defined using the subtracted EE for the null cut $\partial \widetilde{A}$ determined by $\widetilde{X}^-$:
\be
\mathcal{P}_-(A) \equiv \frac{\delta}{\delta x^-(0)} \left( S_{EE}(A, \left| \psi \right>) -  S^\Omega_{EE}( \widetilde{A}, \left| \Omega \right>) \right)
\ee

\subsubsection*{The case with $\mathcal{K}^+ \neq 0$}

We now turn to $X^+ \neq 0$ where we will discover terms that render the QNEC inapplicable. We consider only the case where also $X^- \approx 0$, leaving potential cross terms between the two extrinsic curvatures to future work. There is no limitation to studying these cross terms, they require just a little more work.  We now consider a new replacement surface:
\be
\widetilde{\partial A} \,: \qquad \left( v = \widetilde{X}_A^+(y) \,,\, u = 0 \right) 
\ee
which agrees with the exact cut $ \widetilde{X}_A^+(y)\approx X_A^+(y)$ for a finite number of $y$ derivatives around $y=0$. We additionally require that  $\widetilde{X}_A^+(y)$ smoothly match
onto a flat defect with $\widetilde{X}^+_A(y) =0$ for larger $|y| > y_\star$. We still place a local defect operator $\widehat{O}_j$ elsewhere on the defect along $\widetilde{\partial A}$ but in the region where $\widetilde{X}^+_A(y) =0$. Since we make
this replacement in vacuum we can now use the results in \cite{Faulkner:2016mzt,Faulkner:2015csl} to compute shape deformations of the flat defect. This should allow us to express $\beta^i$ as an expansion in extrinsic curvatures and their derivatives. We should also find that the non-local dependence on
$\widetilde{X}^+_A$, which is arbitrarily chosen, should drop out. 

We illustrate this with an example, where we account for the linear in $\mathcal{K}^+$ terms and derivatives thereof appearing in $\beta^{\widehat{\mathbb{1}}}$ the defect identity operator coefficient. 
As $n \rightarrow 1$ the shape deformed defect will now change the three point function term which we computed around \eqref{unitstress}:
\be
\left< \Sigma_n^0(\widetilde{\partial A}) \widehat{\mathbb{1}}  \overbracket{ \mathcal{O}_{B}\mathcal{O}_{\bar{A}} } \right> 
=\left< \Sigma_n^0(\partial A_0) \widehat{\mathbb{1}}  \overbracket{\mathcal{O}_{B} \mathcal{O}_{\bar{A}} } \right>  - \int d^{d-2} y' \widetilde{X}_A^+(y')  \left< \Sigma_n^0 \widehat{D}_+(y) \overbracket{ \mathcal{O}_{B} \mathcal{O}_{\bar{A}}}  \right> + \ldots
\ee
where we have expanded the shape deformation to linear order using the displacement operator. Recall that this is the defect operator $\widehat{D}_+ \equiv - 2\pi \widehat{T}_{-1}$. 
We know that the displacement operator inserted in a correlator like this simply gives $2\pi(n-1)\times$ the half ANEC integral as we send $n \rightarrow 1$ \cite{twistdisplacement}. The  contribution to the full modular Hamiltonian can be found after applying $ (-\partial_n)/(2\pi)$ and extending the half ANEC integral to the full ANEC integral:
\be
\lim_{n \rightarrow 1} \left< \Sigma_n \widehat{\mathbb{1}} \right> G_{\mathbb{1} \mathbb{1} }^{-1} \delta C_{\mathbb{1}}   = \delta C_{\mathbb{1}}  =  - \int d^{d-2} y' \widetilde{X}^+(y') \int_{-\infty}^\infty d \lambda \left< \mathcal{O}_B T_{++}(0,\lambda,y') \mathcal{O}_{\bar A} \right> 
\ee
and where we have used $\lim_{n \rightarrow 1} \left< \Sigma_n \widehat{\mathbb{1}} \right>=1$
and $G_{\mathbb{1} \mathbb{1} }=1$. 
In order to compute this ANEC integral we will use the follwing three point function:
\be
\left< T_{++}(0,\lambda,y) \mathcal{O}(u_B,v_B,0) \mathcal{O}(u_{\bar A},v_{\bar{A}},0) \right>
= \frac{ c_\Delta^T (y^2 \Delta u + u_B u_{\bar{A}} \Delta v )^2 ( - \Delta u \Delta v))^{-\Delta + h-1} }{4 (\lambda u_B - v_B u_B+ y^2)^{h+1}  (\lambda u_{\bar A} - v_{\bar{A}} u_{\bar A} + y^2)^{h+1} }
\ee
where we have kept higher order terms $v_B u_B$ and $v_{\bar{A}} u_{\bar A}$ that we dropped previously in \eqref{cft3}. These will be important to keep here to regulate some divergences that arise as we send these to zero. After some computation we can write the answer as:
\be
\delta C_{\mathbb{1}} =-  i \frac{\Delta_{\mathcal{O}}}{\Delta v} \int d^{d-2} y' \widetilde{X}_A^+(y') \frac{ c_{d-1} z^d }{((y')^2 + z^2)^{d-1}}  \,,
\qquad z^2 = \frac{u_B u_{\bar{A}} \Delta v}{\Delta u}
\ee
and where $c_{d-1} = (4/\pi)^{h-1/2} \Gamma(h+1/2)/(d-1)$. 
 We see that this will contribute local geometric terms if we expand
as $z^2 \rightarrow 0$:
\be
 \frac{c_{d-1}z^{d}}{ (z^2 + y^2)^{d-1} } =  \left(  \delta^{d-2} (y) + \frac{1}{2(d-2)} z^{2} \partial^2  \delta^{d-2} (y)  + \ldots \right)
\ee
Note that  first term will not contribute because we have chosen $X_A^+(0)=0$.
When we integrate this against the profile we find the expected extrinsic curvature dependence:
\be
\delta C_{\mathbb{1}} =- i \frac{\Delta_{\mathcal{O}}}{\Delta v}\left( \frac{1}{2(d-2)} z^2 \mathcal{K}_A^+ + a_1 z^4 \partial^2 \mathcal{K}_A^+ + \ldots  + z^d \int d^{d-2} y' \frac{ c_{d-1} \widetilde{X}_A^+(y')}{(y')^{2(d-1)}} + \ldots \right)
\label{bad}
\ee
where $\mathcal{K}_A^+$ is the trace of the extrinsic curvature.
However this story is not yet complete - the answer we have so far depends on the full function $\widetilde{X}_A^+(y)$ which was randomly chosen except for it's local behavior around $y=0$. This is because of the last term in \eqref{bad} which has a non-local dependence on $\widetilde{X}^+_A(y)$.  
This term is taken care of by the EE subtraction in \eqref{subtraction} which is now a subtraction using the
deformed replacement cut determined by the profile $\widetilde{X}^+_A(y)$. We call this term $S_{EE}(\widetilde{A},\left| \Omega \right>) =S_{EE}^\Omega\{ \widetilde{X}^+_A\}$ and note that any reasonable regulator for EE will yield a term which cancels the the non-local term in \eqref{bad}. 

To make this clear, let us collect all the terms that appear in $\mathcal{R}$ \eqref{commkk0}. We can write this suggestively as:
\be
\label{newR}
\mathcal{R} =  - \frac{i \Delta_{\mathcal{O}}}{\Delta v} \left( X^+_{RT}(z,0) +\{T_{--} \}{\rm \,-\, terms } \right)
+ \ldots
\ee
where we are suppressing the stress tensor dependence since it is the same as before. Here we have defined the ``Ryu-Takayanagi'' profile:
\be
\label{profile}
X_{RT}^+(z,y) = c_{d-1} \int d^{d-2} y' \frac{z^{d}  \widetilde{X}_A^+(y')}{\left( (y-y')^2+z^2\right)^{d-1}} + \frac{8 G_N  z^{d} }{d}\frac{\delta 
  }{\delta x^-(y)} \left(  S_{EE}^{\psi}(A) - S_{EE}^{\Omega}\{ \widetilde{X}_A^+\} \right)
\ee
This profile has several interesting features. Firstly it only depends on $\widetilde{X}^+_A$ locally at the point $y=0$. As expected one can show the non-local part  cancels appropriately. This is because if we expand $\frac{\delta}{\delta x^-(y)} S_{EE}^{\Omega}\{ \widetilde{X}_A^+\}$ in  $\widetilde{X}_A^+$ we will always find a universal cutoff independent piece which is the second order shape deformation of the vacuum entanglement entropy (sometimes called the ``entanglement density'' \cite{Bhattacharya:2014vja}.) This was studied in depth in \cite{Faulkner:2015csl} where one can check that the CFT entanglement density exactly cancels
the term in \eqref{bad} that was troubling.

Secondly the cutoff dependence $\epsilon$ used to define $S_{EE}$ should be absent as we remove it $\epsilon \rightarrow 0$. The natural regulator for entanglement in our computation is a vacuum subtraction for which we don't expect state dependent divergences to arise as explained in Footnote~\ref{statedep}. This is to be expected since we are ultimately computing a UV finite quantity ($\mathcal{R}$ and $f(s)$). 

Lastly this is exactly the profile of the RT surface that one finds by linearizing the surface equations of motion near the boundary of AdS in a \emph{holographic CFT}. We linearize near the boundary but allow for a totally general bulk. The linearization is necessary because we only include linear terms in $\widetilde{X}^+_A$ in our analysis. The linear in $\widetilde{X}^+_A$ terms are important for the leading terms in the $z$ expansion however at higher orders in the $z$ expansion we expect to see non-linear dependence on $\widetilde{X}^+_A$  . This dependence is in principle computable using this approach and we expect agreement with the non-linear RT functional, perhaps supplemented with the appropriate higher derivative corrections. We leave checking this for future studies. 

We now go back and compute $f(s)$ with this new $\mathcal{R}$ \eqref{newR}. Tracking through the computation in Section~\ref{sec:modflow} the new terms can be grouped in with the displacement operator terms as the transform in the same way under modular flow. The result has the same form as in \eqref{profile} but where we use a slightly different definition of the $z$ coordinate due to the intervening modular flow:
\be
z^2 = - \frac{ \Delta v ( ( \Delta u - \delta x^-)^2 - (\delta x^- e^{-s})^2 )}{ \Delta u - \delta x^-}
\ee
For large $s$ this becomes:
\be
z^2 \rightarrow -\frac{\Delta v(\Delta u - \delta x^-)}{4}
\ee
Using this $z$ coordinate we have for large $s$:
\be
1-f(s)  \rightarrow  e^s \frac{\Delta_{\mathcal{O}} }{(- \Delta v) } \left( X_{RT,A}^+(z,0) -X_{RT,B}^+(z,0)
+ z^d \frac{16 \pi G_N}{d} \int_0^{\delta x^-} du \left< T_{--} (u) \right>_\psi \right)
\ee
where recall that the lightcone expansion is controlled by $z \ll 1$. 
We can now understand why the QNEC might not be satisfied for surfaces with non-vanishing extrinsic curvatures $\mathcal{K}^+$. It is because the leading terms that we might constrain using the sum rule actually trivially satisfy the positivity constraint. These terms are the extrinsic curvature terms:
\be
\label{1fs}
1-f(s) =  e^s \frac{\Delta_{\mathcal{O}} }{ (-\Delta v)} \frac{z^2}{2(d-2) } \left(\mathcal{K}_A^+(0) -
 \mathcal{K}_{B}^+(0) \right) + \ldots + \left( Q_- {\rm \,-\, term} \right) + \ldots
\ee
for which the sum rule is non-negative by the nesting condition \eqref{matpos}.
Here $\mathcal{K}_{A,B}^+(0)$ is the trace of the extrinsic curvature. At this point if we demand
$ \mathcal{K}_A^+(0) -  \mathcal{K}_{B}^+(0) =0$ then we might still succeed in proving the QNEC (with the appropriate subtracted $S_{EE}$'s) in $d\leq 3$, since the next leading term is the $Q_-$ term.

However we should be cautious here because the leading extrinsic curvature terms in \eqref{1fs} actually competes with $1$ (that is $z^2/\Delta v= (\delta x^- - \Delta u)/4 \sim \mathcal{O}(1)$)) so in some sense it spoils the perturbative expansion altogether. Thus we should only trust this analysis for small $\mathcal{K}^+_{A,B}$. Note that even though there is a cancelation for this term if we demand $ \mathcal{K}_A^+(0) = \mathcal{K}_{B}^+(0)$, this cancelation may not be enough to save the break down in the perturbative expansion. We hope to clear up this question in the future.

Actually the correct thing to do for the cases where $\mathcal{K}_{A,B}^+(0)$ are non-zero is to make the dOPE replacement using a spherical defect where the trace of the extrinsic curvature are designed to agree locally with those of $\partial A, \partial B$. For a CFT the spherical defects have known modular Hamilonians \cite{Casini:2011kv,Hislop:1981uh} and there is an obvious path to follow to proving a so called conformal QNEC (see \cite{Koeller:2015qmn} for the original discussion of this in the holographic proof of the QNEC). We leave the conformal QNEC case to future work. 

Either way we have succeeded in proving the necessity of certain local conditions the entangling surface must satisfy in order to claim a QNEC. Most conservatively we should demand that $\mathcal{K}^+_{A,B}$  and enough derivatives vanish at $y=0$. Such local conditions have been extensively studied very recently for the curved space QNEC and QFC \cite{Akers:2017ttv,Fu:2017evt,Leichenauer:2017bmc,Fu:2017lps} and it is obviously interesting to extend our work to that case. 

\subsection{From CFT to QFT}

Most of our arguments have in one way or another relied on working with a CFT. We would like to extend this proof to a relativistic QFT found via a relevant deformation $+ \lambda \Phi$ of the CFT.  Since we are working in the lightcone limit these relevant deformations do not play a very important role - this makes sense from the holographic point of view, since the important physics occurs near the boundary of the dual spacetime and in the UV of the QFT, which is then controlled by the CFT fixed point. In our computation we expect the light cone OPE limit is also essentially controlled by the CFT.
Thus for example the stress tensor and displacement contribution to $\mathcal{R}$ and thus ultimately $f(s)$ will be the same as before. However we have the same issue as in the previous subsection where there might be more leading terms in the $z$ expansion of $f(s)$ due to $\lambda$ dependent effects. The analogous effect in AdS/CFT arises via a Fefferman-Graham expansion of the metric and the RT surface \cite{Koeller:2015qmn}, which are also sensitive to $\lambda$. 

Again we expect this to arise in our computations because the coupling $\lambda$ may appear in the dOPE coefficient for the unit operator $\beta_{\widehat{\mathbb{1}}}(\lambda,\lambda^2, \ldots)$. Only polynomial powers should appear and the analogy with the extrinsic curvature terms is strengthened by taking spacetime dependent couplings $\lambda(x)$ such that $\beta_{\widehat{\mathbb{1}}}$ can depend locally on $\lambda(0)$ and its spacetime derivatives.
We can deal with these new terms using the same idea as above for dealing with $\mathcal{K}^-$ terms. We can simply use the vacuum modular flow for the deformed theory. This still works because these still have known modular Hamiltonians that are constrained by the theory of half-sided modular inclusions.  We can then run the same argument for the QNEC in this case (see the previous subsection on the $\mathcal{K}^-$ extrinsic curvature contributions for all the details.) 

For surfaces where $\mathcal{K}^+$ is not $\approx 0$ then the above argument does not work and we would need to combine the relevant deformation with the $X^+$ shape deformation in perturbation theory. This should be doable, and we basically expect to reproduce any terms one might expect to see in a Fefferman-Graham expansion in this way. We again leave the complete discussion to the future, where likely it would be nice to have a more systematic way to study all of these effects at one time (relevant operator deformations, extrinsic curvature deformations and even metric deformations.) 


\subsection{Higher spin versions of the QNEC}
\label{subsec:higherspin}

It is easy to extend the derivation of QNEC to the case of the higher-spin symmetric traceless operator $\mathcal{J}_{-\ldots -}$ of conformal dimension $\Delta_J$ and even spin $J$, where the twist $\tau_J=\Delta_J-J$ is the minimum among operators of the same spin. Previously we found that in this case there is a family of displacement operators  $\widehat{\mathcal{D}}^{\ell},\;1\leq\ell\leq J-1$ emerging at order $\mathcal{O}\left(n-1\right)$.  Again we can compute $\mathcal{R}$ of \eqref{defR} 
which we can then use to compute $f(s)$.  We leave the details of these functions to Appendix~\ref{app:R}. The result for $f$ is analogous to \eqref{fcrank} in a somewhat obvious extension, see \eqref{fJ}.

In the limit $e^{s}\gg 1$, the null integral is dominated by the interval $0<u<\delta x^-$, and we have:
\begin{eqnarray}
\label{larges}
f(s) &\approx & 1+ \sum_{J} e^{s(J-1)}\left[\frac{ \Delta v\left(u_B-\delta x^-\right)u_{\bar{A}}}{\Delta u- \delta x^-}\right]^{\frac{\Delta_J+J}{2}-1}\left(-\Delta v/4\right)^{1-J} G^{J} \mathcal{Q}_{J} \\
\mathcal{Q}_{J}&=&\int^{\delta x^-}_0 du\; \langle \mathcal{J}_{-\ldots-}(u)\rangle_\psi + \frac{\langle \widehat{\mathcal{D}}^{J-1} H^\psi_A\rangle_\psi}{J-1}-\frac{\langle \widehat{\mathcal{D}}^{J-1} H^\psi_{B}\rangle_\psi}{J-1}\nonumber
\end{eqnarray} 
where the new coupling is:
\be
G^{J}= \frac{c_{\mathcal{J} \mathcal{O} \mathcal{O}} }{c_{\mathcal{J}\mathcal{J}}} \frac{2^{\Delta_J-J+1} \Gamma\left(\frac{\Delta_J+J+1}{2}\right)}{\sqrt{\pi}\Gamma\left(\frac{\Delta_J+J}{2}\right)}
\ee
Note that $G^T = - 4\pi G_N \Delta_{\mathcal{O}}/d$ for the stress tensor. The sign of $G^{J}$ for the other operators is ambiguous if they don't correspond to some conserved currents since we can send $\mathcal{J} \rightarrow - \mathcal{J}$. However, as we will see, $G^{J} \mathcal{Q}_{J}$ does have a definite sign. 

The new ingredient above are the one point functions of the higher spin displacement operators. Again these are only non-zero at order $(n-1)$ where we bring down a factor of the (half) modular Hamiltonian, and for this reason they correspond to some object in the QFT which is non-linear in the state. More specifically the various displacement operators appear as singular terms when we take $\mathcal{J}_{- \ldots -} $ close to the modular Hamiltonian:
\be
\left< H_A^\psi \mathcal{J}_{- \ldots -} (w) \right> \sim \sum_{\ell=1}^{J-1}\frac{ \left< \widehat{D}^\ell  H_A^\psi \right>}{w^{J-\ell}} + {\rm regular}
\label{higherdisp}
\ee
After an application of the first law of entanglement one can interpret these as the variational response of the EE to a deformation with respect to the higher spin field $+ \mu^{\nu_1 \ldots \nu_J} \mathcal{J}_{\nu_1 \ldots \nu_J}$ and picking a particular profile for the $\mu$ close to the entangling surface. However since $\mathcal{J}$ is not a conserved current it is hard to make a precise statement here. Note that in the large $s$ limit \eqref{larges} we only find a contribution from the highest spin displacement operator $\ell = J-1$.

To extract a sum rule we place the operators symmetrically ($v_B = -v_{\bar{A}}$ and $u_B =-u_{\bar{A}} + \delta x^-$) and we use the definitions for $\sigma,z$ in \eqref{oursig} and \eqref{ourz}:
\begin{eqnarray}
F(\sigma) &\approx & 1+\sum_{\mathcal{J}} z^{\tau_J}\left(-i\sigma\right)^{1-J} G^{J}\mathcal{Q}_{J}\nonumber + \ldots
\end{eqnarray}
We can now obtain a sum rule by extracting the higher-order pole $\sigma^{1-J}$ using a new version of the  projection \eqref{eq:sum_rule}, the analytic properties of $F(s)$ then forces out the following constraint: 
\begin{eqnarray}
(-1)^{\frac{J}{2}} G^J \mathcal{Q}_J=\lim_{R \rightarrow 0} \lim_{z \rightarrow 0} \frac{ z^{-\tau_J}}{\pi} \int^R_{-R}d\sigma\;\sigma^{J-2}\left[1-\text{Re} F \right] \geq 0
\end{eqnarray}
This is now a higher spin version of the QNEC. If we integrate this up, by taking $\delta x^-$ to infinity, we recover the higher spin version of the ANEC first studied in \cite{Hartman:2016lgu}. The QNEC is a more local version and indeed gives a local bound on the expectation value of a higher spin field in any state:
\be
(-1)^{J/2} c_{\mathcal{J} \mathcal{O} \mathcal{O} } \left( \left< \mathcal{J}_{- \ldots -}(0) \right>_\psi \frac{d x^-_\lambda(0)}{d\lambda}  - \frac{1}{J-1} \frac{d  }{d \lambda} \left< \widehat{D}^{J-1}  H^\psi_{A_\lambda} \right> \right) \geq 0 
\ee
where $x^-_\lambda(y)$ parameterizes a small null deformation of the entangling surface $A_\lambda$ and with $d x^-(0)/d\lambda > 0$.  It would be interesting to give a gravitational/stringy  interpretation/analog of this bound.  It would also be interesting to study this in free theories extending the proof of the QNEC for free QFTs \cite{Bousso:2015wca}.\footnote{There is now evidence that the higher spin QNEC is violated in free theories. We thank A. Wall for informing us about this result.} 

\section{Discussion}

\label{sec:discussion}

In this paper we have found a way to reconstruct the Ryu-Takayanagi entangling surface in a putative dual gravitational theory in any interacting QFT. The reconstruction happens near the boundary of the space where from the outset one might have expected to make progress using an OPE argument. We found that the correct argument involves working with entanglement in the Replica trick and studying the spectrum of defect operators localized on the $d-2$ entangling surface twist defect. An essential ingredient included the introduction of probe operators (any operators are ok) which can be made to probe the boundary spacetime in a precise way. With this setup we studied the modular Hamiltonian evaluated between matrix elements of the probe operator, which we then bootstrapped into a study of modular flow. The profile of the RT surface appeared due to a shift in the action of the modular Hamiltonian on the probe operators. Analyticity of modular flow was related to causality which we then used to constrain the sign of this shift thus proving the quantum null energy condition, which was the original goal of this study. 

The exact surface that we reconstructed should likely be compared to the quantum corrected extremal surface advocated in \cite{engelhardt2015quantum} which was recently proven, in the context of theories with a gravity dual, to compute the entanglement entropy of the dual QFT in \cite{Dong:2017xht}. One piece of evidence for this comes from studying mixed states, where the entropy of complementary regions is not the same. This means there will be two entangling surfaces, one for $A$ and one for $\bar{A}$, which is indeed what we found. 

We should remark that our proof of the QNEC should be considered as a proof strategy that can be tailored to various situations depending on the details of the entangling surface, the space that the QFT lives on, and any potential relevant operator deformations involved. In this paper we worked exclusively in flat space allowing for uniform relevant deformations, although it is possible to generalize to curved space etc. We expect the main difference in this case comes from studying the local-geometric contributions to the defect OPE coefficients $\beta_i$. We sketched how this works when the entangling surface has extrinsic curvature in Section~\ref{sec:loose}.

Our strongest statements could be made for entangling surfaces that approach non-uniform null cuts of a Rindler horizon, however we managed to also include some leading order effects due to non-stationarity when the extrinsic curvature $\mathcal{K}^+$ did not vanish, although these terms disrupted the QNEC proof in a controllable manner. Understanding exactly when local geometric terms might disrupt a statement of the QNEC is an important avenue for future study, especially in curved space. Recently there have been several approaches to studying this problem \cite{Fu:2017evt,Akers:2017ttv}. One is to study holographic theories and examine the causality of EWN when the boundary theory lives on a curved metric. The second is to assume the Quantum Focusing Condition and check what constraints must be imposed in order to derive a QNEC in the semi-classical limit. Lastly the condition that the QNEC itself must be a UV finite quantity in order for the bound to make sense, puts similar constraints on the background about which one might prove the QNEC. Since we now have a general proof strategy we should be able to find the general set of conditions for any interacting QFT. Our expectation however is that our results will be in line with those already known due to the holographic proof and so one might not expect to learn anything new here. It is still worth pursuing of course. 

We now mention some other more speculative avenues for future pursuit. 

\subsection{Beyond the lightcone limit}

It would be interesting to push this computation beyond the lightcone limit.
In order to have some control we would need to work with a QFT with a large-$N$ limit and a gap in anomalous dimensions to the single trace higher spin fields \cite{Heemskerk:2009pn}. This would involve moving to very large modular flow, still maintaining $ s \ll \ln N$ where we move into a controlled Regge like regime where we expect to reproduce the bulk physics of a theory with a gravitational dual \cite{Maldacena:2015waa,Shenker:2014cwa,Camanho:2014apa,Afkhami-Jeddi:2016ntf}. 
It is not clear what physics we should look for -  presumably it should be related to the causal structure of the entanglement wedges but now deep in the bulk. For example this might be a way to give a proof of the entanglement wedge nesting from a purely boundary point of view and potentially beyond the semi-classical limit. There are several challenges here. For example one needs to both control the Regge limit at the same time as potentially higher order corrections to modular perturbation theory. We think that modular perturbation theory can likely be controlled by working, as we did in Appendix~\ref{sec:appmodflow}, with (double) modular flow directly in the replica trick. It would also be important to figure out the role of double trace operators and their ``displacement operator'' contributions, for which we have very little understanding right now. There have been many recent advances in studying this limit for the ANEC version of this problem \cite{Li:2017lmh,Kulaxizi:2017ixa,Alday:2017gde,Caron-Huot:2017vep} and likely we should make use of this new technology.

\subsection{Meaning of higher spin displacement operators}

It is natural to wonder about the physics of the higher spin displacement operators.
Recall that these only arise out of (symmetric traceless) operators with spin $\geq 2$. 
Their origin suggests they should be interpreted as new fields living on the RT surface, possibly related to higher spin fields or stringy states - although the double trace versions muddy this possible interpretation. 
Ignoring the double trace operators for now, one might speculate that these correspond to new modes on the RT surface which typically have a large mass in theories with gravitational duals for the usual reasons that we expect a gap in anomalous dimensions to the stringy states. Yet they could be important for understanding EE more generally, for example in CFTs dual to Vasiliev gravity \cite{Klebanov:2002ja}. They also might have some relation to higher spin entanglement studied in 2d CFT using the various 3d versions of higher spin theories \cite{Ammon:2013hba,deBoer:2013vca}. 
 
For the double trace versions of the displacement operators it is natural to speculate that these are related to the bulk entanglement contribution to the boundary entropy \cite{Faulkner:2013ana}, although the details of this are not clear to us right now.

\subsection{A new regulator for entanglement entropy}

We have several expressions now that relate the modular evolution of probe operators to
the EE of the underlying state $\psi$. We might then invert these relations to give an independent \emph{definition} of EE.  EE typically suffers from UV divergences and is hence not a good continuum object - this is related to the fact that the algebra of a region in QFT is a type III von Neumann algebra which does not have a trace.  Thus it is important to find a natural UV regulator for this problem.
We actually have in hand a continuum definition where the usual/expected divergences would be controlled using the kinematics of the lightcone limit. That is, while the modular flow correlators are UV finite, the limit we consider has diverging terms parameterized by $z^2 = - \Delta v(\Delta u - \delta x^-)/4$ as $\Delta v \rightarrow 0$. Thus the divergences in EE due to local correlations would be the same or similar to those found using the RT functional in holographic theories. There are a few caveats here - firstly we would only ever by able to extract the null shape deformation of EE and often, as we argued above, this is UV finite anyway. Although in the presence of extrinsic curvature $\mathcal{K}^+$ this is no longer true and thus in this situation this new regulator would be useful. One possible proposal is:
\be
\left. \frac{\delta S_{EE}(A;\psi)}{\delta x^-(0)} \right|_{{\rm reg}(z)}
\equiv - 2\pi \int_0^\infty \left<T_{--}(y=0)\right>_\psi dx^- + \frac{d z^{2-d}}{2\pi G_N \Delta_{\mathcal{O}}}
\int_{-R}^R d\sigma \Re(1-f(\sigma) )
\ee
where we have sent the region $B$ far away by taking $\delta x^- \rightarrow \infty$ holding fixed $u_B -\delta x^-$ fixed and $z,\sigma$ were defined in \eqref{ourz} and \eqref{oursig} and $R$ satisfies $z \ll R \ll 1$. 
Note we could have used single modular flow here instead. We use double modular flow so we can use the more developed formulas for that case. Note that the important thing here is to carefully pick the entangling region $A_0$ which is used in denominator for $f$, which should have a computable modular flow and should come close to the entangling surface $\partial A$. We have dropped the various local geometric terms that we discovered in Section~\ref{sec:locgeo} would then be dropped since they are anyway local to the entangling surface and so could be removed in another regulator using appropriate counter terms. 

This is to be compared and contrasted with the mutual information regulator of \cite{Casini:2015woa}. There are several questions that arise now relating to the properties of this putative EE. 
We know that this quantity is constrained by the QNEC - but does this imply that it satisfies strong subadditivity, and other constraints obeyed by the usual EE? Also can we give a useful definition along these lines\footnote{Clearly lots of modifications will be necessary.} for a non-relativistic QM system, that in this case reduces to the usual definition of EE?
  
\acknowledgments

We would like to thank Alex Belin, Xi Dong, Netta Engelhardt, Zach Fisher, Monica Guica, Tom Hartman, Diego Hoffman, Nima Lashkari,   Rob Leigh, Aitor Lewkowycz,  Daliang Li, Arvin Moghaddam, Marco Meineri, Mark Mezei,  Donald Marolf, Onkar Parrikar, Eric Perlmutter, Steve Shenker,  Aron Wall, Matthew Walters, Sasha Zhiboedov and other (remote) participants of the Aspen Center for Physics workshop on entanglement in quantum field theory.   This work was performed in part at the Aspen Center for Physics, which is supported by National Science Foundation grant PHY-1607611. We are also supported by the DARPA YFA program, contract D15AP00108 and work for v2 was supported by the Department of Energy contract DE-SC0015655.

\appendix

\section{Half-sided modular inclusions}
\label{app:modincl}

In his appendix we would like to derive the form of the modular Hamiltonian for vacuum states and geometric reigion determined by an arbitrary null cut of the Rindler horizon. This form was proven for free theories in \cite{Wall:2011hj}, conjectured in general in \cite{Faulkner:2016mzt}, proven assuming the QNEC is true in \cite{Koeller:2017njr}, and derived in several ways in \cite{Casini:2017roe}. One of the derivations in \cite{Casini:2017roe} has a some overlap with the discussion in this appendix, in particular they also rely heavily on the theorems associated to half-sided modular inclusions. 

Let us start by taking the undeformed null cut $A_0$ to be that of the Rindler cut in vacuum. That is a half
space where the region $u >0 , v<0$ is the associated wedge $\mathcal{D}(A_0)$. The ``deformed'' region $B_0$  we take to be associated to a null cut of the future Rindler horizon of $\partial \mathcal{D}(A_0)$. That is $\partial B_0$ is determined by the equation $u = X_B^-(y)>0$
and $v=0$. In this case, since the action of modular flow for the region $A_0$ is geometric, we have the important inclusion relation: 
\be
\label{inclapp}
 e^{i s K_A^0} \mathcal{D}(B_0) e^{ - i s K_A^0} \subset \mathcal{D}(B_0)
\,, \qquad  s>0
\ee
where the the space-time region $\mathcal{D}(B_0)$ also determines the algebra of operators which thus also has this inclusion property. Under this condition it has been shown that the modular Hamiltonian's satisfy an algebra, which is the same as the obvious algebra that would have applied if $B_0$ was also a uniform Rindler cut. That is \eqref{algincl2} which we reproduce here:
\be
\label{algapp}
\left[ K^0\{ X_B^- \}, K_A^0 \right] = i \left( K_A^0 - K^0\{ X_B^- \} \right)
\ee
Some of the ingredients that go into a proof of this were sketched in Section~\ref{modinclmain}. We will just take this as an input. The other input will be the perturbative results of \cite{Faulkner:2016mzt} which showed that to first order in the shape deformation $X_B^-$ one can show that:
\be
\label{totrunc}
K^0\{ X_B^- \} = K_A^0 - \int d^{d-2} y \int d x^- X^-_B(y) T_{--}(x^-,y) + \ldots
\ee
This result was originally proven for arbitrary non-timelike shape deformations (which then includes an additional $- \int d x^+ X^+ T_{++} $ term), and in this case we expect the higher order corrections to be non-trivial. However it was reasonable to guess that for null deformations the perturbative series truncated. Here we show this by applying the algebra \eqref{algapp}. We note that the geometric action in \eqref{inclapp} implies that we can write this
algebra as a differential equation:
\be
\lambda \frac{d}{d \lambda}\left( K^0\{ \lambda X^-_B \} - K_A^0\right) =  \left( K^0\{ \lambda X^-_B \}-K^0_A\right)
\ee
This is a trivial operator/matrix differential equation (i.e. take matrix elements of both sides) with solution:
\be
K^0\{ \lambda X_B^-\}  - K_A^0 = \lambda \widehat{M}(X_B^-)
\ee 
Taking $\lambda$ small we can fix $\widehat{M}$ via \eqref{totrunc} and the truncation of \eqref{totrunc} follows.

Finally we need to show that this algebra (suitably generalized in \eqref{algboth}) also applies when both region $A_0$ and $B_0$ correspond to non-uniform null cuts of the same Rindler horizon. We simply compute:
\begin{align}
\label{firstline}
\left[ K^0\{ X_A^- \}, K^0\{ X_B^- \} \right]
&= \left[ K^0,K^0\{ X_B \} \right] + \left[ K^0\{ X_A\},K^0 \right] \\
& + \int d^{d-2} y_1 \int d^{d-2} y_2 \left[ \mathcal{E}_-(y_1), \mathcal{E}_-(y_2) \right] X_A^-(y_1) X^-_B(y_2) 
\nonumber
\end{align}
where here $K^0$ is the undeformed Rindler modular Hamiltonian we previously called $K^0_A$.
The first line \eqref{firstline}, after using \eqref{algapp}, gives us the sought after algebra that we quoted in \eqref{algboth}.  So we just have to show that the second line of \eqref{firstline} vanishes.
Here the ANEC operator is:
\be
\mathcal{E}_-(y) = \int_{-\infty}^{\infty} d x^- T_{--}(x^-,y)
\ee
Two such operators commute when $y_1 \neq y_2$ since the null generators are always space-like separated. When they lie on top of each other they commute because they are the same operator. This argument is not really justified since there could be singularities that invalidate these statements. See \cite{Casini:2017roe} for many different approaches to deriving of this algebra in the more general case. 

\section{More details on defect operator spectrum}

\label{app:defect}

In this appendix, we compute the scaling dimensions of  some defect operators that appear in the dOPE of scalars, and the stress tensor. In holographic CFT, we compute them for arbitrary values of the Renyi index $n$. Further, in the $n\to 1$ limit, we provide up to $O(n-1)$, the values of these scaling dimension for the ``minimal twist'' defect operators in the dOPE of scalars valid for any CFT.  We do this analytic computation first, then we turn to the holographic case.

\subsection{Analytic considerations}

We would like to compute more explicitly the leading $(n-1)$ shift in the ambient space scalar two point function. From this we can extract the shifts in the conformal dimensions of the defect primaries. 
We need to analyze the two terms \eqref{bothscalar}. As explained in the main text the first term is actually the stress tensor conformal block with 4 external scalars. We can thus look up the answer. We can also just do the integral over the stress tensor which defines the modular Hamiltonian. Either way it is possible to reduce this block to a single integral representation which looks very similar to the second term in \eqref{bothscalar}. Combining these we have:
\be
(\partial_n -1) \left< \Sigma_n^0 \phi \phi \right>_{n=1}
= c_\Delta (X^2)^{-\Delta} \int_0^{-\infty} \frac{d\lambda}{(\lambda-1)^2}\left( - \left( Y_\lambda^2 \right)^{-\Delta}
- \frac{ \Delta(\lambda^2+1 - h(\lambda+1)^2)}{\lambda (d-1)(d-2)} \left( Y_\lambda^2 \right)^{1-h} \right)
\label{thisresult}
\ee
where
\be
X^2 = (w-z)(\bar{w}-\bar{z}) + y^2 \,, \qquad Y_\lambda^2 = 1 + \frac{w\bar{z}}{X^2} ( 1- \lambda) + \frac{ z\bar{w}}{X^2} ( 1- \lambda^{-1})
\ee
The fact that this result \eqref{thisresult} is single valued as one moves one operator around the entangling surface in Euclidean ($z \rightarrow e^{2\pi i} z, \bar{z} \rightarrow e^{-2\pi i} \bar{z}$) becomes the statement that the residue at $\lambda=1$ vanishes, which can be easily checked. 

We need to analyze the limit $z, \bar{w} \rightarrow 0$. After setting this to zero we find a log divergence coming from the lower $\lambda \approx 0$ integral. This divergence should thus be cut off at $\lambda \approx z\bar{w}/X^2
\approx z\bar{w}/y^2$ as the lightcone limit was not uniform in $\lambda$. The coefficient of the log is easily calculable:
\be
\label{lncalc}
\partial_n \left< \Sigma_n^0 \phi \phi \right>_{n=1} \ni
\frac{\Delta}{2(d-1)} (X^2)^{-\Delta} \left( 1 + \frac{w\bar{z}}{X^2} \right)^{1-h} \int_{z\bar{w}/y^2} \frac{d\lambda}{\lambda} 
\ee
There are other $\log$ terms coming from the upper limit of the integral $\lambda \rightarrow -\infty$, but these always give $\ln (w\bar{z}/y^2)$ terms, there are several sources of such terms. We don't actually need to do the computation however, since we know the coefficient of the $\ln (w\bar{z}/y^2)$ term should be the same as derived from \eqref{lncalc} so that they sum up to a single valued function on the Euclidean section $\propto \ln (w \bar{w} z\bar{z})$. There will be also non-log terms which we do not keep track of. We can combine these into the claimed result in the main text \eqref{n1} where the term multiplying the log in \eqref{lncalc} becomes \eqref{q1}.  

\begin{figure}
\includegraphics[trim={0cm 0cm 0cm 0cm},scale=1]{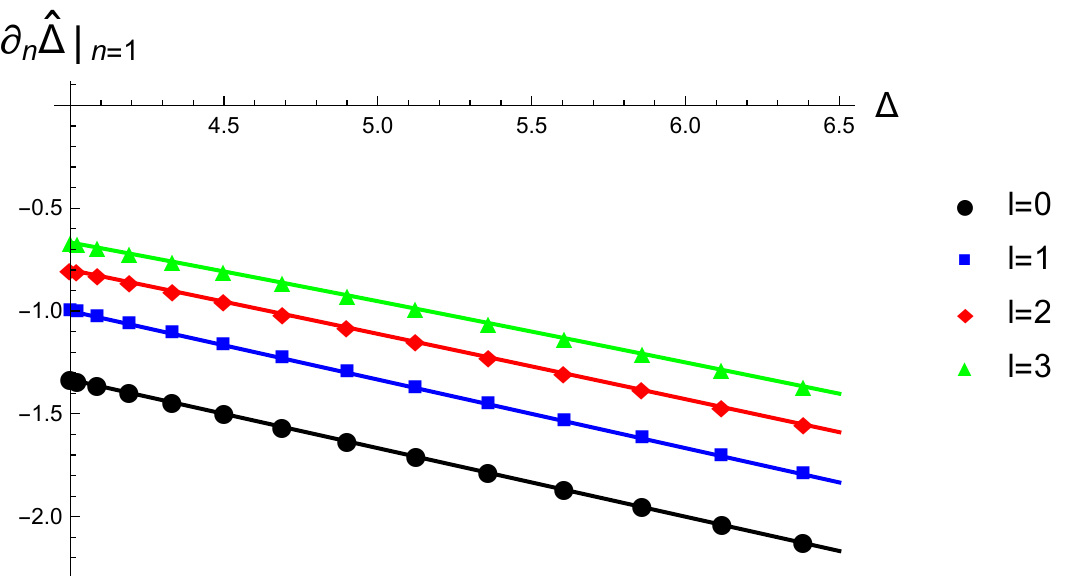}
\caption{Plots of the leading $(n-1)$ correction to the defect operator spectrum. The discrete points are obtained from the numerical method for holographic CFT, and the continuous lines from the anaytic result for arbitrary CFTs (\ref{eq:resultn1})}.
\label{fig:scalaranalytic}
\end{figure}

In $d=4$, we find the following $O(n-1)$ correction to the scaling dimension $\widehat{\Delta}$ of defect operator transverse spin $l$ that appears in the dOPE of an ambient scalar of dimension $\Delta$:
\begin{equation}\label{eq:resultn1}
\widehat{\Delta} =
      (\Delta + l) -\f{\Delta(\Delta-1)}{3(\Delta+l-1)} (n-1) + O((n-1)^2)
\end{equation}
We compare this result with that of the holographic computation outlined in the next section of this Appendix, in the plot Fig.\ref{fig:scalaranalytic}.

\subsection{Holographic computation: scalars}
We consider a probe scalar field in the hyperbolic black hole background \cite{Casini:2011kv} given by 
\be
ds^2 = \f{dr^2}{f(r)} + f(r) d\tau^2 + \f{r^2}{\rho^2}(d\rho^2 + d\vec{y}^2).
\ee
While we can  perform this computation in arbitrary fixed dimension, we restrict to the case of the 5d hyperbolic black hole $\mathcal{B}_n$, which is dual to the twist defect in a 4d holographic CFT. In this case we have
\be
f(r) = -1 + \f{r_h^2-r_h^4}{r^2} + r^2,~~~~~  n = \f{r_h}{-1+2 r_h^2},
\ee
where $\tau \sim \tau + 2\pi n$ and $n$ is the Renyi index of the dual defect CFT. 
Consider the following ansatz for a scalar field of mass $\mu$ in this background
\be
\phi(r,\tau,\rho,\vec{y}) = \rho^{\widehat{\Delta}}  e^{i l \tau} \psi(r),
\ee
where $\widehat{\Delta} \in \mathbb{R}$ has interpretation as scaling dimensions and $l \in \mathbb{Z}$ is the transverse SO(2) spin. Even for non-integer $n$ we still take $l \in \mathbb{Z}$, which is justified since we want this operator to be single valued upon shifting one replica $\tau \rightarrow \tau + 2\pi$. This might seem strange since the ansatz is not compatible with the thermal periodicity of $\tau$  for non-integer $n$, however this is the correct procedure for analytic continuation as pointed out in \cite{Faulkner:2013ana}. Roughly speaking we can think of this as studying $\phi$ in a bulk spacetime defined via the quotient with respect to the replica symmetry $\mathcal{B}_n / \mathbb{Z}_n$. This space time has a conical deficit at $r=r_h$ but is well defined for any value of $n$ and the $\phi$ fluctuations on top of this space time are now single valued. The analytic continuation procedure also fixes a unique boundary condition at the deficit as we will argue below.

Plugging the ansatz into the Klein-Gordan equation this then becomes an ODE in $r$ given by
\be
\left(-\frac{l^2}{f^2}-\frac{\mu ^2}{f}+\frac{\widehat{\Delta}^2-2\widehat{\Delta} }{r^2 f}\right)\psi  +   \left(\frac{f'}{f}+\frac{3}{r}\right)\psi'  +  \psi'' = 0.
\label{eq:diffscalar}
\ee
Near the horizon at $r = r_h$, solutions behave as 
$\psi(r) \sim (r-r_h)^{\pm \f{l n }{2}}$.  We must pick the solution that behaves as $(r-r_h)^{+ \f{l n }{2}}$ so that the scalar field is regular near the horizon.
Normalizable modes have the property $\psi(r) \to 0$ as $r\to \infty$.  Only for specific values of $\widehat{\Delta}$, will the solutions  be both regular near the horizon and normalizable. These values correspond to the scaling dimensions of defect \emph{primary} operators that appear in the dOPE of the scalar operator of dimension  $\Delta =  \f{d}{2} + \sqrt{(d/2)^2 + \mu^2} $.

\begin{figure}
\includegraphics[trim={2cm 0cm 0cm 0cm},scale=0.67]{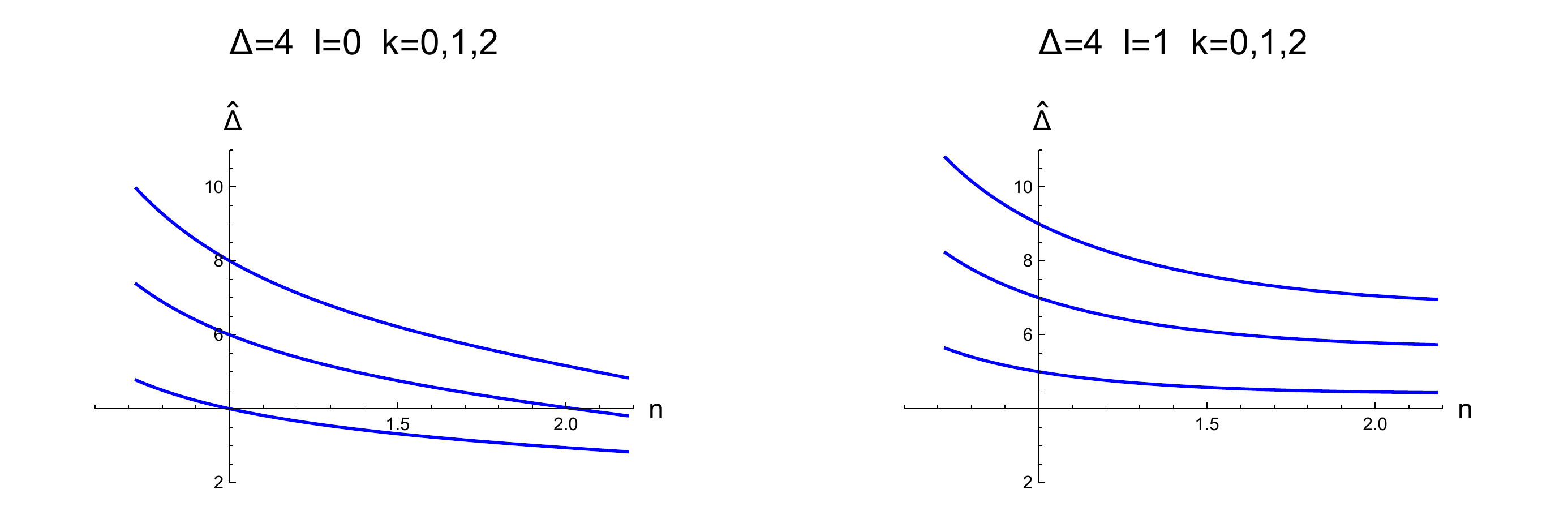}
\caption{Plots of defect operator scaling dimension in a holographic model $\widehat{\Delta} $ as a function of Renyi index $n$. In the bulk we work with a massless scalar ($\mu=0$ and $\Delta=4$).  }
\label{fig:mu0}
\end{figure}

The numerical procedure resulting from the above discussion, is to solve the differential equation (\ref{eq:diffscalar}) by specifying boundary conditions at $r=r_h$. We then check if the solution has the property  $\psi(r)\to 0$ as $r \to \infty$. Proceeding this way, we see that the scaling dimensions of the defect operators at $n=1$ are given by 
\be \widehat{\Delta}  = \Delta + l + 2k ,~~~ k=0,1,2,\ldots \ee
These scaling dimensions at $n=1$ can also been obtained by expanding the usual CFT 2 point function of scalars in the absence of a defect (using the appropriate conformal blocks since the holographic method singles out primary operators).
The scaling dimensions will receive $O(n-1)$ corrections, and above we were able to compute these $O(n-1)$ correction for the twist defect in arbitrary CFTs for a subset of defect operators which satisfy $k=0$, the ``minimal twist'' operators. 
The numerical procedure also gives us the values of $\widehat{\Delta}$ for arbitrary values of $n$, and we have plotted  $\widehat{\Delta}$ vs $n$ in a few examples in Figure~\ref{fig:mu0}.

\subsection{Holographic computation: stress tensor}
The computation in this section is a generalization of  \cite{twistdisplacement} (see also \cite{Dongshaperenyi,Bianchi2016shaperenyi}).
 \cite{twistdisplacement} focused on  the displacement operator, which is a defect operator that appears in the dOPE of stress tensor for $n \neq 1$ and has a protected scaling dimension of $\widehat{\Delta}_{\text{disp}} = d-1$.  Following \cite{twistdisplacement}, we start with the metric fluctuation ansatz
\begin{multline}
ds^2 = \f{dr^2}{f} + f d\tau^2 + \f{r^2}{\rho^2} (d\rho^2 + (d\vec{y}_{d-2})^2)  + \\  +\epsilon e^{i l \tau} \rho^{\widehat{\Delta}} \left( f  k_{\tau \tau} d\tau^2 + \frac{2 f}{\rho} k_{\tau \rho} d\tau d\rho + \f{r^2}{\rho^2} k_{\rho\rho} d\rho^2 + \f{r^2}{\rho^2} (d\vec{y}_{d-2})^2 \right),
\label{eq:metric_ansatz}
\end{multline}
where $k_{\tau \tau},k_{\tau \rho},k_{\rho \rho},k_{yy} $ are functions of $r$. This ansatz will single out defect operators of scaling dimension $\widehat{\Delta}$ under the isometry in the bulk that corresponds to scaling in the dCFT, and transverse  spin $l \in \mathbb{Z}$ for the $SO(2)$, and $0$ spin in the $SO(d-1,1)$. Also, we perform the numerical computations for $d=3$, in which case we have 
\be f(r) =  -1 + r^2 + \f{r_h - r_h^3}{r},~~~~ n =  \f{2 r_h}{-1 + 3 r_h^2}.
\ee

There are three diffeomorphisms that leave the  form of (\ref{eq:metric_ansatz}) fixed.  In terms of infinitesimal parameters $\delta_r, \delta_\tau, \delta_\rho$, these diffeomorphisms act on the ansatz as,
\begin{align}
k_{\tau\tau} \to k_{\tau\tau} + 2 i l \delta_\tau + \left(2 i l X_\tau + \f{f'}{\sqrt{f}} \right)\delta_r,\\
k_{\tau\rho} \to k_{\tau \rho} + \delta_\tau \widehat{\Delta} + \f{i l r^2}{f} \delta_\rho + \left(\f{i l r^2 X_\rho}{f} + \widehat{\Delta} X_\tau\right) \delta_r,\\
k_{\rho\rho} \to k_{\rho \rho} + 2 \widehat{\Delta} \delta_\rho + \left( 2 \widehat{\Delta} X_\rho + 2 \f{\sqrt{f}}{r} \right) \delta_r,\\
k_{yy} \to k_{yy} - 2 \delta_\rho + (2  \f{\sqrt{f}}{r} - 2 X_\rho)\delta_r .
\end{align}
where the functions $X_\tau(r)$ and $X_\rho(r)$ satisfy 
\begin{align}
X_\tau'(r) = -\f{i l }{f^{3/2}},~~~ \lim_{r\to\infty} X_\tau(r) =0\\
X_\rho'(r)= - \f{\widehat{\Delta}}{r^2 \sqrt{f}},~~~\lim_{r\to\infty} X_\rho(r) =0.
\end{align}
We started with 8 first order degrees of freedom , namely $k_{\tau \tau},k_{\tau \rho},k_{\rho \rho},k_{yy} $  and their first derivatives in $r$. Three of the components of Einstein's equations are first order constraint equations.  Further there are 3 gauge transformations. This leaves us with 2 residual first order degrees of freedom. In fact, in this case, we can construct a gauge invariant linear combination,
\be
\sigma =  k_{yy} + \mathcal{A}_{\tau\rho} k_{\tau \rho}+
 \mathcal{A}_{\rho\rho} k_{\rho \rho}+
  \mathcal{A}_{\tau\tau} k_{\tau \tau},
\ee
where
\begin{align}
\mathcal{A}_{\tau\rho} =  -\f{4 i l (\widehat{\Delta}+1) f}{\mathcal{B}},~~~
\mathcal{A}_{\rho\rho} = \f{\widehat{\Delta}-r^2(2 l^2+3\widehat{\Delta})+\widehat{\Delta} f}{\mathcal{B}}, ~~~ \mathcal{A}_{\tau\tau} = \f{2 \widehat{\Delta} (1+\widehat{\Delta}) f}{\mathcal{B}}, 
\end{align}
where $\mathcal{B} = \widehat{\Delta}^2(1-3 r^2+f)+2 r^2 l^2$, so that $\sigma$ and $\sigma'$ are the  two gauge invariant, independent first order degrees of freedom we seek. Thus, $\sigma$ obeys a second order differential equation,  which we can write (schematically) as,
\be
\sigma'' + \mathcal{C}_1( r) \sigma' + \mathcal{C}_0 ( r) \sigma = 0.
\label{eq:diffsigma}
\ee
This procedure for finding \eqref{eq:diffsigma} was inspired by the similar fluctuation computation appearing in \cite{Faulkner:2012gt,Edalati:2010hk}.

The normalizability condition  is that the $r \to \infty$ behavior of the fluctuation  must be pure gauge \cite{twistdisplacement}.  In terms of $\sigma$, this reduces to the condition 
\be
\lim_{r \to \infty} \sigma = 0.
\label{eq:sigma_normalizability}
\ee

\begin{figure}
\includegraphics[trim={2cm 0cm 0cm 0cm},scale=0.55]{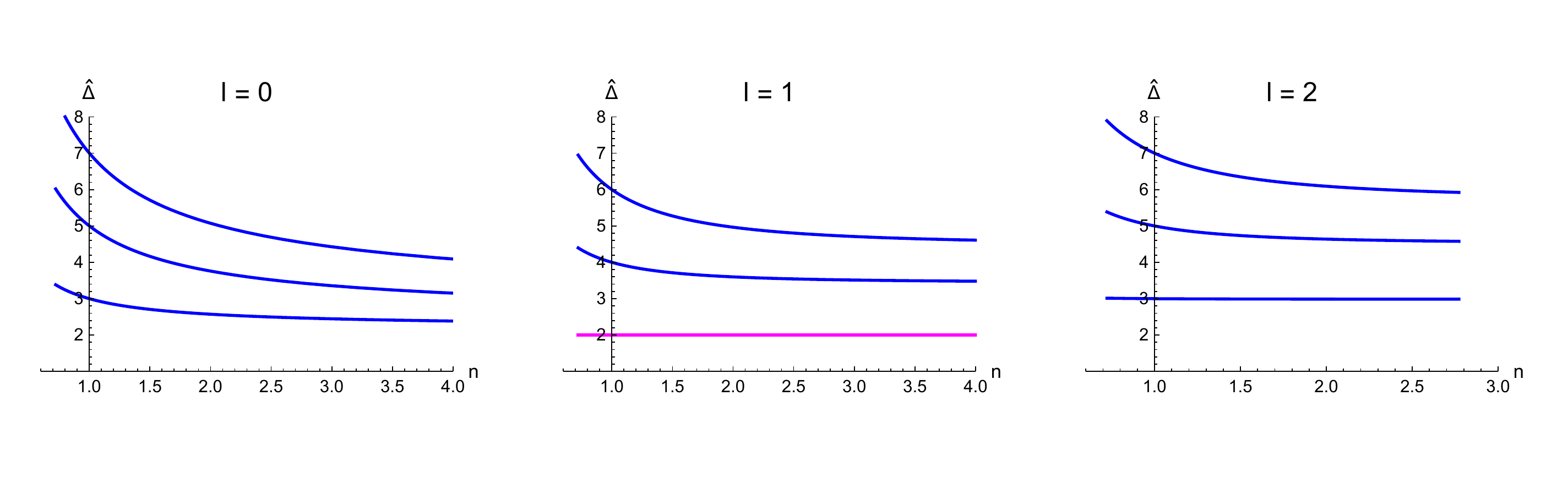}
\caption{Plots of scaling dimension $\widehat{\Delta}$ of defect operators as a function of Renyi index $n$ in the different $\ell$ channels at $d=3$. The displacement operator is shown in pink and has a protected scaling dimension. The lowest dimension operator for fixed $\ell$ is ``minimal twist'' except for $\ell=0$ channel.  }
\label{fig:stressplot2}
\end{figure}

Near the horizon, solutions to (\ref{eq:diffsigma}) behave as  $\sigma(r) \sim (r-r_h)^{\pm \f{l n}{2}}$.  Demanding regularity near horizon leads us to the choice $\sigma \sim (r-r_h)^{+\f{l n}{2} }$. The numerical prescription then involves solving the differential equation (\ref{eq:diffsigma}) with regular boundary condition at horizon . We then seek values of $(\widehat{\Delta}, l)$ such that the solution satisfies the normalizability condition  (\ref{eq:sigma_normalizability}). These values correspond to the scaling dimension and transverse spin of defect operators that appear in the dOPE of the stress tensor (See  Figure~\ref{fig:stressplot} in the main text and Figure~\ref{fig:stressplot2}).

\section{Multi-replica operators}

\label{app:multi}

In this Appendix explain why we left out the more general multi-replica operators \eqref{bulkorb} when we extracted the defect operator spectrum by moving ambient space orbifold operators close to the defect. If multi-replica operators were to give rise to new defect operators we would discover them in the multi-replica two point functions in the presence of $\Sigma_n^0$. However when we bring the multi-replica operators close to the defect, on the branched manifold $\mathcal{M}_n$ the operators on different replicas in some sense are coming close to each other - this is completely clear in $d=2$ where the covering map: $ w = z^{n}$ is a conformal transformation which removes the twist defect leaving the replica operators approaching the origin at the positions $z= w^{1/n} e^{ 2\pi k/n}$ 
for $ k = 0 , \ldots n-1$ with $|w| \rightarrow 0$. We can then simply replace these operators by another local operator using the regular CFT OPE such that the multi-replica two point function is a sum over single replica two point functions. Hence nothing new. 

In dimensions  $d >2$ this is quite a bit more subtle because one cannot remove the defect with a conformal transformation - the best one can do is map to the space $\mathbb{H}_{d-1} \times S^1$ with the twist defect operator being sent to the boundary of this space. Now the cluster of operators coming from a single multi-replica operator are located at the same point close to the boundary of hyperbolic space and distributed around the thermal circle at $\theta_k = 2\pi k$ for $k= 0, \ldots n-1$ and where $\theta \equiv \theta + 2\pi n$. 
The limit towards the defect will then mean the two clusters of multi-replica operators will be separated very far from each other on the hyperbolic factor. In this limit one might hope to again use the ambient space OPE to replace a cluster of operators by a sum over local operators at some fixed $\theta$, say $\theta=0$ and claim victory. However we then might worry about convergence of this OPE since these operators have separation of order the curvature scale of $\mathbb{H}_{d-1}$. This does not cause a huge problem - we can consider deforming each cluster of operators to lie around the thermal circle at $\theta_0 < \theta_1 \ldots < \theta_{n-1} \ll 1$ such that the OPE does converge. This implies that the defect operators we have already discovered (from single replica operators) will give the full answer for these values of $\theta_k$. This allows us to compute the answer in a slightly different way, instead of making use of the ambient space OPE we can now use the defect OPE.  We direclty replace the clusters\footnote{In the orbifold theory the clusters at general $\theta_k$ should now be thought of as non-local operators with appropriate Wilson lines attached. We will suppress this detail here as it does not effect the defect OPE argument, as was the case
for $\overbracket{\mathcal{O}_B \mathcal{O}_{\bar{A}}}$.} with the known defect operators, in particular we can make use of the same defect OPE we used in Section~\ref{sec:dope} to re-write the two point correlator schematically as:
\begin{align}
\nonumber
&\left< \Sigma_n^0 \left( \bigotimes_{k=0}^{n-1} \mathcal{O}_{\alpha_k}^{(k)} (\theta_k, \rho, y)  + \mathbb{Z}_n {\rm symm}  \right)
\left(  \bigotimes_{k=0}^{n-1}  \mathcal{O}_{\beta_k}^{(k)} (\theta'_k, \rho', y') + \mathbb{Z}_n {\rm symm}  \right) \right> \\ \nonumber
& = \sum_{ij} \left< \Sigma_n^0  \widehat{O}_i(y') \Sigma_n^0 \left( \bigotimes_{k=0}^{n-1} \mathcal{O}_{\alpha_k}^{(k)} (\theta_k, \rho,y) + \mathbb{Z}_n {\rm symm}  \right)  \right>   G_{ij}^{-1}(y,y') \\
& \qquad \qquad \qquad \times \left< \Sigma_n^0 \widehat{O}_j(y) \Sigma_n^0 \left( \bigotimes_{k=0}^{n-1} \mathcal{O}_{\beta_k}^{(k)} (\theta_k', \rho',y') + \mathbb{Z}_n {\rm symm}  \right)  \right>
\label{quant}
\end{align}
where $i,j$ sum over the known defect operators.
This answer must agree with the ambient space OPE of the paragraph above. However note that the bulk to defect correlators in \eqref{quant} can be extracted from the appropriate projections onto the known defect operator spectrum acting on one operator inside a $n$ plus $1$ point correlation function in the presence of $\Sigma_n^0$ (or rather on the branched covering $\mathcal{M}_n$). These correlators, and hence their projection, are well defined without requiring a convergent OPE sum. Additionally  there is no obstruction to using the expression \eqref{quant} for any values of $\theta_k$ which can now be continued to the required values of $\theta_k = 2\pi k$ without worrying about convergence. We have effectively re-summed the OPE (with unknown but potentially knowable functions) which then implies that this is the full answer and we have not missed any defect operators in writing \eqref{quant}. Thus we again conclude that no new operators arise.

\section{Modular flow from the replica trick}\label{sec:appmodflow}

In this Appendix we would like to compute the modular flow correlator:
\be
h(s) = \left< \psi \right| \mathcal{O}_B e^{ i K s} \mathcal{O}_{\bar{A}} \left| \psi \right>
\ee
using a different method to the main text. In the main text (Section~\ref{sec:modflow}) we used a perturbative
expansion, which was at the time not fully justified. Here we will find the same results using a more complete method, thus closing the gap on our computation of single modular flow. 

The idea is to compute $h(s)$ directly in the replica trick in combination with the defect OPE. Consider:
\be
Z_{n,p} = {\rm Tr}_A \rho_A^{n-1}  \left[ {\rm Tr}_{\bar{A}}\left( \mathcal{O}_{\bar{A}} \left| \psi \right> \left< \psi \right| \right) \right]  \rho_A^{p}  \mathcal{O}_B \rho_A^{-p} 
\ee
which is well defined for $n,p \in \mathbb{Z}$ and $0\leq p < n$. If we can find an analytic continuation from 
these integers to real $n$ and complex $p$ maintaining the thermal periodicity $0\leq {\rm Re} p < n$.\footnote{Recall that $n$ is like an inverse temperature.} then we can send $n \rightarrow 1$ and $p \rightarrow i s/(2\pi)$ to recover:
\be
h(s) = \lim_{n \rightarrow 1} \left. Z_{n,p} \right|_{p = is/(2\pi)}
\ee
Of course the trick is finding the (natural) analytic continuation which agrees on the integers and has nice properties for large $n,p$. Indeed this later requirement makes it clear that we should search for an analytic continuation first in $n$ and \emph{only then} in $p$. This is because for fixed integer $n$ we only have a finite number of $p = 0, \ldots, n-1$'s to work with and there is certainly no unique analytic continuation of a function from a finite number of values. For example once we have $Z_{n,p}$ for $n \in \mathbb{R}_{>0}$ and $p \in \mathbb{Z}$ with $0<p < n$ then we can consider $Z_p(m) \equiv Z_{m+p,p}$ defined for $p \in \mathbb{Z}$ and $p =0,1 , \ldots \infty$ at fixed $m >0 $. We then seek an analytic continuation of this $Z_p(m)$
to complex $p$.\footnote{ Note that for $n=1$ we have very strong constraints on the analyticity of the resulting function of $s$ due to Tomita-Takasaki theory. It is reasonable to assume these constraints also hold for $n \neq 1$.}

Keeping this in mind, we proceed. We can imagine computing $Z_{n,p}$ for integral values of $n,p$ using the same defect OPE
method we used in the main text. We can write this as an orbifold correlator:
\be
Z_{n,p}  = \left< \Sigma_n \overbracket{ \mathcal{O}_B\left(\circlearrowleft_{p} \right)  \mathcal{O}_{\bar{A}} }  \right>_{CFT^n/\mathbb{Z}_n}
\ee
where the notation $\left(\circlearrowleft_p \right)$ means to move the operator $\mathcal{O}_{B}$ around the defect $p$ times. Again we could write this using an attached Wilson line that circles the defect $p$ times. Since this operator is still local to the defect as we zoom out we expect the defect OPE method to still apply. We replace the bi-local operator with a sum over defect operators $\sum_{i} \beta_i(p,n) \widehat{\mathcal{O}}_i$ . The only difference to our computations in Section~\ref{sec:modham} come from the three point function terms which are used to compute the OPE coefficients $\beta_i(p,n)$.  That is:
\be
C_\ell(p,n) \equiv \left< \Sigma_n^0 \widehat{T}_{\ell}(y)  \overbracket{ \mathcal{O}_B\left(\circlearrowleft_{p} \right)  \mathcal{O}_{\bar{A}} }  \right>
= -\frac{n}{2\pi i} \oint \frac{d \bar{z}}{\bar{z}} \bar{z}^{-\ell+2} \left( \sum_{j=0}^{n-1} \left< T_{++}^{(j)}(z,\bar{z},y) \mathcal{O}_B^{(-p)} \mathcal{O}_{\bar{A}}^{(0)} \right>_{\mathcal{M}_n} \right)
\ee
Note that we would like to send $z \rightarrow 0$ relevant for the light cone operators, but we hold $z$ small for now.
Again we write the $j$ sum using a contour integral:
\be
C_\ell(p,n) = -\frac{n}{(2\pi i)^2} \oint \frac{d \bar{z}}{\bar{z}} \bar{z}^{-\ell+2} \int_{\mathcal{C}_{\bar{A}} \cup \mathcal{C}_B } \frac{d\lambda \lambda}{\bar{z} (\lambda-\bar{z})} \left<T_{++}(z \bar{z}/\lambda , \lambda,y)\mathcal{O}_B^{(-p)} \mathcal{O}_{\bar{A}}^{(0)} \right>_{\mathcal{M}_n}
\ee
where we have pushed the integration contour to surround the four branch cuts coming 
when  $T_{++}$ hits the (two) lightcones of each operator $\mathcal{O}$ (at a fixed $y$ separation.)  
The contour $\mathcal{C}_X$ encircles the $X$ operator branch cuts that lie on different replicas.
These are approximately at the locations: $\lambda = - y^2/ u_{\bar{A}}, \lambda = v_{\bar{A}} z\bar{z}/y^2$ and
$\lambda = - e^{- i 2\pi p} y^2/ u_{B}, \lambda = e^{ -i 2\pi p} v_{B} z\bar{z}/y^2$.

At this point we would like argue for an analytic continuation. The three point function is well defined for any $n$ - and can be thought of as a 3 point function of the CFT on Hyperbolic space $\times \mathbb{S}^1$ with inverse temperature $n$. Thus holding $p$ fixed and integer we have achieved the first required step. 
Note that we do not know this three point function
in general except at $n=1$. We could however compute it, for example, using a holographic CFT but we will only need the answer at $n=1$.

Now it should be possible to continue $p$. Firstly note that it is important that the $\lambda$ integral is not moved around by the analytic continuation - this is so that the $p \rightarrow is$ continuation has the desired analyticity properties which would not be the case if $\lambda$ could be forced to move onto the pole at $\lambda = \bar{z}$. This means we should continue the various contour integrals wrapping the branch cuts 
$\mathcal{C}_B, \mathcal{C}_{\bar{A}}$ differently for each term. The $\mathcal{C}_{\bar{A}}$  contour integral should be left as is, while for the contour $\mathcal{C}_{B}$ we should apply a $ 2\pi p$ rotation to move this integral to the ``first sheet'' (or in other words we are simply relabeling the replicas using a shift by $p$.) This operation effectively moves $\mathcal{O}_{\bar{A}}$ to the $p$'th sheet. Note that for integer $p$ non of these operations have an effect on the answer.  Now continuing $p$ to non-integer is simple, we simply rotate either $\mathcal{O}_{\bar{A}}$ or $\mathcal{O}_{B}$  by an amount $e^{ \mp 2\pi i p} \rightarrow  e^{\pm s}$ respectively. It is important to note that we have arranged things so we only rotate the operator that is \emph{not} surrounded by the $\lambda$ contour integral.

At this stage we can send $n \rightarrow 1$ and plug in the flat space CFT 3 point function. Using this we can then check that the the branch cut contribution for small $\lambda \sim z\bar{z}$ vanishes as we send $z \rightarrow 0$. Thus we are left with two terms:
\begin{align}
2\pi i C_\ell( is/(2\pi) , 1) &= 
 \int_{\mathcal{C}_{\bar{A}}} d\lambda \lambda^{1-\ell}
\left<T_{++}(0 , \lambda,y)\mathcal{O}_B(-s) \mathcal{O}_{\bar{A}} \right> +\int_{\mathcal{C}_{B}} d\lambda \lambda^{1-\ell}
\left<T_{++}(0 , \lambda,y)\mathcal{O}_B \mathcal{O}_{\bar{A}}(s) \right>
\end{align}
If we take a derivative $\partial_s$ and set $s=0$ then we re-derive the three point function results from Section~\ref{sec:modham} (it should be compare with $C_{\ell}^{(1)} + C_{\ell}^{(2)}$ from that section.) Taking higher derivatives gives all the nested commutators. 

Working at finite $s$ it is now a simple task to use the integral representation \eqref{int2} and \eqref{ac12} to sum everything up
to \eqref{togen} which was the result of the perturbative computation that we had wished to put on more firm grounds. 

\section{Double modular flow from single modular flow}\label{sec:appmodflowdouble}

We start by noting that our formulas for single modular flow imply a very strong statement, that:
\be
\eta_A \equiv \frac{\| e^{ i K_A s} \mathcal{O}_{\bar{A}} \left| \psi \right> - \mathcal{O}_{\bar{A}}(s) \left| \psi \right> \|^2}{ \left< \Omega \right| \mathcal{O}_{\bar{A}} (s^\star) \mathcal{O}_{\bar{A}} (s) \left| \Omega \right> }
\,\, \approx 0 
\label{difff}
\ee
up to the order that we work at in the light cone limit (i.e up to and including terms $(uv)^{\tau/2}$
for $\tau = d-2$ the twist of the stress tensor). More explicitly we can show that that $\eta_A \ll (uv)^{\tau/2}$.\footnote{Recall that $u,v$ are proxies for $-u_{\bar{A}},v_{\bar{A}}$ as well as later for $u_{B},-v_{B}$.}
Recall that $\mathcal{O}_{\bar{A}}(s) $ is simply the boosted operator acting with $K_A^0$. 
This result can be shown by expanding out the norm, resulting in four terms which can be computed using our expression for single modular flow \eqref{sameside} (when the two operators act in the same wedge). At the order we compute all the terms cancel.  We should take $\Im s <0$ in order to avoid the two operators coming close which would necessitate smearing. 

This does not mean that $e^{ i K_A s} \mathcal{O}_{\bar{A}} \left| \psi \right> \mathop{=}^? \mathcal{O}_{\bar{A}}(s) \left| \psi \right> $ which is clearly false because the norms of these states individually are different. However this difference in norms might only be seen in \eqref{difff} at higher orders in the light cone expansion (at least before $ \mathcal{O}(uv)^{\tau}$), which can be explicitly shown using a Cauchy-Schwarz argument. 

Let us now define:
\begin{align}
\left| a \right> & = e^{ i K_A s} \mathcal{O}_{\bar{A}} \left| \psi \right> /\sqrt{\mathcal{N}_A}
\qquad \left| \alpha \right> = \mathcal{O}_{\bar{A}}(s_A) \left| \psi \right>  /\sqrt{\mathcal{N}_A} 
\qquad \mathcal{N}_A = \left< \Omega \right| \mathcal{O}_{\bar{A}} ((s^\star)_A) \mathcal{O}_{\bar{A}} (s_A) \left| \Omega \right>   \\
\left| b \right> & = e^{ i K_B s^\star} \mathcal{O}_{B} \left| \psi \right>  /\sqrt{\mathcal{N}_B}
\qquad \left| \beta \right> = \mathcal{O}_{B}((s^\star)_B) \left| \psi \right>   /\sqrt{\mathcal{N}_B}
\qquad \mathcal{N}_B = \left< \Omega \right| \mathcal{O}_{B} (s_B) \mathcal{O}_{B} ((s^\star)_B) \left| \Omega \right>
\end{align}
where we use notation established in Section~\ref{sec:dmodflow} for modular flow with respect
to $K_A^0$ and $K_B^0$. In addition to $\eta_A = \| \left|  a \right> - \left| \alpha \right> \|^2 \ll (uv)^{\tau/2}$ also have that $\eta_B \equiv \| \left|  b \right> - \left| \beta \right> \|^2 \ll (uv)^{\tau/2}$ still maintaining $\Im s <0$. Applying Cauchy-Schwarz:
\be
\left| \left( \left< b \right| - \left< \beta \right|\right) \left( \left| a \right> - \left| \alpha \right> \right) \right|
\leq (\eta_A \eta_B)^{1/2} \ll (uv)^{\tau/2}
\ee 
But since we only want to compute $f(s) \propto \left< b \right|\left. a \right>$ up to order $\mathcal{O}(uv)^{\tau/2}$ we can ignore the $(\eta_A \eta_B)^{1/2}$ term and simply set:
\be
\left< b \right|\left. a \right> \approx \left< b \right|\left. \alpha \right>+ \left< \beta \right|\left. a \right>
-  \left< \beta \right|\left. \alpha \right>
\ee
This formula relates double modular flow on the left hand side to single modular flow on the right hand side. So the right hand side can be computed using \eqref{togen}. Indeed combining all the terms we find exactly the same answer as the perturbative method for double modular flow applied in Section~\ref{sec:modflow}.

We can roughly understand these result as telling us the following:
\be
K_A \mathcal{O}_{\bar{A}} \left| \psi \right>  =\left[ K_A^0, \mathcal{O}_{\bar{A}}\right] \left| \psi \right> 
+ \epsilon_{\parallel} \widehat{\Phi}_{\parallel } \left| \psi \right> +  \epsilon_{\perp} \widehat{\Phi}_{\perp} \left| \psi \right>
\ee
where the action of $\Phi_{\parallel} \left| \psi \right>$ is in the code subspace (See Footnote~\ref{foot} for the context of this discussion). Schematically $\Phi_{\parallel} \left| \psi \right> = \int_{LC} du' dv' K(u',v') \mathcal{O}(u',v') \left| \psi \right>$ for $LC$ satisfying $-v' \ll 1$ and $u'$ not too large.  This kernel should be constrained to give the known answer for the perturbative correction
$ \left< \psi \right| \mathcal{O}_B \epsilon_{\parallel} \widehat{\Phi}_{\parallel } \left| \psi \right> \sim (uv)^{\tau/2}$. We are agnostic to the action of $\widehat{\Phi}_\perp$ except that 
it should have no overlap with the various light cone states $\left< \psi \right| \mathcal{O}_B$ and 
the small parameter should satisfy $\epsilon_{\perp}^2 \ll (uv)^{\tau/2}$. 
This is a weaker form of the half sided projecting onto the code subspace than was advocated in Footnote~\ref{foot}, however it is sufficient to sketch why the perturbative  expansion methods  in Section~\ref{sec:modflow} worked. If the action of $K_A$ where to move us out of the light cone limit, via $\widehat{\Phi}_{\perp}$, then if we were computing lightcone matrix elements of $(K_A)^2$ then we would need the second action of $K_A$ to move us back which again can only happen via $\widehat{\Phi}_{\perp}$. So we get two factors of $\epsilon_A^2 \ll (uv)^{\tau/2}$ much smaller than the terms we are interested in. This works at higher order in $K_A$ also. 

\section{Higher spin QNEC}\label{app:R}

In our proof of the QNEC, we computed the lightcone limit of the quantity $\mathcal{R}\left(x_{2},x_{1};A\right)$ defined in~(\ref{defR}), which we then used to compute the single modular flow correlator~(\ref{modflwu}) and the double modular flow correlator $f(s)$ in~(\ref{affs}). Specifically, we computed the contribution to these correlators coming from stress-tensor exchange in the OPE of the probe operators $\mathcal{O}$. As discussed in \ref{subsec:higherspin}, it is straightforward to generalize our methods to derive a higher spin version of the QNEC for the symmetric traceless operator $\mathcal{J}_{-\dots-}$ of conformal dimension $\Delta_{J}$, even spin $J$, and minimal twist $\tau_{J}=\Delta_{J}-J$ among operators of the same spin. This requires computing the contribution to the same set of correlators now coming from $\mathcal{J}$-exchange. In this appendix, we provide some of the details of these computations. 

The lightcone limit of $\mathcal{R}$ due to $\mathcal{J}$-exchange is written succintly in terms of the following function:
\be
F_{J}\left(x_{2},x_{1};u\right)\equiv-G_{J}\left(4 \Delta u\right)^{J-1}\left(-\Delta u\Delta v\right)^{\frac{\tau}{2}}\left(\frac{\left(u_{2}-u\right)\left(u-u_{1}\right)}{\left(u_{2}-u_{1}\right)^{2}}\right)^{\frac{\Delta_{J}+J}{2}-1}
\ee
\be
G_{J}\equiv\left(\frac{c_{\mathcal{JOO}}}{c_{\mathcal{JJ}}}\right)\frac{2^{\Delta_{J}-J+1}}{\sqrt{\pi}}\frac{\Gamma\left(\frac{\Delta_{J}+J+1}{2}\right)}{\Gamma\left(\frac{\Delta_{J}+J}{2}\right)}
\ee
This generalizes the function $F$ defined in~(\ref{Fx2x1u}). In particular, for the stress tensor $G_{T}=-4\pi G_{N}\Delta_{\mathcal{O}}/d$. Using the notation and conventions of \ref{sec:singmodflow}, the $\mathcal{J}$-contribution to $\mathcal{R}$ is:
\begin{eqnarray}
\mathcal{R}\left(x_{2},x_{1};A\right) & = & i\left(\int_{u_{1}}^{0}du\left\langle \mathcal{J}_{-\dots-}\left(u\right)\right\rangle _{\psi}u^{J}\partial_{u}\left(u^{1-J}F_{J}\left(x_{2},x_{1};u\right)\right)\right.\nonumber \\
 &  & \left.\phantom{\frac{\left\langle \hat{\mathcal{D}}^{\ell}H_{A}^{\psi}\right\rangle _{\psi}}{\Gamma\left(J-\ell\right)}}-\sum_{\ell=1}^{J-1}\left(-1\right)^{J-\ell}\frac{\left\langle \widehat{\mathcal{D}}^{\ell}H_{A}^{\psi}\right\rangle }{\Gamma\left(J-\ell\right)}\partial_{u}^{J-1-\ell}F\left(x_{2},x_{1};u\right)_{u=0}\right)+\dots,
\end{eqnarray}
This generalizes the stress-tensor contribution found in~(\ref{Ranswer}). Note that there are now contributions from a family of displacement operators, as discussed around~(\ref{higherdisp}). 

With the result for $\mathcal{R}$ in hand, we can proceed to the single flow correlator. The stress tensor contribution was given in~(\ref{togen}). The general $\mathcal{J}$-contribution looks similar:
\begin{equation}
\frac{\left\langle \mathcal{O}_{B}e^{isK_{A}}\mathcal{O}_{\bar{A}}\right\rangle _{\psi}}{\left\langle \mathcal{O}_{B}\mathcal{O}_{\bar{A}}\left(s\right)\right\rangle _{\Omega}}=1-\int_{u_{\bar{A}}}^{u_{B}}du\left\langle \mathcal{J}_{-\dots-}\left(u\right)\right\rangle _{\psi}A_{J,s}\left(u\right)+\sum_{\ell=1}^{J-1}\frac{\left(-1\right)^{J-\ell}\mathbb{J}^{J-\ell}A_{J,s}\left(0\right)}{\Gamma\left(J-\ell\right)\ell}\left\langle \widehat{\mathcal{D}}^{\ell}H_{A}^{\psi}\right\rangle _{\psi}+\dots
\label{singleflowJ}
\end{equation}
Here
\begin{eqnarray}
A_{J,s}\left(u\right) & = & F_{J}\left(x_{B},x_{\bar{A}}\left(s\right);u\right),\qquad0<u<u_{B}\nonumber \\
 & = & F_{J}\left(x_{B}\left(-s\right),x_{\bar{A}};u\right),\qquad u_{\bar{A}}<u<0
\end{eqnarray}
generalizes the piecewise function $A_{s}\left(u\right)$ in~(\ref{region}), with higher-order jump discontinuity defined by
\begin{equation}
\mathbb{J}^{n}A_{J,s}\left(u\right)\equiv\lim_{\epsilon\rightarrow0}\partial_{u}^{n-1}\left[A_{J,s}\left(u+\epsilon\right)-A_{J,s}\left(u-\epsilon\right)\right].
\end{equation}
In particular,
\begin{equation}
\mathbb{J}^{J-\ell}A_{J,s}\left(0\right)=\left(1-e^{-s\ell}\right)\partial_{u}^{J-1-\ell}F_{J}\left(x_{B},x_{\bar{A}}\left(s\right);u\right)_{u=0}.
\end{equation}

Next, we turn to the double flow correlator $f(s)$. The stress tensor contribution was given in~(\ref{fcrank}), and the $\mathcal{J}$-contribution is again a straightforward generalization:
\begin{align}
f(s)  = 1 &- \int_{u^s_{\bar{A}} }^{u^s_B} d u \left< \mathcal{J}_{-\dots-}(u) \right>_\psi B_{J,s}(u) \nonumber \\
& + \sum_{\ell=1}^{J-1} \frac{(-1)^{J-\ell}}{\Gamma(J-\ell)\ell} \left( \mathbb{J}^{J-\ell} B_{J,s}\left(0\right) \left\langle \widehat{\mathcal{D}}^{\ell} H^\psi_A\right\rangle_\psi  + \mathbb{J}^{J-\ell} B_{J,s}\left(\delta x^-\right) \left\langle \widehat{\mathcal{D}}^{\ell} H^\psi_B\right\rangle_\psi\right).  
\label{fJ}
\end{align}
where the generalization of $B_{s}\left(u\right)$ in~(\ref{Bsu}) is
\begin{align}
B_{J,s}(u) &= F_J(x^s_B(s_{B},-s_A),x^s_{\bar{A}}; u)  \qquad  u^s_{\bar{A}} < u < 0 \nonumber \\
&= F_J(x^s_B(s_{B}),x^s_{\bar{A}}(s_A); u)  \qquad   0 < u < \delta x^- \nonumber \\
&= F_J(x^s_B,x^s_{\bar{A}}(s_A,-s_{B}); u)  \qquad \delta x^- < u < u^s_B.
\end{align}

With these ingredients in hand, the next step is to consider the function $F(s)$ in~(\ref{Fs}). Note that $F(s)$ is defined as the double flow correlator $f(s)$ with two single flow correlators subtracted out. Let $\tilde{F}_{J}\left(s\right)$ denote the lightcone contribution to $F(s)$ coming from $\mathcal{J}$ exchange. The expression for $\tilde{F}_J(s)$ is obtained by using~(\ref{fJ}) for $f(s)$ and~(\ref{singleflowJ}) for each single flow correlator, evaluated for the appropriate operator coordinates. Rather than writing out the full expression for $\tilde{F}_J(s)$ here, we will instead discuss an important property of the result. 

In particular, as in the case of the stress tensor (see~(\ref{Fstrip})), a crucial property is that 
\be
\tilde{F}_{J}\left(t+i\pi/2\right)+\tilde{F}_{J}\left(t-i\pi/2\right)=0,\qquad t\in\mathbb{R}.
\label{StripProperty}
\ee
This fact, along with the general analytic properties of $F(s)$, are what allow us to extract a higher spin QNEC constraint, as detailed in \ref{subsec:higherspin}.
\begin{figure}[t]
\centering
\includegraphics[scale=.3]{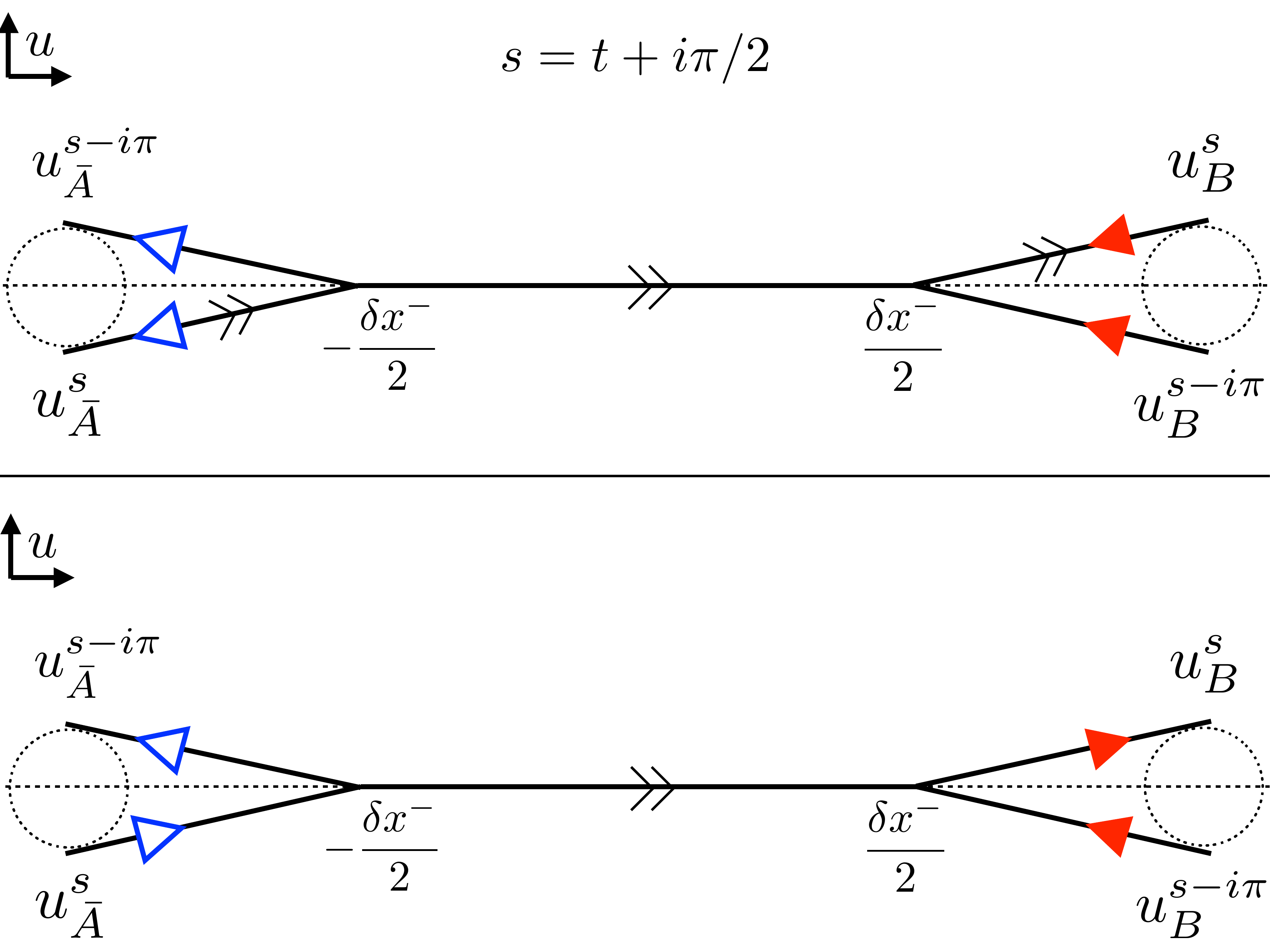}
\caption{\textbf{Top}: The piecewise integral contributions to $\tilde{F}_J(t+i\pi/2)$. The black double arrows are the contribution from the double flow correlator $f(s)$, the blue open arrows are the contribution from the single flow correlator $h_{\bar{A}}(s)$, and the red solid arrows are the contribution from the single flow correlator $h_B(s)$. The arrows point in the direction of integration. The dashed circles have radius $\frac{\delta x^-}{2}e^{-s}$, which is the size of the $s$-dependent shift we have introduced in the coordinates $u_{B,\bar{A}}$. \textbf{Bottom}: The net integral contributions to $\tilde{F}_J(t+i\pi/2)$.
\label{app-plot}}
\end{figure}
%
%
The property~(\ref{StripProperty}) is a consequence of several precise cancellations, as we now discuss.  

First note that $\tilde{F}_J(s)$ has an `integral part' involving piecewise integration in the complex $u$ plane and also a `displacement part' due to the displacement operator contributions. Let us focus on the integral part first. Fig.~\ref{app-plot} is useful for visualizing qualitatively how the cancellations happen. The top figure shows the piecewise nature of the integral part of $\tilde{F}_J(t+i\pi/2)$: the black double arrows are the contribution from the double flow correlator $f(s)$, the blue open arrows are the contribution from the single flow correlator $h_{\bar{A}}(s)$, and the red solid arrows are the contribution from the single flow correlator $h_B(s)$. A cancellation happens along the two legs of the figure that have contributions from both double and single flow.\footnote{One must use the fact that $F_J(x_2,x_1;u)=(-1)^{J-1}F_J(x_1,x_2;u)$.} The result is the bottom diagram in Fig.~\ref{app-plot}. In particular, the net contributions are such that the integral along a leg below the real axis is equal to the negative complex conjugate of the integral along the mirror image leg above the real axis. We depict this by using the same arrows in the upper and lower half-planes. Verifying this property explicitly in the expression for $\tilde{F}_J(t+i\pi/2)$ requires using reflection positivity, 
\be
\left(\left\langle \mathcal{J}_{-\cdots-}\left(u\right)\right\rangle _{\psi}\right)^{*}=\left\langle \mathcal{J}_{-\cdots-}\left(u^{*}\right)\right\rangle _{\psi}.
\ee 
Finally, when we sum all the legs, it follows that $\tilde{F}_J(t+i\pi/2)$ is purely imaginary.\footnote{The contribution along $-\frac{\delta x^-}{2}\leq u \leq \frac{\delta x^-}{2}$ is purely imaginary because of an overall factor of $e^{s(J-1)}$.} 

The last step in understanding~(\ref{StripProperty}) is to note that the left hand side picks out the real part of $\tilde{F}_J(t+i\pi/2)$, which we have just determined to vanish. This follows because $\tilde{F}_J(s)$ is real for real $s$ (a consequence of the probe operators $\mathcal{O}$ commuting for modular flow by real $s$), so by the Schwarz reflection principle $\tilde{F}(s^*)=(\tilde{F}(s))^*$. This completes our discussion of the integral part of $\tilde{F}_J(t+i\pi/2)$.

Now we turn to the displacement part of $\tilde{F}_J(s)$. For arbitrary $s$ and even spin $J$, these contributions combine to give
\begin{eqnarray}
\tilde{F}_J\left(s\right)|_{\text{disp}} & = & \sum_{\ell=1}^{J-1}\frac{\left(-1\right)^{J-\ell}}{\Gamma\left(J-\ell\right)\ell}\left(e^{s\ell}-\frac{1+\left(-1\right)^{\ell}}{2}\right)\times\nonumber \\
 &  & \left\{ \partial_{u}^{J-1-\ell}F\left(x_{B}^{s}\left(s_{B},-s_{A}\right),x_{\bar{A}}^{s};u\right)_{u=-\frac{\delta x^{-}}{2}}\left\langle \hat{\mathcal{D}}^{\ell}H_{A}^{\psi}\right\rangle _{\psi}\right.\nonumber \\
 &  & \left.-\partial_{u}^{J-1-\ell}F\left(x_{B}^{s},x_{\bar{A}}^{s}\left(s_{A},-s_{B}\right);u\right)_{u=+\frac{\delta x^{-}}{2}}\left\langle \hat{\mathcal{D}}^{\ell}H_{B}^{\psi}\right\rangle _{\psi}\right\}.
\end{eqnarray}
From this expression, one can verify~(\ref{StripProperty}) by keeping track of the overall $s$-dependence. The first observation is that the $\partial_u^n F_J(x_2,x_1;u)$ appearing in this formula have the following structure for even and odd number of derivatives (we are suppressing all $s$-independent prefactors):
\begin{eqnarray}
\partial_{u}^{\text{odd}}F_J\left(x_{B}^{s}\left(s_{B},-s_{A}\right),x_{\bar{A}}^{s};u\right)_{u=-\frac{\delta x^{-}}{2}} & \sim & e^{-s}\left(1+e^{-2s}+e^{-4s}+\dots\right) \nonumber\\
\partial_{u}^{\text{even}}F_J\left(x_{B}^{s}\left(s_{B},-s_{A}\right),x_{\bar{A}}^{s};u\right)_{u=-\frac{\delta x^{-}}{2}} & \sim & \left(1+e^{-2s}+e^{-4s}+\dots\right)
\end{eqnarray}
It follows that the $s$-dependence of the displacement contribution has the schematic form 
\begin{eqnarray}
\left.\tilde{F}_J\left(s\right)\right|_{\text{disp}}  \sim  \left[ e^{\left(J-1\right)s}\left\langle \hat{\mathcal{D}}^{(J-1)}H\right\rangle +\sum_{\text{odd }\ell\leq J-3}e^{\ell s}\left(\left\langle \hat{\mathcal{D}}^{(\ell)}H\right\rangle +\left\langle \hat{\mathcal{D}}^{(\ell+1)}H\right\rangle \right)\right] \times && \nonumber \\
 \left(1+e^{-2s}+e^{-4s}+\dots\right). &&
\end{eqnarray}
The strictly odd powers of $e^s$ within the square brackets ensure~(\ref{StripProperty}).

With~(\ref{StripProperty}) and the analytic properties of $F(s)$ in hand, one can prove a higher spin QNEC for $\mathcal{J}$. We refer the reader back to the main text in \ref{subsec:higherspin} for that discussion.

\bibliographystyle{JHEP}
\bibliography{QNEC}

\end{document}